\documentclass[a4paper,fleqn,usenatbib]{mnras} % usenatbib

\usepackage{latexsym}
\usepackage{amssymb}
\usepackage{amsfonts}
\usepackage{amsmath}
\usepackage{bm}
\usepackage{graphicx}
\usepackage{subfigure}
\usepackage{times}
\usepackage{units}
\usepackage{hyperref}
\usepackage{multirow}
\usepackage{comment}
\usepackage{ulem}
\usepackage{hyperref}
\usepackage{graphicx}
\usepackage{acronym}
\usepackage{xcolor}
\usepackage{booktabs}

\usepackage{pifont} % \cmark \xmark % http://ctan.org/pkg/pifont

\usepackage{enumitem}
\setlist[itemize]{noitemsep, topsep=0pt}

\usepackage{multirow}
\usepackage{morefloats}
\usepackage{mathrsfs}
\usepackage{latexsym}
\usepackage{dcolumn}
\usepackage{lipsum}
\usepackage{mathtools}
\usepackage{cuted}
\renewcommand{\cite}[1]{\citep{#1}}
\renewcommand\eqref[1]{(\ref{#1})}

\newcommand{\GW}{GW170817}
\newcommand{\AT}{AT2017gfo} 
\newcommand{\GRB}{GRB170817A}
\newcommand{\swind}{spiral-wave wind}

\newcommand{\mr}{mass ratio} 
 
\newcommand{\bnc}{bounce} 

\newcommand{\rproc}{$r$-process}

\newcommand{\surname}[1]{\text{#1}}

\newcommand{\ie}{\textit{i.e.}}
\newcommand{\eg}{\textit{e.g.}}

\newcommand{\be}{\begin{equation}}
    \newcommand{\ee}{\end{equation}}
\newcommand{\bea}{\begin{eqnarray}}
    \newcommand{\eea}{\end{eqnarray}}
\newcommand{\bel}{\begin{align}}
    \newcommand{\eel}{\end{align}}

\def\Msun{{\rm M_{\odot}}}
\def\GMc2{{\rm G M_{\odot} c^{-2}}}

\def\eps{\epsilon}

\def\ccm{\,\text{cm}^{-3}}

\newacro{MRI}{magnetorotational instability}

\newacro{BH}{black hole}
\newacro{BNS}{binary neutron star}
\newacro{EM}{electromagnetic}
\newacro{EOS}{equation of state}
\newacroplural{EOS}[EOSs]{equations of state}
\newacro{GR}{general relativity}
\newacro{HD}{hydrodynamics}
\newacro{MHD}{magnetohydrodynamics}
\newacro{PIC}{particle-in-cell}
\newacro{BPL}{broken power law}
\newacro{SNR}{supernova remnant}
\newacro{BM}{Blandford \& McKee}
\newacro{ST}{Taylor-von Neumann-Sedov}
\newacro{FBOT}{fast blue optical transient}
\newacro{SSA}{synchrotron self-absorption}
\newacro{SED}{spectral energy distribution}
\newacro{GRHD}{general-relativistic hydrodynamics}
\newacro{GRMHD}{general-relativistic magnetohydrodynamics}
\newacro{GW}{gravitational wave}
\newacro{BZ}{Blandford–Znajek}
\newacroplural{GW}[GWs]{gravitational waves}
\newacro{LES}{large-eddy simulation}
\newacroplural{LES}[LES]{large-eddy simulations}
\newacro{GRLES}{general-relativistic large-eddy simulation}
\newacro{NR}{numerical relativity}
\newacro{NS}{neutron star}
\newacroplural{NS}[NSs]{neutron stars}
\newacro{GRB}{gamma-ray burst}%{$\gamma$-ray burst}
\newacroplural{GRB}[GRBs]{gamma-ray burst}
\newacro{kN}{kilonova}
\newacroplural{kN}[kNe]{kilonovae}
\newacro{SGRB}{short \ac{GRB}}%{short $\gamma$-ray burst}
\newacro{ISM}{interstellar medium}
\newacro{MNS}{massive neutron star}
\newacro{NSBH}{neutron star-black hole}
\newacro{NSE}{nuclear statistical equilibrium}
\newacro{SN}{supernova}
\newacroplural{SN}[SNe]{supernovae}
\newacro{CCSN}{core-collapse supernova}
\newacroplural{CCSN}[CCSNe]{Core-collapse supernovae}
\newacro{MP}{metal-poor}
\newacro{UFG}{ultra-faint dwarf galaxy}
\newacroplural{UFG}[UFGs]{ultra-faint dwarf galaxies}
\newacro{NRN}{nuclear reaction network}
\newacroplural{NRN}[NRNs]{nuclear reaction networks}
\newacro{RR}{reaction rate}
\newacroplural{RR}[RRs]{reaction rates}
\newacro{ODE}{ordinary differential equation}
\newacroplural{ODE}[ODEs]{ordinary differential equations}
\newacro{IR}{infrared}
\newacro{NIR}{near-infrared}
\newacro{FIR}{far infrared}
\newacro{LC}{light curve}
\newacroplural{LC}[LCs]{light curves}
\newacro{UV}{ultraviolet}
\newacro{LTE}{local thermodynamic equilibrium}
\newacro{TOV}{Tolman-Oppenheimer-Volkoff}
\newacro{RMS}{root mean square}
\newacro{MM}{multi-messenger}
\newacro{LK}{light curve}
\newacro{CBM}{circumburst medium}
\newacro{LF}{Lorentz factor}
\newacro{EFE}{Einstein’s field equations}
\newacro{ADM}{Arnowitt, Deser and Misner}
\newacro{IVP}{initial value problem}
\newacro{RHS}{right hand side}
\newacro{RMF}{relativistic mean-field}
\newacro{PDE}{partial differential equation}
\newacroplural{PDE}[PDEs]{partial differential equations}
\newacro{EM}{electromagnetic}
\newacro{HMNS}{hyper-massive neutron star}
\newacro{SMNS}{supra-massive neutron star}
\newacro{EATS}{equal time arrival surface}
\newacroplural{EATS}[EATSs]{equal time arrival surfaces}
\newacro{JWST}{James Webb Space Telescope}
\newacro{VLA}{Very Large Array}
\newacro{SKA}{Square Kilometre Array}
\newacro{DSA}{Diffusive Shock Acceleration}
\newacro{PN}{post-Newtonian}
\newacro{EOB}{effective-one-body}
\newacro{PC}{prompt collapse}
\newacro{LIGO}{Laser Interferometer Gravitational-Wave Observatory}
\newacro{MF}{magentic field}
\newacroplural{MF}[MFs]{magnetic fields}
\newacro{AGN}{active galactic nucleus}
\newacroplural{AGN}[AGNs]{active galactic nuclei}
\newacro{LOS}{line of sight}
\newacro{DE}{dynamical ejecta}
\newacro{SWW}{\swind{}}

\newacro{BW}{blast wave}
\newacroplural{BW}[BWs]{blast waves}

\renewcommand{\cite}[1]{\citep{#1}}
\defcitealias{Margalit:2021kuf}{\scshape MQ21} % \citetalias{Margalit:2021kuf}
\defcitealias{Nedora:2021eoj}{\scshape N21} % \citetalias{Nedora:2021eoj}

\title[Kilonova afterglow modeling]{
    Modeling
    kilonova afterglows: Effects of the thermal electron
    population and interaction with GRB outflows
}

\author[V.\ Nedora et al.]{%V. Nedora et al.
    Vsevolod \surname{Nedora}$^{1,2}$, 
    Tim \surname{Dietrich}$^{1,2}$,
    Masaru \surname{Shibata}$^{1,3}$,
    Martin \surname{Pohl}$^{2,4}$, 
    Ludovica \surname{Crosato Menegazzi}$^{1}$ % Ludovica Crosato Menegazzi
    \\
    ${}^1$Max Planck Institute for Gravitational Physics (Albert Einstein Institute), Am M{\"u}hlenberg 1, Potsdam 14476, Germany\\
    ${}^2$Institute for Physics and Astronomy, University of Potsdam, Potsdam 14476, Germany\\
    ${}^3$Center for Gravitational Physics and Quantum Information, Yukawa Institute for Theoretical Physics, Kyoto University, Kyoto, 606-8502, Japan,\\
    ${}^4$Deutsches Elektronen-Synchrotron DESY, Platanenallee 6, 15738 Zeuthen, Germany
}

\date{%\today
    Accepted XXX. Received YYY; in original form ZZZ
}

\begin{document}
\label{firstpage}
\maketitle

\date{\today}

\begin{abstract}
    Given an increasing number of gamma-ray bursts accompanied by 
    potential kilonovae there is a growing importance to 
    advance modelling of kilonova afterglows. 
    In this work, we investigate how the presence of two electron 
    populations that follow a Maxwellian (thermal) and a power-law
    (non-thermal) distributions affect kilonova afterglow light curves. 
    We employ semi-analytic afterglow model, 
    \texttt{PyBlastAfterglow}. 
    We consider 
    kilonova ejecta profiles 
    from ab-initio numerical relativity binary neutron star merger 
    simulations, targeted to GW170817. 
    We do not perform model selection. 
    We find that the emission from thermal electrons dominates 
    at early times. If the interstellar medium density is 
    high (${\simeq}0.1\,\ccm$) it adds an early time peak to the light curve. 
    As ejecta decelerates the spectral and temporal indexes change 
    in a characteristic way that, if observed, can be used to
    reconstruct the ejecta velocity distribution. 
    For the low interstellar medium density, inferred for GRB 170817A, 
    the emission from the non-thermal electron population generally dominates. 
    We also assess how kilonova afterglow light curves 
    change if the interstellar medium has been partially removed 
    and pre-accelerated by laterally expanding gamma-ray burst ejecta.  
    For the latter we consider properties informed by observations of GRB170817A. 
    We find that the main effect is the emission suppression at early time 
    ${\lesssim}10^{3}\,$days, and at its maximum it reaches ${\sim}40\%$ 
    when the fast tail of the kilonova ejecta moves subsonically through the wake of laterally 
    spreading gamma-ray burst ejecta. 
    The subsequent rebrightening, when these ejecta break through and 
    shocks form, is very mild (${\lesssim}10\%$), 
    and may not be observable. 
\end{abstract}


\section{Introduction}
\label{sec:intro}

Formed in a binary, compact objects, \eg, \acp{NS} and \acp{BH}, inspiral 
and merge due to emission of \acp{GW}. Compact binary mergers in which at 
least one of the constituents is a \ac{NS} can
lead to ejection of matter with varying properties and at 
various timescales 
\citep[\eg][]{Shibata:2019wef,Radice:2020ddv,Bernuzzi:2020tgt}. 
Given the high neutron fraction of this material, such outflows allow for a 
rapid neutron capture (\rproc{}) nucleosythesis \citep[\eg][]{Wanajo:2014wha,Barnes:2016umi,Kasen:2017sxr,Tanaka:2017qxj,Miller:2019dpt,Bulla:2019muo}. 
Heavy nuclei produced in this process are unstable to the 
$\beta$-decay \cite{Rolfs:1988}. Before reaching the valley of stability they 
release energy that, with a certain efficiency, thermalises 
and can be observed as a quasi-thermal counterpart to \ac{BNS} or 
\ac{NSBH} mergers, called \ac{kN} \cite{Arnett:1982,Metzger:2010,Metzger:2016pju,Metzger:2019zeh}. 
For decades \ac{NR} simulations with various complexity allowed us to 
assess the properties of the ejected matter \cite{Hotokezaka:2013b,Bauswein:2013yna,Sekiguchi:2015dma,Sekiguchi:2016bjd,Dietrich:2016hky,Radice:2016dwd,Radice:2018pdn,Nedora:2020pak,Fujibayashi:2020dvr,Camilletti:2022jms,Fujibayashi:2022ftg}, 
and establish a tenuous link between the binary parameters and 
ejecta properties \cite{Dietrich:2016fpt,Kruger:2020gig,Nedora:2020qtd}.

Additionally, \ac{BNS} merger remnants are expected to 
be able to launch a relativistic jet. 
Possible mechanisms for jet launching include magnetic 
field–mediated energy extraction from a remnant spinning \ac{BH}
\cite{Blandford:1977ds,Komissarov:2009,Ruiz:2016rai},
magnetized winds from a remnant magnetar \cite{Bucciantini:2011kx,Zhang:2000wx}
or neutrino/antineutrino-powered fireballs \cite{Eichler:1989ve}. 
However, self-consistent, ab-inito \ac{NR} simulations of  
jet-formation are extremely challenging and so far were 
not able to produce jets with properties consistent with 
cosmological \acp{GRB}. 

For a subset of cosmological \acp{GRB},
the \ac{kN} emission, \ie, the \ac{IR} and \ac{NIR} excess, was 
found in the afterglow
\cite{Tanvir:2013pia,Berger:2013wna,Yang:2015pha,Jin:2016pnm,Jin:2017hle,Troja:2018ybt,Lamb:2019lao,Jin:2019uqr,Rastinejad:2022zbg} 
(see \eg, \citet{Fong:2017ekk,Klose:2019amd} for compiled data).  
However until $2017$ the observational data on the \ac{kN} ejecta 
was sparse due to large distances.
\GRB{}, accompanied by the \ac{GW} event, \GW{} and the \ac{kN} \AT{} was 
the closest short \ac{GRB} with the best sampled \ac{kN} until now
\cite{Savchenko:2017ffs,Alexander:2017aly,Troja:2017nqp,Monitor:2017mdv,Nynka:2018vup,Hajela:2019mjy}. 
Detected by Fermi \cite{TheFermi-LAT:2015kwa} and INTEGRAL 
\cite{Winkler:2011}, the \GRB{} was later followed up by a number of 
observatories across the world and across the \ac{EM} spectrum 
\cite{Arcavi:2017xiz,Coulter:2017wya,Drout:2017ijr,Evans:2017mmy,Hallinan:2017woc,Kasliwal:2017ngb,Nicholl:2017ahq,Smartt:2017fuw,Soares-santos:2017lru,Tanvir:2017pws,Troja:2017nqp,Mooley:2018dlz,Ruan:2017bha,Lyman:2018qjg}. 
Both numerical and semi-analytic models of \GRB{} hinted towards a 
non-trivial lateral structure of the \ac{GRB} ejecta  
\cite{Fong:2017ekk,Troja:2017nqp,Margutti:2018xqd,Lamb:2017ych,Lamb:2018ohw,Ryan:2019fhz,Alexander:2018dcl,Mooley:2018dlz,Ghirlanda:2018uyx}, 
created, at least in part, when the relativistic jet was drilling 
through the \ac{kN} ejecta \cite{Lamb:2022pvr}.

Kilonova models, both semi-analytic and based on the radiation 
transport, when applied to \AT{}, showed that several ejecta 
components with different properties are required to explain 
the observations 
\citep{Shibata:2017xdx,Siegel:2019mlp,Perego:2017wtu,Kawaguchi:2018ptg}. 
Specifically, the emission in high frequency bands, peaking 
within a day after the \ac{GW} trigger (\ie, ``blue kilonova''), 
requires low opacity, fast ejecta. Such ejecta 
is typically found in \ac{NR} simulations as a part of so-called 
dynamical ejecta, that forms shortly prior and during the merger 
\citep[\eg][]{Hotokezaka:2013b,Bauswein:2013yna,Radice:2016dwd,Radice:2018pdn,Fujibayashi:2022ftg}
and in secular ejecta (post-merger winds)  \citep[\eg][]{Beloborodov:2008nx,Lee:2009uc,Fernandez:2013tya,Dessart:2008zd,Perego:2014fma,Just:2014fka,Fernandez:2015use,Aasi:2013wya,Radice:2018xqa,Nedora:2020pak,Fujibayashi:2020qda}. 
The properties of these ejecta are set by a range of entangled 
phyical processes operating in a strong-field regime and at 
densities many times the nuclear saturation density. 
Importantly, the properties of matter in such conditions are 
not well understood and present one of the biggest 
multidisciplinary open questions.

\ac{NR} simulations show that within the velocity distribution of 
dynamical ejecta, there is ${\sim}(10^{-6}-10^{-5})\,\Msun$ of matter 
ejected at very high velocities (${\gtrsim}0.8\,c$) 
\citep{Hotokezaka:2013b,Metzger:2014yda,Hotokezaka:2018gmo,Radice:2018pdn,Radice:2018ghv,Nedora:2021eoj}.
The mechanisms behind this fastest eject include the 
shocks launched at core bounces \citep{Hotokezaka:2013b,Radice:2018pdn} and 
shocks generated at the collisional interface \cite{Bauswein:2013yna}. 
% shocks produced when spiral arms, formed after merger, collide \cite{Fujibayashi:2023}
Thus, properties of this ejecta component encode the information 
about early postmerger dynamics that is of particular interest  
for determining the remnant fate and \ac{EOS} properties. 
However, given the small amount of this ejecta component  
it is difficult to obtain its properties in \ac{NR} simulations. 
Moreover, being low mass and fast, 
it is affected by the presence of artificial atmosphere in a 
\ac{NR} simulation domain \cite{Fujibayashi:2022ftg}.

Additional ejecta from the postmerger disk can occure on longer
timescales~\citep{Perego:2014fma,Just:2014fka,Kasen:2014toa,Metzger:2014ila,Wu:2016pnw,Siegel:2017nub,Fujibayashi:2017puw,Miller:2019dpt,Nedora:2020pak}; 
Neutrino irradiation can lead to the ejection of ${\sim}5$\% 
of the disk with velocities ${\lesssim}0.08\,c$ from the polar
region~\citep{Perego:2014fma,Martin:2015hxa}.
A large fraction of the disk, ${\lesssim}40\%$, can become 
unbound on time scales ${\gtrsim}100\,$ms due to 
magnetic-field induced viscosity and/or nuclear recombination \citep{Dessart:2008zd,Fernandez:2014bra,Wu:2016pnw,Lippuner:2017bfm,Siegel:2017nub,Fujibayashi:2017puw,Radice:2018xqa,Fernandez:2018kax,Miller:2019dpt}.
Spiral density waves, driven by dynamical instabilities in the 
postmerger remnant can generate a characteristic wind, so-called \swind{} 
\citep{Nedora:2019jhl,Nedora:2020pak}.  
These secular ejecta are expected to have velocities ${\lesssim}0.05{-}0.2\,$
and thus contribute to a very late afterglow, ${\sim}10^4\,$days. However, 
if present, the secular ejecta can give the dominant contribution to the
\ac{kN} \citep[\eg][]{Fahlman:2018llv}.

When the dynamical ejecta moves through the \ac{ISM}, shocks 
are generated and, in turn, non-thermal afterglow emission 
is produced. This \ac{kN} afterglow is phenomenologically  
similar to \ac{GRB} afterglows and \acp{SNR}. 
Behind shocks, the synchrotron radiation is produced by 
electrons gyrating around the magnetic field lines 
\citep[\eg][]{Kumar:2014upa,Nakar:2019fza}.
For non-relativistic shocks, the emission is expected to 
peak in radio band on a timescale of years, 
\ie, the deceleration timescale on which the ejecta slows 
down, accreting matter from the \ac{ISM}  \citep[\eg][]{Nakar:2011cw,Piran:2012wd,Hotokezaka:2015eja,Radice:2018pdn,Hotokezaka:2018gmo,Kathirgamaraju:2018mac,Desai:2018rbc,Nathanail:2020hkx,Hajela:2021faz,Nakar:2019fza}. 
For ejecta with non-uniform velocity distribution, 
however, the \ac{kN} afterglow is more complex and 
is defined by the collective dynamics of various fluid elements 
\cite{Hotokezaka:2015eja}. 
For instance, in the presence of a fast tail, the \ac{kN} afterglow 
emission may be detectable early, on a \ac{GRB} 
afterglow timescale, (\eg, tens-to-hundred of days) 
\cite{Hotokezaka:2018gmo,Nedora:2021eoj}.

So far, no \ac{kN} afterglow has been unambiguously detected  
despite the increasing number of \ac{GRB} observations, 
afterglow of which contains \ac{NIR} excess. 
Difficulties in detecting a \ac{kN} afterglow include very 
low luminosities and long timescales over which the transient 
evolves. For instance, even for the closest short \ac{GRB} 
detected so far, \GRB{}, the latest observations made 
$4.5\,$years after the burst with 
one of the most sensitive radio observatories, \ac{VLA}, 
showed that the radio emission has gone below the detection threshold  
\cite{Balasubramanian:2022sie}.  
However, the ability to detect \ac{BNS} and \ac{NSBH} 
mergers without relaying on the bright on-axis \ac{GRB}, \ie, 
via \acp{GW}, as well as new radio facilities with increasing sensitivity, 
such as ng\ac{VLA} \cite{Lloyd-Ronning:2017dci,Selina:2018vla,Corsi:2019} 
and \ac{SKA} \cite{Carilli:2004nx,Aharonian:2013av,Leung:2021ebv}, 
will potentially make the first \ac{kN} afterglow detection 
a reality within this decade. 
It is thus important to improve \ac{kN} afterglow modelling 
and update the expectations regarding future observations.

In this work, we study two aspects related to the afterglow. 

The first aspect we investigate relates to the presence 
of two electron populations, thermal and power-law populations, 
behind the shock. 
This is motivated by first principles \ac{PIC} simulations, 
which predict that most of the electrons behind a mildly 
relativistic shock follow a quasi-thermal energy distribution 
\cite{Park:2014lqa,Crumley:2018kvf,Pohl:2019nvw,Ligorini:2021lbj}. 
Additionally, recently discovered new type of transients, 
\acp{FBOT} \cite{Margalit:2021kuf,Ho:2021bjl} %\cite{Ho:2018emo,Ho:2020hwf} 
that are at least in part 
attributed to the emission from mildly relativistic shocks, 
displayed signatures of thermal electron population 
(\ie, steep spectrum \cite{Ho:2018emo}).

The second aspect that we investigate is how the \ac{kN} 
afterglow changes if the medium into which the \ac{kN} ejecta 
moves, has been modified by a 
passage of \ac{GRB} \ac{BW}.  
In this case we consider \ac{GRB} model that fits the 
observations of \GRB{} and the parameters of which lie 
within tolerance ranges inferred 
by other studies for this burst. 
Such \ac{kN}-\ac{GRB} \ac{BW} interaction is expected 
to produce observable features, such as late-time 
radio-flares \cite{Margalit:2020bdk}.

Regarding the initial \ac{kN} ejecta profile, we focus on those, 
inferred from ab-inito \ac{NR} simulations with advanced 
input physics that have both angular- and velocity dependence 
of ejecta properties. 
We neglect the change in \ac{kN} ejecta properties 
due to \ac{GRB} jet break out and we do no consider pollution 
of the polar region due to jet wall dissipation. 

We employ a semi-analytic model to describe the afterglow. 
This model is an extension of the one presented in 
\citet{Nedora:2021eoj} (hereafter \citetalias{Nedora:2021eoj}), 
called \texttt{PyBlastAfterglow}.
Thus, we focus the discussion on qualitative and limited 
quantitative analysis and leave a more rigorous numerical 
exploration to future work. 

The paper is organized as follows. 
In Sec.~\ref{sec:method} we describe the semi-analytic 
afterglow model and methods that we employ to compute the 
\ac{BW} dynamics and synchrotron radiation. 
In Sec.~\ref{sec:results} we describe the \ac{kN} afterglow 
spectra in the presence of two electron populations behind the 
shock, the observed \acp{LC} and spectral indexes. Then, we 
consider the \ac{CBM} density profile behind a \ac{GRB} 
\ac{BW} and the dynamics of the \ac{kN} \ac{BW} moving 
through it. 
Finally, in Sec.~\ref{sec:conclusion} we summarize and conclude 
the work.
Additionally, we compare \ac{GRB} and \ac{kN} afterglow 
\acp{LC} computed with \texttt{PyBlastAfterglow} with those 
available in the literature in App.~\ref{sec:app:comp_grb} and 
App.~\ref{sec:app:comp_kn} respectively.

\section{\ac{GRB} and kN afterglow model}\label{sec:method}

The key components of both \ac{GRB} and \ac{kN} afterglow 
modelling are 
(i) dynamics of the fluid; 
(ii) electron distribution and radiation;
(iii) evaluation of the observed emission. 
In this section we describe the formulations and 
methods we implement in \texttt{PyBlastAfterglow}, 
introducing them first in a general, model-independent 
way. 

We consider \ac{GRB} and \ac{kN} \acp{BW} separately. 
For the former, the static, constant density \ac{ISM} 
is always assumed. 
For a \ac{kN} \ac{BW} the medium into which it 
propagates has properties that depend on the angle \ie,
whether it is inside or outside the \ac{GRB} opening angle, 
and the distance to the \ac{GRB} \ac{BW} if it is inside. 
We call this medium \ac{CBM} to differentiate it 
from static \ac{ISM}, that the \ac{kN} \ac{BW} encounters 
if it moves outside the \ac{GRB} jet opening angle. 

For the sake of generality, we first derive the evolution 
equations for a \ac{kN} \ac{BW} that moves into the \ac{CBM} 
in Sec.~\ref{sec:method:dynamics:kn} and then for a  
laterally expanding \ac{GRB} \ac{BW} that moves 
into static \ac{ISM} in Sec.~\ref{sec:method:dynamics:grb}. 
Further, in Sec.~\ref{sec:method:dens_prof} we describe the 
exact form of the \ac{CBM} density profile we use. 
Then, in Sec.~\ref{sec:method:synch} we describe methods we 
use to compute comoving synchrotron emission from a power-law 
electron distribution only that we adopt for \ac{GRB} afterglow 
(Sec.~\ref{sec:method:synch_pl}) and from a combined Maxwell plus 
power-law electron distributions that we use for \ac{kN} afterglow 
(Sec.~\ref{sec:method:synch_th}). 
In Sec.~\ref{sec:method:discret} we introduce the specific coordinate 
system we employ, and how we discretize the \ac{GRB} and \ac{kN} 
ejecta (in Sec.~\ref{sec:method:grb} and Sec.~\ref{sec:method:ejecta} 
respectively). 
Finally, in Sec.~\ref{sec:method:eats} we describe how the radiation 
in the observer frame is computed, taking into account relativistic 
effects.

\subsection{Dynamics} \label{sec:method:dyn_kn}

The interaction between two fluids can be treated as a relativistic Riemann problem, 
in which shocks (rarefraction waves) are produced when the required 
conditions for velocities, densities, and pressures are satisfied; 
cf.~\citet{Rezzolla:2013} for a textbook discussion. 

This problem has been extensively studied semi-analytically 
with different levels of approximation 
\citep[\eg][]{Huang:1999di,Uhm:2006qk,Peer:2012,Nava:2013,Zhang:2018book,Ryan:2019fhz,Guarini:2021gwh,Miceli:2022efx}. 
Most models implicitly assume the uniform and static medium into which 
\ac{BW} is moving. 
In order to model the dynamics with a pre-accelerated and non-uniform medium 
in front of the \ac{BW}, modifications to standard 
formulations are required. 
Here, we briefly outline the derivation of the evolution equation. 
Notably, such formulation can be used for modelling the early 
\ac{GRB} afterglows, where the radiation front pre-accelerates 
\ac{ISM} in front of the 
shock \cite{Beloborodov:2002nkf,Nava:2013}. 
In the following we neglect the presence of the reverse shock 
for simplicity.
Also, it was shown than the reverse shock does not significantly 
alter the \ac{kN} afterglow \acp{LC} \cite{Sadeh:2022enp}.

The stress energy tensor for a perfect fluid in flat space-time 
reads 
\begin{equation} \label{eq:method:tmunu}
    T^{\mu\nu} = (\rho' c^2 + e' + p')u^{\mu}u^{\nu} + p'\eta^{\mu\nu}\, ,
\end{equation}
where $u^{\mu} = \Gamma(1,\beta)$ is the fluid four-velocity with 
$\Gamma$ being the \ac{LF} and $\beta=\sqrt{1-\Gamma^{-2}}$ is the  
dimensionless velocity (in units of $c$), 
$p' = (\hat{\gamma}-1)e'$
is the pressure, and 
$e'$ % $e'=\Gamma\rho'c^2$ % From Hang!! 
is the internal energy density,  
$\hat{\gamma}$ is the adiabatic index (also 
called the ratio of specific heats), and 
$\eta^{\mu\nu}$ is the metric with signature 
$\{-1,1,1,1\}$. 
Hereafter, prime denotes quantities in the comoving frame.

For the perfect fluid considered here, we assume $\hat{\gamma}=4/3$ 
if the fluid is ultra-relativistic and $\hat{\gamma}=5/3$ if it 
is non-relativistic. We employ the following, simplified 
relation between $\hat{\gamma}$ and $\Gamma$ \citep[\eg][]{Kumar:2003yt} 
\begin{equation}\label{eq:method:eos}
    \hat{\gamma} \approxeq \frac{4 + \Gamma^{-1}}{3} \, ,
\end{equation}
which satisfies these limits. 
A more accurate prescription can be inferred 
from numerical simulations \cite{Mignone:2005ns}. 

The $\mu=\nu=0$ component of the stress-energy tensor 
Eq.~\eqref{eq:method:tmunu}, then reads 
\begin{equation}
    T^{00} = \Gamma^2(\rho'c^2 + e' + p') - p' = \Gamma^2 \rho' c^2 + (\hat{\gamma}\Gamma^2-\hat{\gamma}+1)e' \, .
\end{equation}
Integrating it over the entire \ac{BW} (assuming it is 
uniform, \ie, is represented by a sufficiently thin shell; 
the so-called thin-shell approximation), one obtains 
\begin{equation} \label{eq:method:Etot}
    E_{\rm tot} = \int T^{00} dV = \Gamma c^2 \rho' V' + \Gamma_{\rm eff} e' V' = \Gamma c^2 m + \Gamma_{\rm eff}E_{\rm int}' \, ,
\end{equation}
where we introduced the effective \ac{LF} 
$\Gamma_{\rm eff} = (\hat{\gamma}\Gamma^2 - \hat{\gamma} + 1)/\Gamma$, 
(see also \citet{Nava:2013,Zhang:2018book,Guarini:2021gwh}),
the enclosed mass $m=\rho'V'$ with $V'$ being the comoving volume, 
and the co-moving internal energy, $E_{\rm int}' = e'V'$.

Similarly, the volume integral of the $\mu=i,\,\nu=0$ 
component of Eq.~\eqref{eq:method:tmunu} gives the total momentum 
\begin{equation} \label{eq:method:Ptot}
    P^i = \frac{1}{c}\int T^{i0}dV = c\Gamma\beta\Big(m+\hat{\gamma}\frac{E_{\rm int}'}{c^2}\Big) \, .
\end{equation}
If there are two colliding \acp{BW}, $1$ and $2$, the 
energy and momentum conservation give the properties of the 
final \ac{BW} as, 
\begin{equation}
    E_{\rm tot;f} = E_{\rm tot 1} + E_{\rm tot 2}; \hspace{3mm} P_{\rm f} = P_{1} + P_{2} \, .
\end{equation}
These equations are non-linear and have an analytic solution only in the 
case of relativistic \acp{BW}. In \citet{Guarini:2021gwh} they were used 
to predict the flares in \ac{GRB} afterglows.

\subsubsection{Dynamics of a kN BW} \label{sec:method:dynamics:kn}

\def\cbm{\rm CBM}

As ejecta moves through the medium it accumulates mass 
$dm$ and losses a fraction of its energy to radiation, 
$dE_{\rm rad}'$. 
Then, the change of the total energy of a \ac{BW} is,  
\begin{equation}\label{eq:method:dEtot}
    d[\Gamma(M_0 + m) c^2 + \Gamma_{\rm eff} E_{\rm int}'] = \Gamma_{\cbm} dm c^2 + \Gamma_{\rm eff}dE_{\rm rad}'\, ,
\end{equation}
where $M_0$ is the initial mass of the \ac{BW} and 
$\Gamma_{\cbm}$ is the \ac{LF} of the \ac{CBM} medium. 
We recall here that if \ac{kN} ejecta moves behind the 
\ac{GRB} \ac{BW} it encounters the \ac{CBM} with 
density profile that depends on the properties of 
the \ac{GRB} \ac{BW} (see Sec.~\ref{sec:method:dens_prof}).

The internal energy $dE_{\rm int}'$ of the fluid behind the 
forward shock changes according to
\begin{equation} \label{eq:method:dEdEdE}
    dE_{\rm int}' = dE_{\rm sh}' + dE_{\rm ad}' + dE_{\rm rad}'\, ,
\end{equation}
where $ dE_{\rm ad}'$ is the energy lost to adiabatic expansion, 
$dE_{\rm sh}'$ is the random kinetic energy produced at the shock 
due to inelastic collisions \cite{Blandford:1976} 
with element $dm$ of the \ac{CBM}. 
From the Rankine-Hugoniot jump conditions for the cold upstream 
medium it follows that in the post-shock frame the average 
kinetic energy per unit mass 
is constant across the shock and equals $(\Gamma_{\rm rel}-1)c^2$, 
where $\Gamma_{\rm rel} = \Gamma\Gamma_{\cbm}(1-\beta\beta_{\cbm})$
is the relative \ac{LF} between upstream and downstream. 
Thus, we have
\begin{equation} \label{eq:method:dEsh}
    dE_{\rm sh}' = (\Gamma_{\rm rel} - 1)c^2 dm \, . 
\end{equation}
Adiabatic losses, $dE_{\rm ad}'$, 
can be obtained from the first law of thermodynamics, 
$dE_{\rm int}' = TdS - pdV'$, for an adiabatic process, \ie, $TdS=0$. 
Recalling that $p' = (\hat{\gamma}-1)E_{\rm int}'/V'$, we write 
\begin{equation}\label{eq:method:dEad}
    dE_{\rm ad}' = -(\hat{\gamma}-1)E_{\rm int}' d \ln V' \, . 
\end{equation}
As $V'\propto R^3 \Gamma_{\cbm}/\Gamma_{\rm rel}$, 
the radial derivative $d \ln V'/dR$ reads 
\begin{equation}\label{eq:method:dlnV}
    \frac{d\ln V'}{dR} = \frac{1}{m}\frac{dm}{dR} - \frac{1}{\rho}\frac{d\rho}{dR} - \frac{1}{\Gamma_{\rm rel}}\frac{d\Gamma_{\rm rel}}{d\Gamma}\frac{d\Gamma}{dR} + \frac{1}{\Gamma_{\cbm}}\frac{d\Gamma_{\cbm}}{dR} \, .
\end{equation}
The equation for the internal energy, Eq.~\eqref{eq:method:dEdEdE},
can then be obtained using Eq.~\eqref{eq:method:dEsh} and Eq.~\eqref{eq:method:dEad} 
(with Eq.~\eqref{eq:method:dlnV} plugged in). 
Notably, the internal energy can also be computed 
integrating the momenta of hadrons and leptons 
\citep{Dermer:2000gu,Nava:2013,Miceli:2022efx}. 

Combining the result with Eq.~\eqref{eq:method:dEtot}, 
we obtain the evolution equation for the \ac{BW} \ac{LF}
\begin{equation} \label{eq:method:dGdR}
    \begin{aligned}
        \frac{d\Gamma}{dR} &= \frac{ -(\Gamma-\Gamma_{\cbm} + \Gamma_{\rm eff} (\Gamma_{\rm rel}-1)) }{ (M_0 + m)c^2 + \frac{d\Gamma_{\rm eff}}{d\Gamma}E_{\rm int}' + \Gamma_{\rm eff}(\hat{\gamma}-1)E_{\rm int}' 
            \frac{d\Gamma_{\rm rel}}{d\Gamma}\frac{1}{\Gamma_{\rm rel}} } \\
        &+ 
        \frac{\Gamma_{\rm eff}(\hat{\gamma}-1)E_{\rm int}' \big( \frac{dm}{dR}\frac{1}{m} - \frac{d\rho_{\cbm}}{dR}\frac{1}{\rho_{\cbm}} -
            \frac{d\Gamma_{\cbm}}{dR}\frac{1}{\Gamma_{\cbm}} \big)}{ (M_0 + m)c^2 + \frac{d\Gamma_{\rm eff}}{d\Gamma}E_{\rm int}' + \Gamma_{\rm eff}(\hat{\gamma}-1)E_{\rm int}' 
            \frac{d\Gamma_{\rm rel}}{d\Gamma}\frac{1}{\Gamma_{\rm rel}} }\, .
    \end{aligned}
\end{equation}
In our implementation, in Eq.~\eqref{eq:method:dGdR} 
the internal energy term, $E_{\rm int}'$, is evaluated according to 
Eq.~\eqref{eq:method:dEdEdE}, neglecting the radiative losses $dE_{\rm rad}'$, 
as they are not of prime importance for the problem we consider. 
However, the radiative losses can easily be added, 
as $dE_{\rm rad} = -\eps_{\rm rad}\epsilon_e dE_{\rm sh}$,
where $\epsilon_e$ is the fraction of energy dissipated by the 
shock, that is gained by leptons 
which radiate a fraction $\eps_{\rm rad}$ of their internal energy 
\cite{Nava:2013,Miceli:2022efx}. 

Equation~\eqref{eq:method:dGdR} describes the evolution of the \ac{BW} \textit{bulk} \ac{LF}\footnote{
    Also sometimes labelled as $\Gamma=\Gamma_{21}$, 
    the relative Lorentz factor of plasma in region 
    behind the shock (region $2$) with respect to region ahead of the shock, 
    (region $1$) in commonly used notations \cite{Kumar:2014upa,Nava:2013,Zhang:2018book}. 
}, \ie, ``dynamical'' average of \acp{LF} at which different 
regions (behind the shock) are moving \cite{Blandford:1978}.  
Using the expression for $\hat{\gamma}$ (Eq.~\eqref{eq:method:eos})  
the derivative, $d\Gamma_{\rm eff}/d\Gamma$, can be obtained analytically 
as $d\Gamma_{\rm eff}/d\Gamma = (\hat{\gamma} \Gamma^2 + \hat{\gamma} - 1) / \Gamma^2$.

The amount of mass that the \ac{BW} sweeps, is 
\begin{equation}\label{eq:method:dmdr_kn}
    \frac{dm}{dR} = 2 \pi \rho_{\cbm} \big(1-\cos(\omega)\big) R^2 \, , %_{\rm ISM}
\end{equation}
where $\omega$ is the \ac{BW} half-opening angle around its symmetry axis, 
\ie, $2\pi(1-\cos{(\omega)})$ is the fraction of the 
$4\pi$ solid angle that the \ac{BW} occupies. 
For the \ac{kN} \ac{BW} $\omega$ is constant throughout the evolution 
and is determined by the \ac{kN} ejecta discretization  
(see Sec.~\ref{sec:method:discret}). 

Solving together Eqs.~\eqref{eq:method:dEdEdE}, Eq.~\eqref{eq:method:dGdR}, and 
Eq.~\eqref{eq:method:dmdr_kn}, we obtain the dynamical evolution of the 
\ac{kN} \ac{BW}. 
Expressions for 
$\rho_{\cbm}$, $\Gamma_{\rm REL}$, and $\Gamma_{\cbm}$ 
are discussed later, in Sec.~\ref{sec:method:dens_prof}.

\subsubsection{Dynamics of a GRB BW} \label{sec:method:dynamics:grb}

For a \ac{GRB} \ac{BW} we assume that the medium, into 
which these ejecta is moving is at rest and uniform, \ie, 
the \ac{ISM} with $\rho_{\rm ISM} = n_{\rm ISM} m_p$, 
where $n_{\rm ISM}$ is the number density and $m_p$ is 
the proton mass. 
Then in Eq.~\eqref{eq:method:dGdR} we have 
$\Gamma_{\cbm} = 1$, 
$\Gamma_{\rm rel}=\Gamma$, 
$d\Gamma_{\rm rel}/d\Gamma = 1$,
$d\Gamma_{\cbm}/dR = 0$, and 
$d\rho_{\cbm}/dR = 0$; 
and the evolution equation for $\Gamma$ becomes, 
\begin{equation} \label{eq:method:dGdRold}
    \frac{d\Gamma}{dR} = \frac{ -(1+\Gamma_{\rm eff})(\Gamma-1) + \Gamma_{\rm eff}(\hat{\gamma}-1)E_{\rm int}'   \frac{dm}{dR}\frac{1}{m}   }{ (M_0 + m)c^2 + \frac{d\Gamma_{\rm eff}}{d\Gamma}E_{\rm int}' + \Gamma_{\rm eff}(\hat{\gamma}-1)E_{\rm int}' 
        \frac{1}{\Gamma} }\, .
\end{equation}
Equation~\eqref{eq:method:dGdRold} is similar to the 
equation~(8.66) of \citet{Zhang:2018book} 
and equation~(7) of \citet{Nava:2013}. 
We compare the \ac{BW} $\Gamma$ evolution computed with 
Eq.~\eqref{eq:method:dGdRold} with the model of \citet{Peer:2012} 
and \citet{Ryan:2019fhz} in App.~\ref{sec:app:dyn} for completeness.

Within a radially evolving collimated \ac{GRB} \ac{BW}, 
the pressure gradient perpendicular to the normal to the \ac{BW} surface 
leads to its lateral expansion \citep[\eg][]{vanEerten:2009pa,Granot:2012,Duffell:2018iig}. 
Indeed, as the transverse pressure gradient adds the velocity 
along the tangent to the surface, the \ac{BW}'s lateral expansion 
sets in. 
The spreading is negligible when the \ac{BW} is relativistic, but as  
it decelerates, more fluid elements come into casual contact with 
each other redistributing energy and pressure gradient; the spreading 
accelerates.

Several prescriptions for a \ac{BW} lateral spreading exist 
in the literature. For instance, \citet{Granot:2012} parameterized 
the lateral expansion as 
\begin{equation} \label{eq:method:dthetadr_GP12}
    \frac{ d \omega }{d R} = R^{-1}\Gamma^{-1-a}\, .%\tan(\theta)^{-a},%\beta\, ,
\end{equation}
In our implementation we use $a=1$, following \citet{Fernandez:2021xce}. 
The spreading is computed after the \ac{BW} starts to decelerate, \ie, 
$R > R_{\rm d}$, where the deceleration radius, $R_d$, is 
\begin{equation}
    R_{\rm d} = \Big(\frac{3E_0}{4\pi \rho_{\rm ISM} \Gamma^2 c^2} \Big)^{1/3}\, ,
\end{equation}
$E_0$ and $\Gamma_0$ are the initial kinetic energy and \ac{LF} of the \ac{BW}. 
Once the \ac{BW} become spherical, $\omega=\pi/2$, 
the spreading is stopped.
For completeness we also compare this prescription with 
others available in the literature in 
App.~\ref{sec:app:spread}.

As the \ac{BW} laterally spreads, the amount of mass 
it sweeps increases. %, \ie, $dm/dR\propto\theta$. 
We follow \citet{Granot:2012} and write
\begin{equation}\label{eq:method:dmdr_grb}
    \frac{dm}{dR} = 2 \pi \rho_{\rm ISM} \Big[ \big(1-\cos(\omega)\big) + \frac{1}{3}\sin(\omega)R\frac{d\omega}{dR} \Big] R^2 \, .
\end{equation}

Solving Eqs.~\eqref{eq:method:dEdEdE},
\eqref{eq:method:dGdRold},
\eqref{eq:method:dthetadr_GP12},
\eqref{eq:method:dmdr_grb}
we obtain the dynamical evolution of the \ac{GRB} \ac{BW}. 

\subsubsection{Density profile behind the GRB \ac{BW}} \label{sec:method:dens_prof}

For both \ac{kN} and \ac{GRB} \acp{BW} the conditions 
at the shock are obtained using the Rankine-Hugoniot conditions 
(mass, energy, momentum conservation).
For the strong shock and cold \ac{ISM}, the downstream density reads 
$    \rho' = (\hat{\gamma} \Gamma + 1) / (\hat{\gamma} - 1) \rho $
where $\rho$ is equal to $\rho_{\cbm}$ for \ac{kN} \acp{BW} that 
move behind the \ac{GRB} \ac{BW} or it is equal to $\rho_{\rm ISM}$ 
otherwise. 
The shock front \ac{LF}\footnote{ 
    denoted as $\gamma_{1s}$ in \citet{Zhang:2018book} } 
is
\begin{equation}
    \Gamma_{\rm sh} = \frac{ (\Gamma + 1) [ \hat{\gamma}(\Gamma - 1) + 1 ] }{ \hat{\gamma}(2-\hat{\gamma})(\Gamma-1)+2 }\, .
\end{equation}
In the ultra-relativistic case the shock compression ratio, 
$\rho'/\rho_{\cbm} = 4 \Gamma$, and the shock \ac{LF} then 
is $\Gamma_{\rm sh} = \sqrt{2}\Gamma$, \ie, 
the shock front travels slightly faster than the downstream fluid. 
In turn, the radius of the shock can be obtained from   
$    dR / dt_{b} = \beta_{\rm sh} c $
where $t_{b}$ is the time in the burster's static frame 
and $ dR / dt_{\rm comov} = dR/dt' = \beta_{\rm sh} \Gamma c $
is the time in the frame comoving with the fluid, 
where $\beta_{\rm sh}$ is the shock velocity in the 
progenitor frame. 

\begin{figure}
    \centering 
    \includegraphics[width=0.49\textwidth]{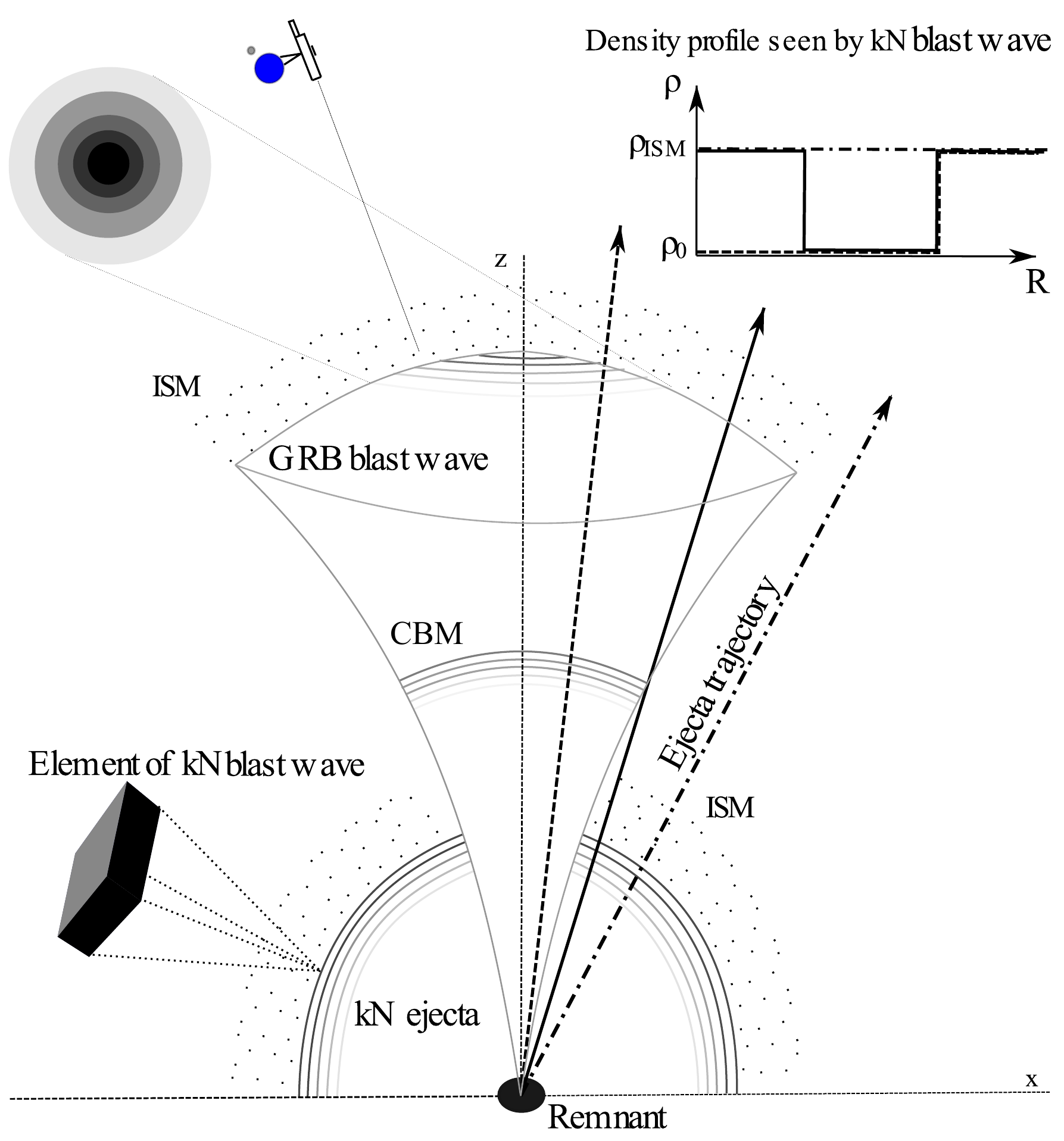}
    \caption{
        Schematic illustration of the model with 
        \ac{GRB} and \ac{kN} \acp{BW}. 
        Concentric circles in the top left part of the figure indicate the axial 
        symmetry of the \ac{GRB} and \ac{kN} \acp{BW}
        The black box in the bottom left part of the figure indicates the 
        discretization of the both ejecta types.  
        % each of which is evolved independently. 
        The little black dots arranged along the \ac{GRB} and \ac{kN} 
        \ac{BW} outer surfaces indicate the constant density, static 
        \ac{ISM}. 
        The possible trajectory for an elementary \ac{kN} \ac{BW} 
        depend on whether it 
        (i) avoids the \ac{CBM} medium entirely (dotted line), 
        (ii) moves behind the \ac{GRB} \acp{BW} from the start (dashed line) 
        interacting with the \ac{CBM}, 
        (iii) or enters the \ac{CBM} region 
        during/after \ac{GRB} \ac{BW} lateral expansion. 
        For all three cases the schematic \ac{kN} \ac{BW} 
        upstream density profile is shown in the upper right 
        part of the figure, normalized to the \ac{ISM} value. 
        The system is observed off-axis. 
    }
    \label{fig:method:structure}
\end{figure}

When considering the interaction between \ac{kN} and \ac{GRB} \acp{BW}, 
we assume that the reverse shock has already crossed the \ac{GRB} ejecta 
when the interaction starts. In other words, the density profile that 
\ac{kN} \ac{BW} encounter is generated by the forward 
shock within the \ac{GRB} \ac{BW}. 
We reiterate that we neglect the effect of \ac{GRB} ejecta break out 
from the \ac{kN} ejecta on the properties of the latter. 
Currently, such processes are studied with numerically expensive \ac{GRMHD} 
simulations \citep[\eg][]{Gottlieb:2022sis} and are not well understood. 
We leave it to future work to assess how the \ac{GRB} shock breakout change 
the \ac{kN} afterglow. 

The \ac{CBM} density profile that \ac{kN} \ac{BW} interacts with depends 
on the properties of the \ac{GRB} \ac{BW}, as shown in 
Fig.~\ref{fig:method:structure}. Specifically, when 
\ac{GRB} \ac{BW} are ultra-relativistic, the profile behind the shock front 
follows the \acl{BM} solution \cite{Blandford:1976}. When the \ac{BW} 
decelerates to $\Gamma\simeq1$, the downstream profile may be 
approximated with the \acl{ST} solution \cite{Sedov:1959}.  
Since the \ac{kN} \ac{BW} is at most mildly relativistic, any interactions 
with the \ac{GRB} \ac{BW} will happen when the latter is slower, 
\ie, also mildly relativistic at most. 
Thus, we assume that the density profile that the \ac{kN} \ac{BW} 
encounters, moving behind the \ac{GRB} \ac{BW} is given by the \acl{ST} 
and reads%, \citet{Book:1994} 
\begin{equation}\label{eq:method:st_rho_prof}
    \rho_{\cbm}(r) = \rho' \mathcal{D}(\eta), \hspace{2mm}
    \beta_{\cbm}(r) = \frac{2 \beta_{\rm sh} \mathcal{V}(\eta)}{(\hat{\gamma}+1)}, \hspace{2mm}
    P_{\cbm}(r) = p' \mathcal{P}(\eta)\, ,
\end{equation}
where $\eta = r / R_{\rm sh}$, $\mathcal{D}$, $\mathcal{V}$, and $\mathcal{P}$ 
are given by equations (9), (10), and (11) in \citet{Book:1994}. 
Here $R_{\rm sh}$, $\rho'$, and $\beta_{\rm sh}$ 
denote the radius, density, and velocity at the shock  
computed with the formalism discussed above. 
We turn the \acl{ST} profile on when the \ac{GRB} \ac{BW} is slowed down to 
$\Gamma\sim2$. Otherwise, if the \ac{kN} \ac{BW} moves behind the 
\ac{GRB} one it experiences negligible upstream density, $\rho_{\cbm}\sim0$. 
Since the \ac{GRB} \ac{BW} spreads laterally, it is possible that the 
\ac{kN} one would enter the evacuated region later. For numerical reasons 
we assume that from the point of entry the $\rho_{\cbm}$ decreases 
exponentially, until the \acl{ST} profile takes over.  
Importantly, in our model we neglect the tail-on shock-shock collision 
itself, when two \acp{BW} catch up with each other.

Numerically, we solve the system of \acp{ODE} using 
explicit Runge-Kutta method of order $8(5,3)$ \cite{Prince:1981}. 
We include the adaptive step-size control as the system of 
\acp{ODE} becomes stiff, once \ac{kN} ejecta enters 
the low-density environment.

\subsection{Comoving synchrotron} \label{sec:method:synch}

In the previous derivation we implicitly assumed that \acp{BW} are not 
magnetized. However, as a \ac{BW} moves through the \ac{ISM} with small 
but finite magnetization, the magnetic fields may become amplified 
via several instabilities \eg, 
the current-driven instability \cite{Reville:2006px}, 
the Kelvin-Helmholtz shear instability \cite{Zhang:2011}, the
Weibel (filamentation) instability \cite{Medvedev:1999tu,Lemoine:2010,Tomita:2016yib}
the \v{C}erenkov resonant instability \cite{Lemoine:2010}, 
the Rayleigh-Taylor instability \cite{Duffell:2013tha}, 
the magneto-rotational instability \cite{Cerda-Duran:2011}, 
or the pile-up effect \cite{RochadaSilva:2014ehi}. 
These processes are very complex and require high resolution, 
computationally expensive \ac{PIC} and \ac{MHD} simulations to study. 
In the \ac{GRB} literature it is common to assume that a fixed 
fraction $\epsilon_B$ of the \ac{BW} internal energy,  
$e' = E_{\rm int}' / V'$, 
is deposited in random magnetic fields behind the shock, 
\ie, $B'=\sqrt{8\pi \epsilon_B e'}$. We assume $B'$ to be 
constant behind the shock.

The incoming electrons gain energy while reflecting off 
and scattering on \ac{MHD} instabilities present in collisionless shocks. 
At the scale of the electron's gyro-radius, \ac{PIC} simulations 
are employed to study particle dynamics \citep[\eg][]{Sironi:2015oza}. 
At larger scales a coupled \ac{MHD}-\ac{PIC} approach is employed. However, 
the spatial and temporal extent of such simulations are still limited 
to a few $10^3$ of proton gyro-scales and few milliseconds 
\cite{Bai:2014kca,Mignone:2018per}. 
These studies show that the main process responsible for electron 
acceleration at collisionless shocks is the first-order Fermi acceleration 
\cite{Spitkovsky:2008fi,Sironi:2009,Sironi:2011,Park:2014lqa}.  
Due to the complexity and computational cost of these simulations 
it is common to assume that a fixed fraction, $\epsilon_e$, 
of the internal energy is used for particle acceleration, while 
electrons, after the acceleration, follow a power-law distribution in 
energy, $dn_{e}/d\gamma_{e}\propto\gamma_{e}^{-p}$ with $\gamma_{e}$ being the 
electron \ac{LF}, and $p$ being the spectral index 
\cite{Dermer:1997pv,Sari:1997qe}.

First-principle simulations provide constraints on the microphysics 
parameters, $\epsilon_B$, $\epsilon_e$, and $p$. Specifically, 
for relativistic shocks $p{\simeq}2$, while for non-relativistic ones 
$p{\simeq}2.2$ \cite{Kirk:1999km,Keshet:2004ch}. 
(See \citet{Sironi:2015oza,Marcowith:2020vho} for recent reviews).
Observations of \ac{GRB} afterglows also provide constraints on these 
parameters, but the range is generally very broad \cite{Kumar:2014upa}. 
We treat them as free parameters of the model.

\subsubsection{Comoving synchrotron from a GRB BW} \label{sec:method:synch_pl}

The \ac{BPL} electron spectrum has the following characteristic \acp{LF}. 
The maximum \ac{LF} $\gamma_{e; \,\rm max}'$ depends on how quickly 
an electron can gain energy in the acceleration process and how 
quickly it radiates it. 
In order to accelerate to a \ac{LF} $\gamma_e'$, an electron 
should not lose more than half of its energy to synchrotron 
radiation during the time required for acceleration. 
As the minimum time needed for electron acceleration is 
of the order of the Larmor time, 
$t_L=m_e c \gamma_{e}' / q_e B'$ \cite{Kumar:2014upa} 
\begin{equation}\label{eq:method:gamma_max}
    \gamma_{e;\, \rm max}' \simeq \sqrt{\frac{9 m_e^2 c^4}{8 B' q_e^3}} \, ,
\end{equation}
where $q_e$ and $m_e$ are the electron charge and mass.

Most of the electrons, however, are injected with 
$\gamma_{e;\,\rm min}'$, 
which can be obtained from the normalization of 
the electron distribution function. 
For the case of a simple \ac{BPL} and if 
$\gamma_{e;\,\rm max}' \gg\gamma_{e;\,\rm min}' $ as considered here, 
it can be obtained analytically \citep[\eg][]{Kumar:2014upa} 
\begin{equation}\label{eq:method:gm_pl_only}
    \gamma_{e; \, \rm min}' = \frac{p-2}{p-1} \frac{\epsilon_e e'}{n' m_e c^2}\, .
\end{equation}

The cooling of electrons is driven by radiation losses and adiabatic 
expansion \citep[\eg][]{Chiaberge:1998cv,Chiang:1998wh}. 
Thus, at any point in time behind the shock there is a population 
of newly injected, ``hot'', electrons and already partially cooled, 
``cold'', electrons. 
The exact evolution of the electron distribution function can be 
obtained by solving the continuity equation, the Fokker-Planck-type equation. 
This is however computationally expensive and in \ac{GRB} afterglow 
literature it is common to consider the ``fast'' and ``slow'' cooling regimes of 
the electron spectrum approximated with \acp{BPL}, depending on whether 
$\gamma_{e;\, \rm min}'$ is smaller or larger than a cooling \ac{LF} 
$\gamma_{e;\, \rm c}'$ 
defined as 
\begin{equation}\label{eq:method:gc_pl_only}
    \gamma_{e;\, \rm c}' = \frac{6 \pi m_e c }{\sigma_T t_{e} B'^2 \Gamma} \, ,
\end{equation}
where $\sigma_T$ is the Stefan-Boltzmann constant and $t_e$ is the emission time.
Using Eqs.~\eqref{eq:method:gamma_max}, \eqref{eq:method:gm_pl_only} and 
\eqref{eq:method:gc_pl_only}, we compute the time evolution of the electron 
spectrum, approximated with the \ac{BPL}. 
This spectrum, in turn, 
can be convolved with the synchrotron function \cite{RybickiLightman:1985} 
to derive analytically the instantaneous synchrotron spectrum which itself is a 
\ac{BPL} with critical frequencies: 
$\nu_{\rm min}'(\gamma_{e;\, \rm min}')$, $\nu_{\rm c}'(\gamma_{e;\, \rm c}')$, 
and $\nu_{\rm max}'(\gamma_{e;\,\rm max}')$ with varying degree of simplification  
\citep[\eg][]{Sari:1997qe,Dermer:1997pv,Wijers:1998st,Johannesson:2006zs}. 
We adopt the derivation of \citet{Johannesson:2006zs} that approximates the 
synchrotron spectrum as a smooth \ac{BPL} (their equations~A1, A2, A6 and A7), 
that we recall here for completeness,
\begin{equation} \label{eq:method:synch_joh06_spec}
    \begin{aligned}
        j_{\rm pl}'(\nu') =& j_{\rm pl;\, max;\, f}' \Bigg[\Big(\frac{\nu'}{\nu_c'}\Big)^{-\frac{\kappa_1}{3}} + \Big(\frac{\nu'}{\nu_c '}\Big)^{\frac{\kappa_1}{2}}\Bigg]^{-\frac{1}{\kappa_2}} \\
        & \times \Bigg[1 + \Big(\frac{\nu'}{\nu_m '}\Big)^{\frac{(p-1)\kappa_2}{2}}\Bigg]^{-\frac{1}{\kappa_2}}, \\
        j_{\rm pl}'(\nu') =& j_{\rm pl;\, max;\, s}' \Bigg[\Big(\frac{\nu'}{\nu_m'}\Big)^{-\frac{\kappa_1}{3}} + \Big(\frac{\nu'}{\nu_m '}\Big)^{\frac{\kappa_3(p-1)}{2}}\Bigg]^{-\frac{1}{\kappa_3}} \\
        & \times \Bigg[1 + \Big(\frac{\nu'}{\nu_c '}\Big)^{\frac{1}{2}\kappa_4}\Bigg]^{-\frac{1}{\kappa_4}},
    \end{aligned}
\end{equation}
for the fast and slow cooling, respectively. 
Here $j_{\rm pl}'(\nu')$ is the comoving emissivity from the
power-law electron population at comoving frequency $\nu'$. 
The characteristic frequencies are 
\begin{equation} \label{eq:method:synch_joh06_nu}
    \nu_i ' = \chi_p \gamma_{e;\,i}'^2 \frac{3 B'}{4 \pi m_e c}\, ,
\end{equation} 
and the 
$j_{\rm pl;\, max;\, f}'$ and $j_{\rm pl;\, max;\, s}'$ 
are the peak values of the spectrum for the fast and slow cooling 
regimes respectively, expressed as  
\begin{align} \label{eq:method:synch_joh06_pmax}
    j_{\rm pl;\, max;\, f}' &= 2.234 \phi_p \frac{q_e^3 n' B'}{m_e c^2}, \\
    j_{\rm p;\, max;\, s}' &= 11.17 \phi_p \frac{p-1}{3p-1}\frac{e^3 n' B'}{m_e c^2},
\end{align}
where, $\phi_p$, $\chi_p$, and $\kappa_i$ are fitting polynomials that 
capture the $p$-dependence \cite{Johannesson:2006zs},
and $n'$ is the number density behind the shock front 
computed from the shock jump conditions (Sec.~\ref{sec:method:dens_prof}). 

Using this formulation, we compute the synchrotron emission from 
a relativistic \ac{GRB} \ac{BW}. 
For completeness we compare it with other formulations 
available in the literature in App.~\ref{sec:app:synch}.

\subsubsection{Comoving synchrotron from a kN BW}\label{sec:method:synch_th}

When a shock is ultra-relativistic $\Gamma_{\rm sh} \gg 1$ or 
non-relativistic $\beta_{\rm sh}\ll 1$ the synchrotron emission 
from a non-thermal population of electrons can explain 
observations of \acp{GRB} afterglows and \acp{SNR}, respectively 
\cite{Sari:1997qe,Chevalier:1982}. 
However, in the case of mildly relativistic shocks, 
$\Gamma_{\rm sh}\beta_{\rm sh}\sim1$, numerical studies of electron 
acceleration at shocks show that most of the energy resides in the 
\textit{thermal electron population}, \ie, electrons that follow thermal, 
Maxwell-J{\"u}ttner distribution function, and that the non-thermal 
(power-law) tail only contains a small fraction of 
the total post-shock energy \cite{Park:2014lqa,Crumley:2018kvf}. 
Thermal electrons were shown to be important in explaining the 
peculiar steep optically-thin radio and mm spectra of the 
\ac{FBOT} AT2018cow \cite{Ho:2018emo}. 
But even before that, 
the thermal electron population was considered in application to 
\acp{GRB} afterglows \cite{Warren:2018lyx,Samuelsson:2020upt,Giannios:2009,Ressler:2017qjo}, 
and hot accretion flows \cite{Ozel:2000}. 
Recently, \citet{Margalit:2021kuf} (hereafter \citetalias{Margalit:2021kuf}) 
presented an analytic formulation 
of the synchrotron radiation arising from the combined thermal and 
non-thermal populations of electrons taking into account the 
\ac{SSA} in both populations and low-frequency corrections of 
emissivities. 
\citetalias{Margalit:2021kuf} considered a Maxwellian 
distribution function for thermal 
electrons and a power law for the non-thermal electrons. 

The pitch-angle averaged emission and absorption coefficients can be 
expressed in terms of $x_{\rm M} = \nu'/ \nu_{\Theta}'$, where  
$\nu_{\Theta}' = 3\Theta^2 e B' / 4\pi m_e c$.
For the thermal electron population emissivity and 
absorption coefficient read 
\begin{equation} \label{eq:method:j_th}
    j'_{\nu', \, \rm th} = 
    \frac{\sqrt{3}q_e^3 n' B'}{8\pi m_e c^2} 
    \times \frac{2\Theta^2}{K_2(1/\Theta)}
    x_{\rm M}
    I (x_{\rm M})\, ,
\end{equation}
\begin{equation}
    \alpha'_{\nu',\, \rm th} = 
    \frac{\pi q_e n'}{3^{3/2}\Theta^5 B'} 
    \times \frac{2\Theta^2}{K_2(1/\Theta)}
    x_{\rm M}^{-1} 
    I (x_{\rm M})\, ,
\end{equation}
where $\Theta$ is the dimensionless electron temperature, 
$\Theta=k_B T_e/m_e c^2$, 
$K_2(1/\Theta)$ is the modified Bessel function of second order, 
and $I(x_{\rm M})$ is the fitting function 
introduced in \citet{Mahadevan:1996cc} 
\begin{equation}\label{eq:method:Ixm}
    I(x_{\rm M}) = \frac{4.0505 a}{x_{\rm M}^{1/6}}
    \Big( 1 + \frac{0.40 b }{x_{\rm M}^{1/4}} +\frac{0.5316 g}{x_{\rm M}^{1/2}} \Big) \exp(-1.8899 x_{\rm M}^{1/3})\, ,
\end{equation}
which describes the emissivity of the thermal population 
of electrons for small and large $x_{\rm M} $ \cite{Pacholczyk:1970,Petrosian:1981}. 
The temperature-dependent coefficients $a$, $b$, $g$ 
are tabulated in \citet{Mahadevan:1996cc} for $\Theta\in(0.084, 5.40)$
or, equivalently, for $T\in(5\times10^8,3.2\times10^{10})\,$K. 
These coefficients deviate from unity for $\Theta < 5$ which 
is of relevance for the low-velocity elements of the \ac{kN} ejecta 
or after the ejecta deceleration.
Thus, we include this dependence in our implementation.

For the non-thermal electron population \citetalias{Margalit:2021kuf} 
considered the standard power-law spectrum $dn_e'/d\gamma_e'\propto\gamma_e'^{-p}$ 
with injection \ac{LF}, $\gamma_{e,\,\rm min}'$, equal to the mean 
\ac{LF} of thermal electrons, 
$\gamma_{e; \, \rm min}' = 1 + a(\Theta)\Theta$, where $a(\Theta)$ 
is the coefficient that varies between $3/2$ for non-relativistic 
electrons and $3$ for ultra-relativistic electrons and 
can be approximated as 
$a(\Theta) = 6 + 15\Theta / (4 + 5\Theta)$ \cite{Ozel:2000}. 
Thus, the power-law distribution contains only supra-thermal electrons.

As ejecta continues to decelerate and $\gamma_{e; \, \rm min}'\rightarrow 1$, 
it enters the so-called deep-Newtonian regime \cite{Sironi:2013tva}, 
that commences when $\beta_{\rm sh} \lesssim 8\sqrt{m_p/m_e}\bar{\epsilon}_e$, 
where $\bar{\epsilon}_e = 4\epsilon_e(p-2)/(p-1)$ \cite{Margalit:2020bdk}. 
Synchrotron emission from electrons accelerated at lower velocity shocks 
is dominated by electrons with \ac{LF} ${\simeq}2$, instead of 
those with $\gamma_{e;\, \rm min}'$. % as is the case for higher velocity shocks. 
This manifests as flattening of the \ac{LC} at late times 
\cite{Sironi:2013tva}. 
Thus, when $\gamma_{e;\,\rm min}'$ gets close to $1$, 
additional adjustments are needed. 
Specifically, we set that only a fraction of injected electrons, 
$\xi_{\rm DN}$, can contribute to the observed emission. 
The $\xi_{\rm DN}$ is computed according to \citet{Ayache:2021six} as
\begin{equation} \label{eq:method:gm_lim}
    \xi_{\rm DN} = \frac{\gamma_{e;\,\rm max}'^{2-p} - \gamma_{e;\,\rm min}'^{2-p}}{\gamma_{e;\,\rm max}'^{2-p} - 1} 
    \times  \frac{\gamma_{e;\,\rm max}'^{1-p} - 1}{\gamma_{e;\,\rm max}'^{1-p} - \gamma_{e;\,\rm min}'^{1-p}}\, ,
\end{equation}
where $\gamma_{e;\, \rm max}'$ is evaluated using 
Eq.~\eqref{eq:method:gamma_max}.

The pitch-angle averaged synchrotron emissivity from 
non-thermal electrons reads (\citetalias{Margalit:2021kuf})
\begin{equation}
    j_{\nu';\,\rm pl}' = C_j\frac{\epsilon_e}{\epsilon_T}\frac{q_e^3 n' B'}{m_e c^2} g(\Theta)x^{-\frac{p-1}{2}}\, ,
\end{equation}
and the self-absorption coefficient is 
\begin{equation}
    \alpha_{\nu';\,\rm pl}' = C_{\alpha} \frac{\epsilon_e}{\epsilon_T} \frac{q_e n'}{\Theta B'}g(\Theta)x^{-\frac{p+4}{2}}\, ,
\end{equation}
where $C_j$ and $C_{\alpha}$ are $p$-dependent coefficients 
\cite{RybickiLightman:1985,Mahadevan:1996cc,Margalit:2021kuf}, 
and $\epsilon_T$ is the fraction of shock energy that goes 
into thermal electrons.  

We also implement the low-frequency corrections to the 
$j_{\nu';\rm pl}'$ and effect of the electron cooling 
following \citetalias{Margalit:2021kuf}.

Thermal emissivity, $j'_{\nu', \rm th}$, decreases faster with 
velocity. Thus, post-deceleration spectrum 
is expected to be dominated by $j'_{\nu', \rm pl}$. 

The total emissivity and absorption then read 
$j_{\nu'}' = j_{\nu';\rm pl}' + j_{\nu';\rm th}'$ and 
$\alpha_{\nu'}' = \alpha_{\nu';\rm pl}' + \alpha_{\nu';\rm th}'$,
respectively.

\subsection{Coordinate system} \label{sec:method:discret}

Both \ac{GRB} and \ac{kN} ejecta have 
angle dependent mass and velocity.   
We assume azimuthal symmetry, ie, 
ejecta properties depend on the polar angle only. 

\ac{GRB} ejecta is discretized into non-overlapping layers 
each of which has its own polar angle and 
initial \ac{LF}, mass and energy. The polar angle, however, 
is not constant and evolves as \acp{BW} laterally expand. 

\ac{kN} ejecta is discretized into elements each of which has 
its own constant polar angle, initial \ac{LF} and mass. 
They comprise shells of equal polar angle (\ie, they overlap) 
and layers of equal initial \ac{LF}.

The coordinate system is implemented as follows. 

Consider a spherical coordinate system $(R,\,\theta,\,\phi)$ 
where $R$ is the distance from the coordinate origin, and $\theta$ 
and $\phi$ are the latitudinal and azimuthal angles respectively. 
The central engine (post-merger remnant) is located at the coordinate 
origin, and the system's symmetry axis ($z$-axis) lies along $\theta=0$. 
The observer is located in the $\phi=\pi/2$ plane and  
$\theta_{\rm obs}$ is the angle between the \ac{LOS} and the $z$-axis.
Thus, the unit vector of the observer is given by 
$\vec{n}_{\rm obs} = \big( 0,\, \sin(\theta_{\rm obs})\vec{y},\, \cos(\theta_{\rm obs}\big)\vec{z} )$.

We follow \citet{Lamb:2017ych,Lamb:2018ohw,Fernandez:2021xce} and 
discretize each hemisphere into $k=\{1,2,...n-1\}$ rings 
centered on the symmetry axis plus the single central spherical cap, $k=0$. 
The spherical cap opening angle is $\theta_{l=1}$% 
between two concentric circles on the sphere with $\theta_{l=i}$ and 
$\theta_{l=i+1}$. Setting the uniform distribution in terms of $\cos{(\theta_{l})}$, 
the $\theta_{l=i} = 2 \sin^{-1} \big( \sqrt{k/n} \sin( \theta_{\rm w} / 2 ) \big)$, 
where $\theta_{\rm w}$ is the initial opening angle of the ejecta. 
For \ac{GRB} ejecta it corresponds to the \ac{GRB} opening angle 
(see Sec.~\ref{sec:method:grb}). For \ac{kN} ejecta it is set to 
$\pi/2$. 
%\mpo{outflow}.
%
Each ring of index number $j$ is 
discretized into $2 i + 1$ 
azimuthal regions bounded by $\phi_{ij} = 2\pi j/ (2 i + 1)$, where 
$j=\{0,1,2...i\}$. 
%4
Overall, each ejecta shell is discretized into $\sum_{i=0}^{i=n-1}(2 i+1)=n^2$ 
\textit{elements}, each of which has a solid angle 
$2\pi \big( 1-\cos{(\theta_{\rm w})} \big) / n^2$ \cite{Beckers:2012}.  
A specific element ``c'' then has coordinates 
$\theta^{\rm c}_{i}, \phi^{\rm c}_{ij}$  
with $\theta^{\rm c}_{i} = (\theta_i + \theta_{i+1})/2$ and 
$\phi^{\rm c}_{ij} = \phi_{ij} + \phi_{ij-1}/2$. 
The coordinate vector of the element is given by
$\vec{v}_{ij} = R_{ij}\big( \sin{(\theta_{i})}\cos{(\phi_{ij})}\vec{x},\, \sin{(\theta_{i})}\cos{(\phi_{ij})}\vec{y},\, \cos{(\theta_{i})}\vec{z} \big)$,
where $R_{ij}$ is the radius of the element.
The angle between the \ac{LOS} and the coordinate vector 
of the element %\mpo{follows as

\begin{equation}  \label{eq:method:mu_obs}
    \cos{(\theta_{ij, \mathrm{LOS}})} = 
    \sin{(\theta_{i})}\sin{(\phi_{ij})}\sin(\theta_{\rm obs}) +  
    \cos{(\theta_{i})}\cos(\theta_{\rm obs}) \, .
\end{equation}

Within this discretization, the \ac{GRB} lateral spreading implies 
that each layer 
laterally expands with its own velocity given by 
Eq.~\eqref{eq:method:dthetadr_GP12}. The interaction between 
layers is neglected, and the gradual pressure gradient expected for 
a lateral structure is approximated with a step-like function. 
This approximation leads to an overestimation of the lateral expansion.
More importantly, since each of the layers interacts with the same 
upstream medium,
collecting mass independently, the slowest \ac{BW} will fall behind the 
faster ones. 
This method has been successfully applied to structured jet 
afterglow modelling \cite{Lamb:2018qfn,Ryan:2019fhz,Fernandez:2021xce}.  
However, its accuracy against numerical simulations of structured 
jets remains to be quantified in full detail.

\subsubsection{\ac{GRB} ejecta structure} \label{sec:method:grb}

Numerical simulations of jets, breaking out from either 
a stellar envelope (in the case of long \acp{GRB}) or \ac{BNS} merger 
ejecta (in the case of short \acp{GRB}) show the presence 
of lateral structure, \ie, the flow properties depend on the angle 
from the polar axis \cite{De:2012,Xie:2018vya,Gottlieb:2020mmk,Lamb:2022pvr}. 
Such jets have a non-trivial afterglow behaviour, that depends strongly 
on the viewing angle \cite{Granot:2002me,Wei:2002gc,Zhang:2002jt,Rossi:2004dz,Granot:2002me,Salafia:2015vla,Lamb:2017ych,Beniamini:2020eza,Takahashi:2020jcc}. 
Observations of \GRB{} also point towards a 
structured jet that was observed off-axis \cite{Fong:2017ekk,Troja:2017nqp,Margutti:2018xqd,Lamb:2017ych,Lamb:2018ohw,Alexander:2018dcl,Mooley:2018dlz,Ghirlanda:2018uyx,Ryan:2019fhz}. 
And among possible structure types, %the one described by 
a Gaussian function is able to provide a good fit to \GRB{} 
(see however \citet{Lamb:2020ccz,Takahashi:2020jcc}). 
In a Gaussian jet, the initial energy per solid angle  
and \ac{LF} of the jet read 
\begin{equation}
    E_0(\theta) = E_{\rm c} e^{-\theta^2 / \xi_1\theta_{\rm c}^2}, \, 
    \Gamma_0(\theta) = 1+(\Gamma_{\rm c} - 1)e^{-\theta^2 / \xi_2\theta_{\rm c}^2}, \, 
\end{equation}
where $E_{\rm c}$, $\Gamma_{\rm c}$, and $\theta_{\rm c}$ are the 
energy, \ac{LF}, and half-opening angle of the jet core, 
$\xi_1=1$ and $\xi_2=2$ are constants, set following 
\citet{Resmi:2018wuc,Lamb:2017ych,Fernandez:2021xce}.

\subsubsection{\ac{kN} ejecta structure} \label{sec:method:ejecta}

We consider dynamical ejecta profiles from a 
large set of \ac{NR} \ac{BNS} merger simulations targeted to
\GW{} \cite{Perego:2019adq,Endrizzi:2019trv,Nedora:2019jhl,Bernuzzi:2020txg,Nedora:2020pak,Nedora:2021eoj,Cusinato:2021zin}. 
For all our simulations the ejecta data are publicly available\footnote{
    Data are available on Zenodo: \url{https://doi.org/10.5281/zenodo.4159620}
}. 
We focus on the list of simulations given in the Table~(2) of 
\citetalias{Nedora:2021eoj}. 
These simulations were performed with the \ac{GRHD} code 
\texttt{WhiskyTHC}~\cite{Radice:2012cu,Radice:2013xpa,Radice:2013hxh,Radice:2015nva}. 
They include leakage and M0 neutrino schemes in optically thick and thin 
regimes respectively~\cite{Radice:2016dwd,Radice:2018pdn}, 
and accounting for the turbulent viscosity of magnetic origin via 
an effective subgrid scheme~\cite{Radice:2017zta,Radice:2020ids}. 
The importance of viscosity and advanced neutrino transport for 
obtaining more accurate dynamical ejecta properties is discussed in 
\citet{Radice:2018ghv,Radice:2018pdn,Bernuzzi:2020txg,Nedora:2020pak}. 
Simulations are classified with their 
reduced tidal deformability $\tilde{\Lambda}$ and \mr{} $q$. 
The former is defined as \citep{Favata:2013rwa}, 
\begin{equation}
    \tilde\Lambda = \frac{16}{13}\frac{(M_A+12 M_B)M_A^4 \Lambda_A}{M^5}+(A\leftrightarrow B)\, ,
    \label{eq:intro:Lambda}
\end{equation}
where $\Lambda_i \equiv 2/3\, C_i^{-5} k^{(2)}_i$ are the quadrupolar tidal parameters, 
$k_i^{(2)}$ are the dimensionless gravitoelectric Love numbers \citep{Damour:2009vw}, 
$C_i \equiv GM_A/(c^2R_A)$ are the compactness parameters, and $i=A,B$. 
Here $A$, $B$ subscripts are used to label individual stars with 
individual gravitational masses $M_A$ and $M_B$, 
baryonic masses as $M_{b;\,A}$ and $M_{b;\,B}$.
The total mass is $M = M_A + M_B$, and the \mr{} $q=M_A/M_B\geq1$. 
Masses and velocities are given in units of $\Msun$ and $c$, respectively.
All simulations were performed using finite temperature and 
composition-dependent nuclear \acp{EOS}. In particular, the following 
set of \acp{EOS} was considered: 
DD2 \citep{Typel:2009sy,Hempel:2009mc},
BLh \citep{Logoteta:2020yxf}, 
LS220 \citep{Lattimer:1991nc},
SLy4 \citep{Douchin:2001sv,daSilvaSchneider:2017jpg}, and
SFHo \citep{Steiner:2012rk}. 
Among them, DD2 is the stiffest (larger \ac{NS} radii, 
larger tidal deformabilities and larger \ac{NS} maximum supported masses), 
while SFHo and SLy4 are the softest.

As in \citetalias{Nedora:2021eoj} the ejecta kinetic energy distribution, 
$E_{k}=f(\Gamma,\theta)$ (that in turn depends on the binary parameters,  
$q$ and $\tilde{\Lambda}$) is used as the initial data for 
%$\Gamma$-shells $\theta$-layers and for 
the afterglow calculation.

\begin{figure*}
    \centering 
    \includegraphics[width=0.49\textwidth]{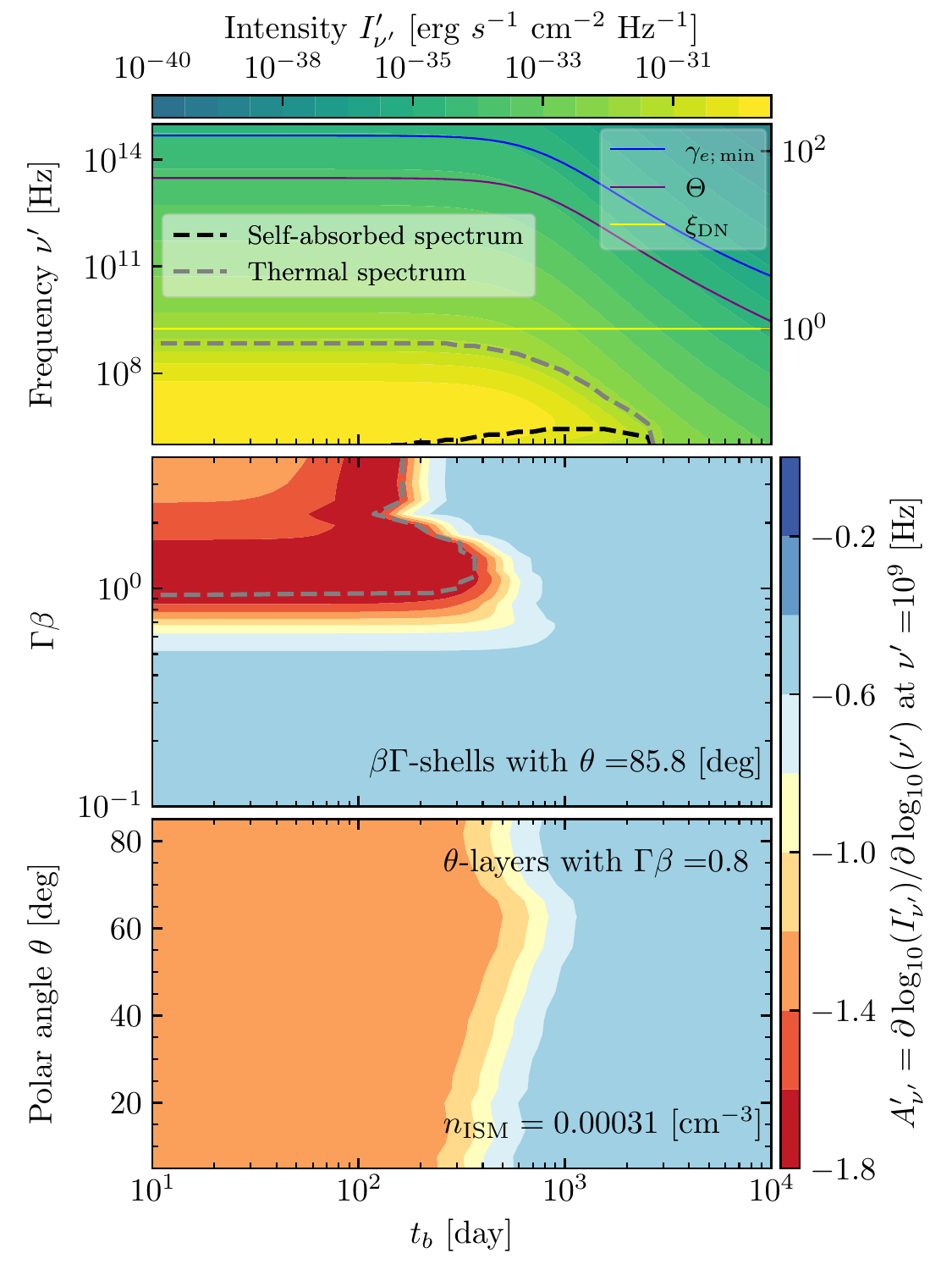}
    \includegraphics[width=0.49\textwidth]{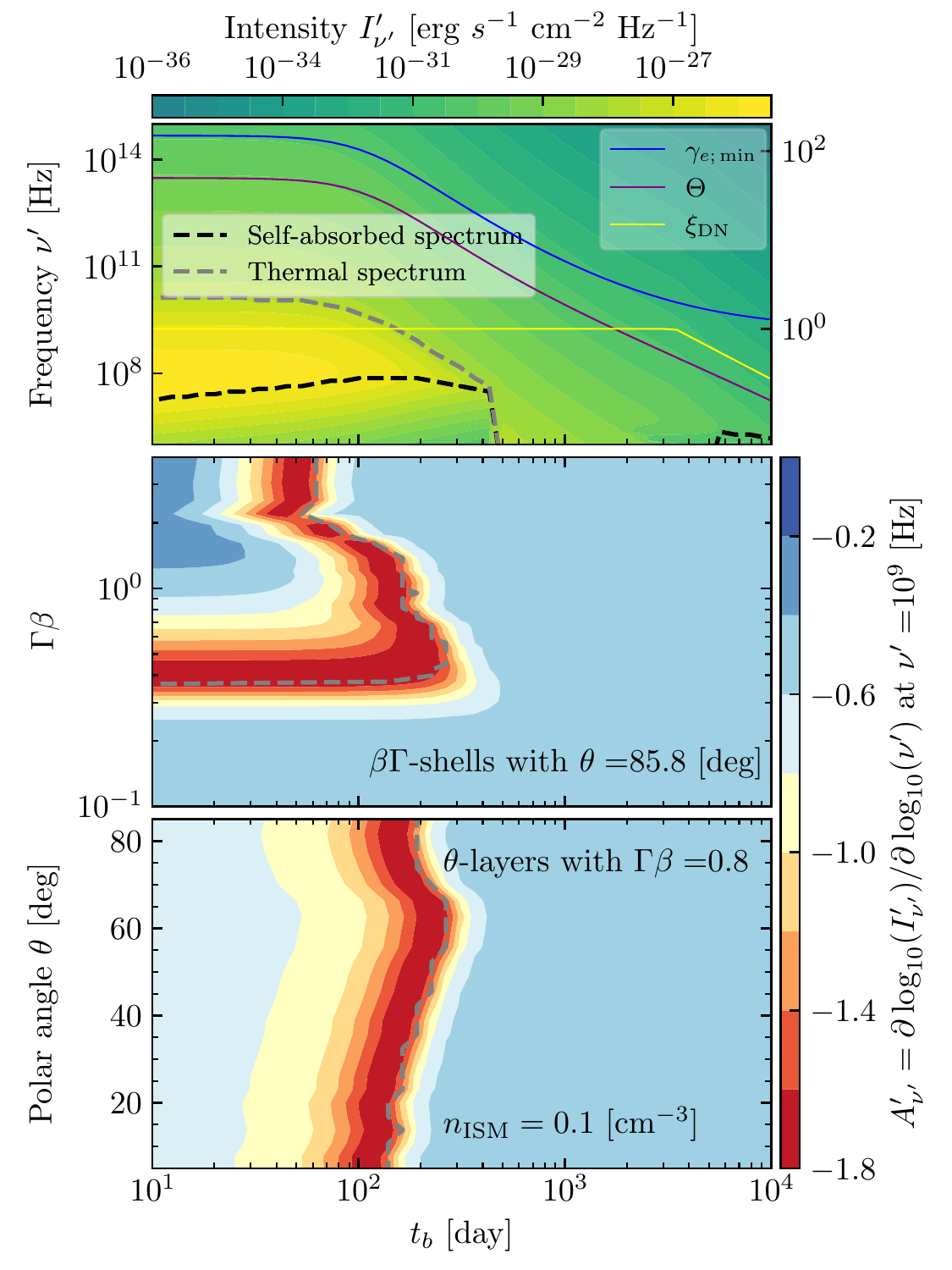}
    \caption{
        \textit{Top panels} display the time evolution of the 
        intensity in the frame comoving with the fluid %comoving intensity 
        produced by a single \ac{BW} with initial momentum 
        $\Gamma\beta=0.8$ and polar angle %from the polar axis 
        $\theta=85.8\,$deg 
        (color-filled contours). 
        The gray dashed line marks the location where 
        the emission from thermal electron population 
        is equal to that from the non-thermal.   
        The black dashed line marks $\tau_{\nu'} = 1$. 
        Also shown is the characteristic \ac{LF} of the non-thermal %power-law 
        electron distribution,  
        %distribution 
        $\gamma_{e;\,\rm min}'$ (blue line); the fraction of electrons that 
        are accelerated to the power-law distribution, $\xi_{\rm DN}$ (yellow line); 
        and the dimensionless electron temperature, $\Theta$ (magenta line).
        Here $\xi_{\rm DN}=1$ imples that all injected electrons that 
        are accelerated to the power-law distribution in energy 
        contribute to the emission.
        \textit{Middle panels} show the %comoving intensity 
        spectral index, 
        $A'_{\nu'}$ at $1\,$GHz for all \acp{BW} with fixed  
        $\theta=85.8\,$deg but varying initial momentum.
        \textit{Bottom panels} show $A'_{\nu}$ at $1\,$GHz for all 
        \acp{BW} with different polar angles, $\theta$, but with the same   
        %angular layers 
        initial momentum $\Gamma\beta=0.8$. 
        The difference between the left and right panels is the \ac{ISM} 
        density, which is $0.00031\,\ccm$ and $0.1\,\ccm$ respectively. 
        The plot shows that for fast ejecta with $\Gamma\beta>1$, 
        the radio part of the spectrum, $\nu'\simeq1\,$GHz, is dominated 
        by the emission from the thermal electron population. 
        Meanwhile, at higher 
        \ac{ISM} density the contribution from thermal electron  
        population is found in \acp{BW} with lower initial velocity. 
    }
    \label{fig:results:therm_spec}
\end{figure*}

\subsection{Observed radiation} \label{sec:method:eats}

After all \acp{BW} corresponding to angular and velocity 
elements of \ac{GRB} and \ac{kN} ejecta are evolved, 
and comoving emissivities and absorption coefficients 
are obtained, the observed radiation is computed 
via \ac{EATS} integration \citep[\eg][]{Granot:1998ek,Granot:2007gn,Gill:2018kcw,vanEerten:2009pa}. 
For simplicity we first 
consider a given \ac{BW} ($ij$) with its own angular 
position is computed. The retardation necessary 
for computing the emission from all \acp{BW} at a given 
observer time is discussed later in the section. 

We consider plane parallel rays of varying impact parameters 
(perpendicular distances of rays to the central line of sight) 
through the emitting region
Solving the radiation transport equation along these rays, 
we obtain \cite{Mihalas:1978} 
\begin{equation}
    \frac{\partial I_{\nu}}{\partial s} = j_{\nu} - \alpha_{\nu} I_{\nu}\, ,
\end{equation}
where $s$ is the line element along the ray. 

The conversions of comoving emissivity and absorption coefficient 
into the observer frame read \cite{vanEerten:2009pa}:  
$j_{\nu} = j_{\nu}' / ( \Gamma (1 - \beta\mu) )^2$,  
$\alpha_{\nu} = \alpha_{\nu}' ( \Gamma ( 1-\beta\mu ) )$, 
where $\mu = \cos(\theta_{ij,\rm LOS})$ for a given \ac{BW}.  
% $\beta$ is the time dependent dimensionless velocity.
The transformation for the frequency reads 
$\nu'=\nu (1+Z) \Gamma ( 1-\beta\mu )$, 
where $Z$ is the source redshift. 

For the uniform plane-parallel emitting region 
the equation has an analytic solution 
\begin{equation} \label{eq:method:intensity}
    I_{\nu} = \frac{j_{\nu}}{\alpha_{\nu}}( 1 - e^{-\tau}) \approxeq 
    j_{\nu} \frac{3}{\tau} \Big[ \frac{1}{2} + \frac{e^{-\tau}}{\tau} - \frac{1-e^{-\tau}}{\tau^2} \Big]\, ,
\end{equation}
where $\tau_{\nu} \approxeq -\alpha_{\nu} \Delta R / \mu'$ is the optical depth with  
\begin{equation}
    \mu' = \frac{\mu - \beta}{1 - \beta \mu},
\end{equation} 
being the parameter relating the angle of emission in local frame to that in the 
observer frame \cite{Granot:1998ek}, accounting for cases when rays
cross the homogeneous slab (ejecta) along directions different from radial. 
In the last equality in Eq.~\eqref{eq:method:intensity} 
we expressed the absorption coefficient as attenuation, 
following the equation $7.122$ in \cite{Dermer:2009}.

The thickness of the emitting region, \ie, the region between the 
forward shock and the contact discontinuity of the \ac{BW} 
in the observer frame reads, 
$\Delta R = \Delta R'/(1 - \mu \beta_{\rm sh})$ %accounting for beaming \red{bulshit. Beaming of what?}
where $\Delta R' = m_2 / ( 2 \pi m_p R^2 ( 1 - \cos(\omega) ) \Gamma n' )$ 
is obtained under the assumption of a homogeneous shell, 
but relaxing the assumption of the uniform upstream medium %\ac{ISM} 
\cite{Johannesson:2006zs}.  Notably, if $1-\cos(\omega)=2$ and 
the swept-up mass $m_2 = 4 \pi R^3 n' m_p / 3$, 
we recover the \ac{BM} shock thickness,  
$\Delta R'=R/12\Gamma^2$ 
\citep[\eg][]{Johannesson:2006zs,vanEerten:2009pa}.

\begin{figure*}
    \centering 
    \includegraphics[width=0.49\textwidth]{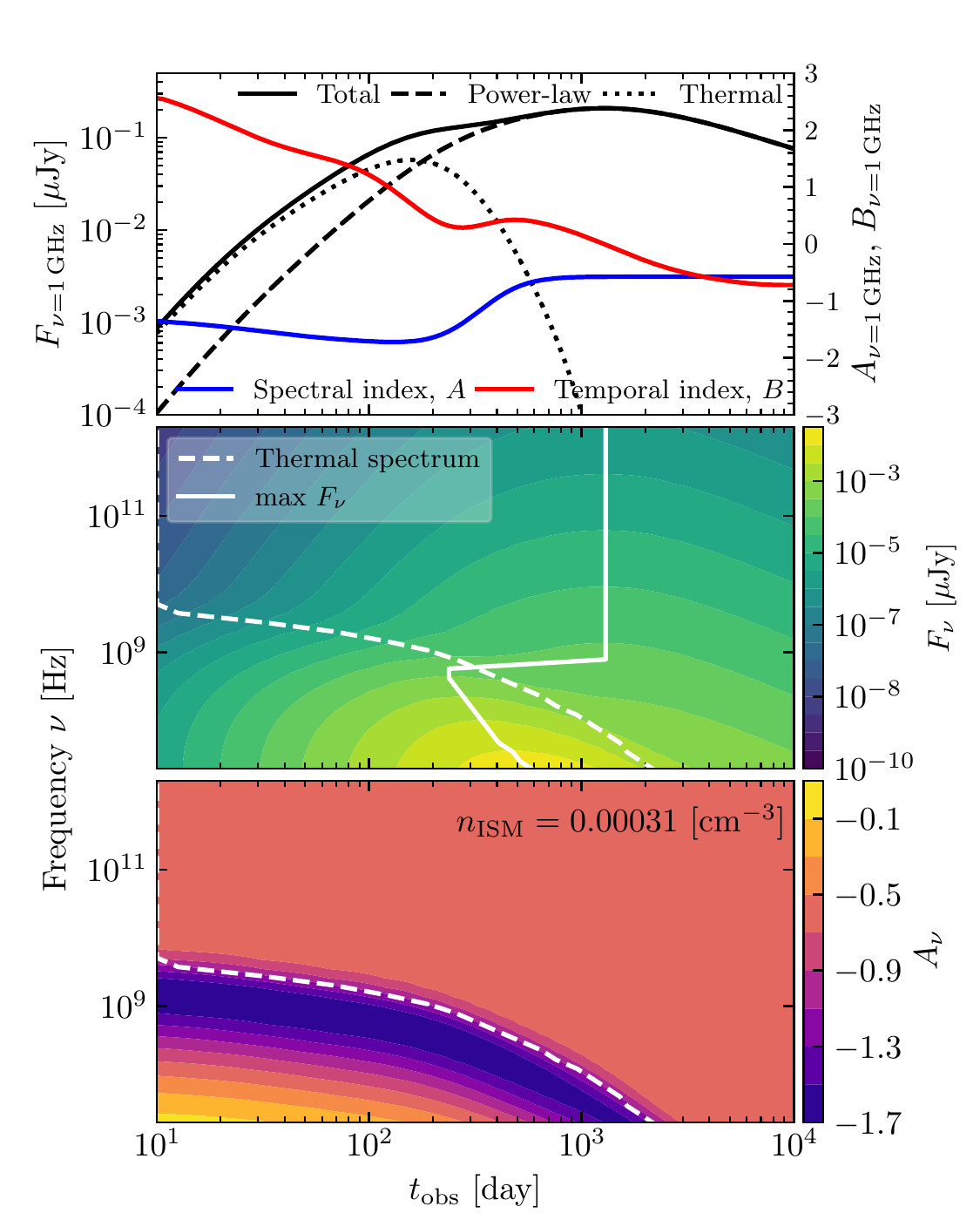}
    \includegraphics[width=0.49\textwidth]{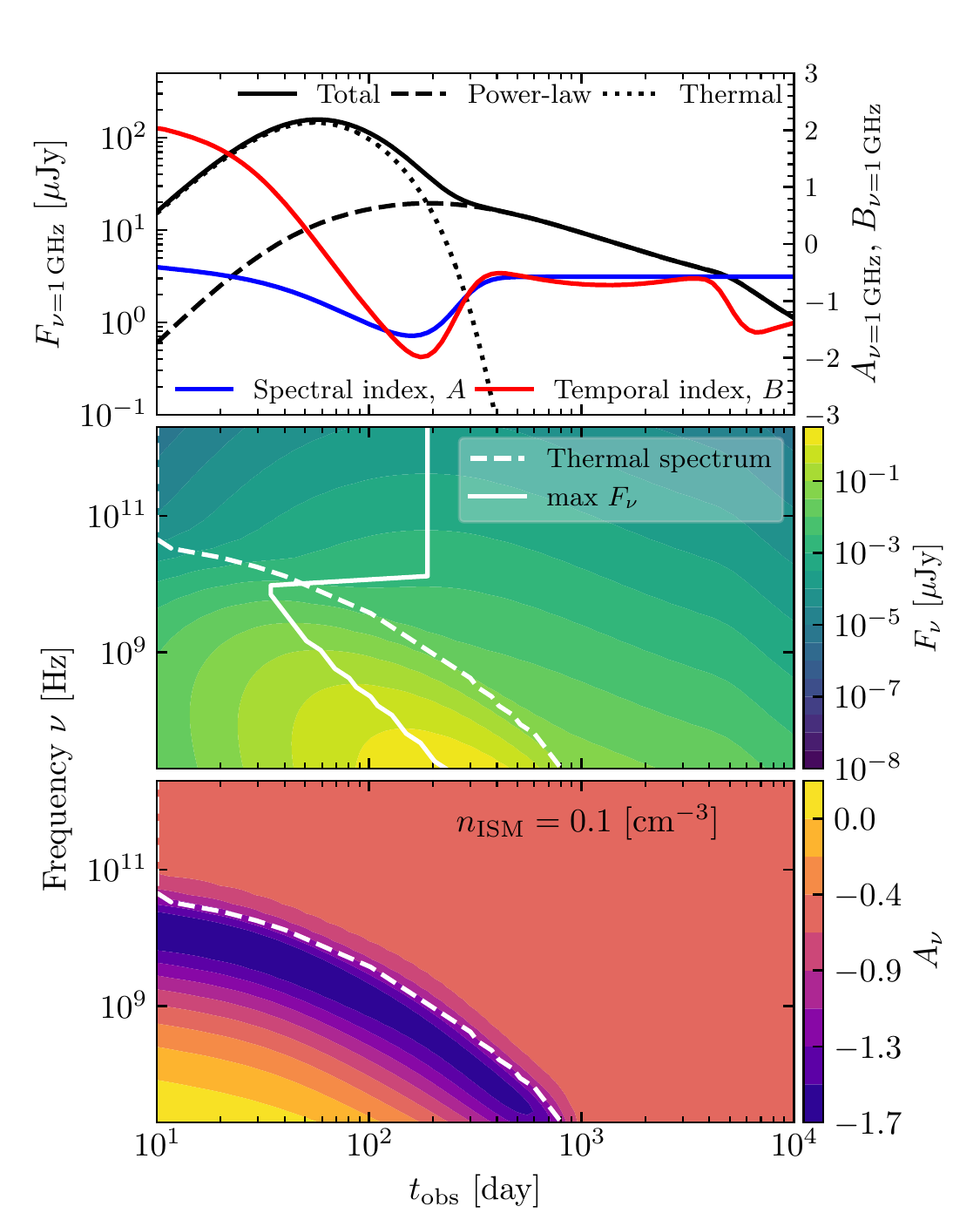}
    \caption{
        \textit{Top panels}: afterglow \acp{LC} at $1\,$GHz, (black line)
        and the contributions from thermal electrons (dotted line) and 
        non-thermal electrons (dashed line). Also shown is the spectral index, 
        $A_{\nu}$, and the temporal index, $B_{\nu}$. 
        \textit{Middle panels}: evolution of the observed spectrum. Below the dashed 
        white line the spectrum is predominantly thermal, \ie, is dominated 
        by the emission from thermal electrons. 
        The solid white line marks the frequency of peak flux.
        The intersection between the two white lines corresponds to the 
        spectrum transition frequency, $\nu_t$.
        \textit{Bottom panels}: time evolution of the spectral 
        index, $A_{\nu}$, across all frequencies. 
        The left and right panels are for low and high $n_\mathrm{ISM}$  
        respectively. 
    }
    \label{fig:results:therm_lc}
\end{figure*}

For a geometrically extended, evolving source, the observed radiation at 
a given frequency $\nu$ and at a given time $t_{\rm obs}$ is composed of 
many contributions from fluid elements emitting at various frequencies $\nu'$ 
and at different times. 

We compute the flux in the observer frame as piece-wise sum 
\begin{equation} 
    F_{\nu} = \frac{1+Z}{2\pi d_L^2} \sum_{ij} R_{ij}(t_{ij,\rm obs})^2 \Delta R_{ij} I_{ij,\nu}(t_{ij,\rm obs})\, ,
\end{equation}
where the arrival time, $t_{ij,\rm obs}$, for a given \ac{BW} 
($ij$) that corresponds to the $t_{\rm obs}$, is obtained via 
equation~(4) of \cite{Fernandez:2021xce}, and 
$d_L$ is the luminosity distance.

\section{Results}\label{sec:results}
\subsection{Effects of thermal electrons on kN afterglow} \label{sec:results:therm_ele}
\subsubsection{Comoving emission}

Here, we examine how the presence of thermal electrons affects 
the \ac{kN} afterglow. 
We consider static, constant density \ac{ISM}, 
$\rho_{\rm ISM} = n_{\rm ISM} m_p$, 
\ie, we neglect the presence of the \ac{GRB}. 
We focus on the equal-mass \ac{BNS} merger simulation with 
BLh \ac{EOS}, as its ejecta profile has a 
fast tail that was closely examined in \citetalias{Nedora:2021eoj} 
(see their figure~3).  
For the remainder of this section we fix the following model 
parameters: $p=2.15$, $\epsilon_e=0.2$, and $\epsilon_B=0.005$.
Following \citet{Margalit:2021bqe}, we set $\epsilon_T=1$. 
The distance to the source is assumed to be $D_{\rm L} = 41.3\,$Mpc, 
and it is observed at an angle of $\theta_{\rm obs}=30\,$deg.
We consider two values for the \ac{ISM} density, $n_{\rm ISM}=0.00031\,\ccm$ 
and $n_{\rm ISM}=0.1\,\ccm$.

In Fig.~\ref{fig:results:therm_spec} we show the evolution 
of the intensity in the \ac{BW} frame, $I_{\nu'}'(t_b)$, 
for the two values of $n_{\rm ISM}$. 
%(evaluated with Eq.~\eqref{eq:method:intensity}), 
Both thermal and non-thermal electron distributions are  included. 
In the top panels of Fig.~\ref{fig:results:therm_spec} we show  
$I_{\nu'}'(t_b)$ 
for a single \ac{BW} that corresponds to the ejecta element 
with polar angle $\theta = 85.8\,$deg.
and initial momentum $\Gamma_0\beta_0=0.8$.  
The choice is motivated by the fast tail angular distribution  
which is largely equatorial. 
At frequencies $\nu'\gtrsim 1\,$GHz, $\nu'\in(\nu_{\rm min}',\nu_{\rm c}')$, 
the spectrum is dominated by the emission from the 
non-thermal electron population. The spectral index, 
$A'_{\nu'}$, defined here as 
$A'_{\nu'}=d\log_{10}(I'_{\nu'})/d\log_{10}(\nu')$\footnote{For the sake of convenience and clarity we denote the spectral index with 
    capital $A_{\nu}$ instead of commonly used $\alpha$ to distinguish it 
    from the absorption coefficient. 
}, is $-0.575$ which corresponds to 
the electron spectral index  
$p=2.15$ and slow cooling regime. 
At very late times, $I_{\nu'}'(t_{b})$ declines as $\gamma_{e;\,\rm min}'$ 
approaches unity and  
the fraction of electrons accelerated to the power-law
distribution and contributing to emission, $\xi_{\rm DN}$, decreases. 
This decline in $I'_{\nu'}$ is seen at all frequencies, 
and it commences earlier for high upstream density. 

\begin{figure*}
    \centering 
    \includegraphics[width=0.49\textwidth]{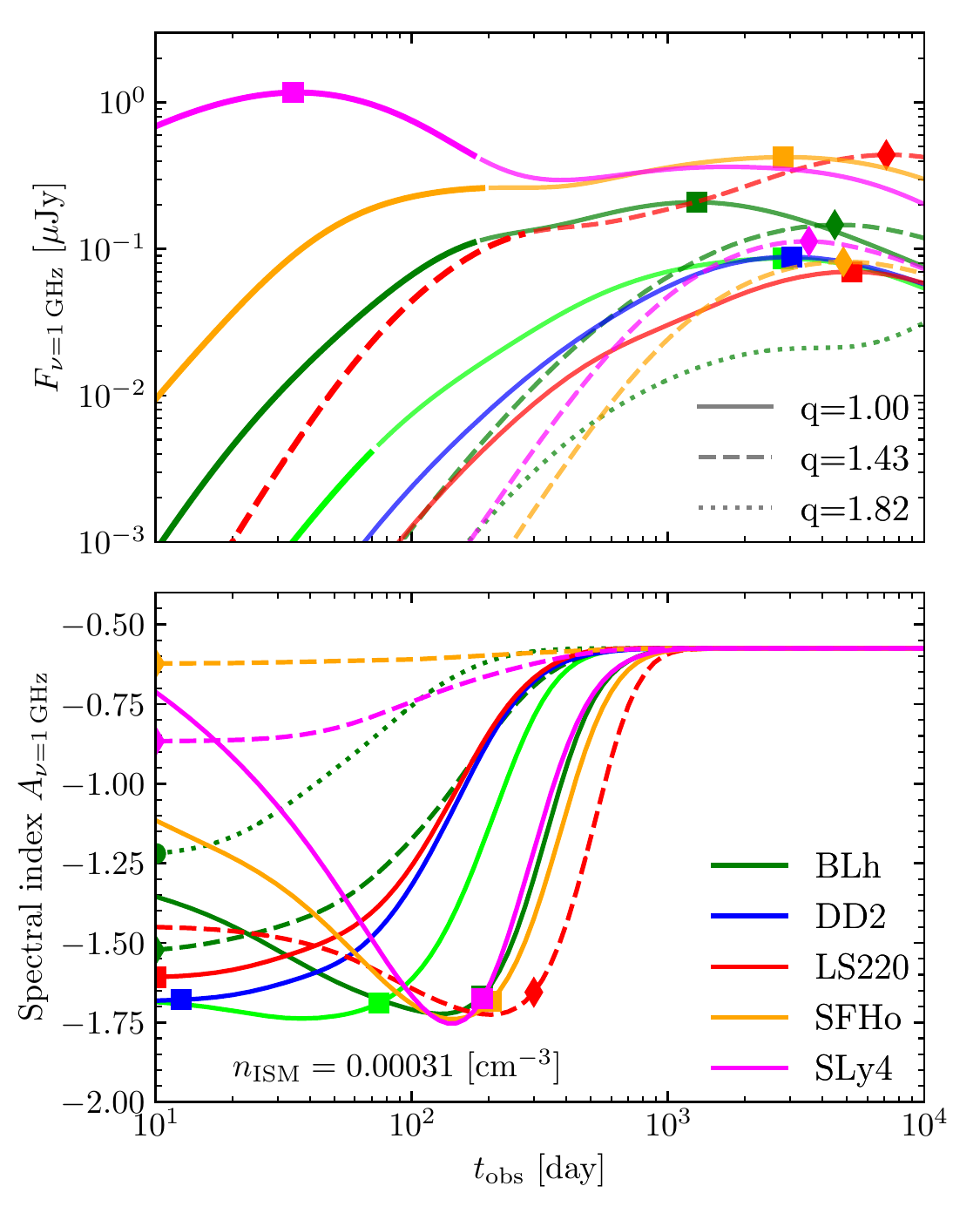}
    \includegraphics[width=0.49\textwidth]{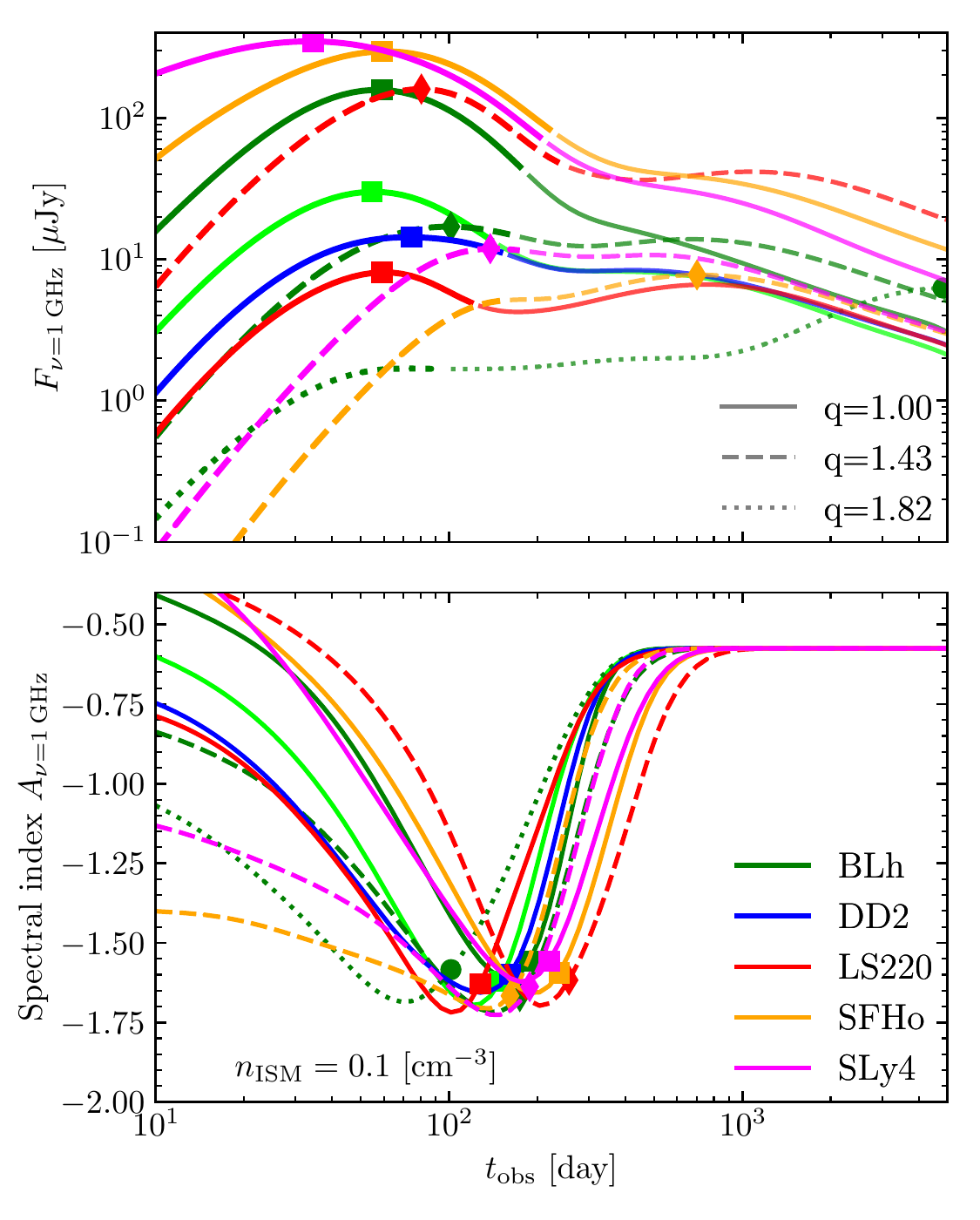}
    \caption{
        \textit{Top panels}: 
        Radio ($1\,$GHz) \acp{LC} for all simulations. 
        Colors indicate different  
        \ac{NS} \acp{EOS}, and linestyles 
        stand for various \ac{BNS} \mr{}.  
        Markers indicate the \ac{LC} peak. 
        \textit{Bottom panels}: spectral index evolution. 
        Markers indicate the spectral transition, 
        $F_{\nu;\,\rm th}=F_{\nu;\,\rm pl}$. 
        The difference between left and right panels is the 
        same as in Fig.\ref{fig:results:therm_spec}, 
        the observer angle $\theta_{\rm obs}=30\,$deg. 
    }
    \label{fig:results:therm_lc_all}
\end{figure*}

At early times and at low frequencies, 
$\nu'\lesssim1\,$GHz, $\epsilon_{\rm th}'$ is larger 
than $\epsilon_{\rm pl}'$. 
They are equal at the frequency marked by 
the dashed gray line, $\nu_t'$, below which 
$\epsilon_{\rm th}' > \epsilon_{\rm pl}'$. 
We call this regime thermal. 
The frequency at which $\epsilon_{\rm th}'=\epsilon_{\rm pl}'$ depends 
primarily on the ejecta velocity, $n_{\rm ISM}$ and 
microphysical parameters, as illustrated in figure~$2$ in 
\citetalias{Margalit:2021kuf}.

Most known short \acp{GRB} with detected \ac{kN} 
signatures occurred in low-density environments, 
$n_{\rm ISM} \ll 1\,\ccm$ \citep[\eg][]{Fong:2017ekk,Klose:2019amd}. 
Thus, under the assumption that $\epsilon_T=1$, 
we expect the transition in the 
spectrum to occur in the radio band. We focus the 
subsequent discussion on this part of the spectrum. 
Notably, for lower $\epsilon_T$, $\epsilon_{\rm th}'$ 
and the transition frequency $\nu_t'$ decreases.  
% even at $n_{\rm ISM}=0.1\,\ccm$. 
This behaviour is generic. We observe it in \ac{kN} 
afterglows from other \ac{BNS} merger simulations.   
At even lower frequencies, $\nu'<\nu_t'$, \ac{SSA} becomes important. 
The region where $\tau_{\nu'}>1$ is marked with 
black dashed line. Notably, even at high $n_{\rm ISM}$, 
\eg, $n_{\rm ISM}=0.1\,\ccm$, 
the self-absorbed part of the spectrum lies below $100\,$MHz.

After the \ac{kN} \ac{BW} starts to decelerate and the 
electron temperature $\Theta$ decrease, 
the spectrum begins to change due to the 
steep dependence of $K_2(1/\Theta)$ on $\Theta$ 
(Eq.~\eqref{eq:method:j_th}). 
%steep dependency of 
%$\epsilon_{\rm th}'$ on the 
%\ac{BW} velocity. 
%
When $\Theta$ drops below ${\simeq}1$, at very late times, 
the corrections added to $I(x_{\rm M})'$ (Eq.~\eqref{eq:method:Ixm}) 
become important and the decrese in $\epsilon_{\rm th}'$ 
becomes even steeper. Subsequently, the radio spectrum 
sharply transitions from thermal to non-thermal. 
This is seen in the top right panel of Fig.~\ref{fig:results:therm_spec} 
as a cut-off of the gray curve at $t_b\simeq3\times10^2\,$days. 
At this time, the non-thermal electrons dominate the 
emission at all frequencies. 
The velocity dependence of $\epsilon_{\rm th}'$ implies that different \ac{kN} 
\acp{BW} with different initial momenta 
and energy produce different spectra that also evolves in time.  
In the middle panels of Fig.~\ref{fig:results:therm_spec} the comoving spectral 
index, $A_{\nu'}'$ is shown as a function of the initial ejecta momentum. 
% (fixing its angle from the binary plane). 
Notably, at $1\,$GHz, and $n_{\rm ISM} = 0.00031\,\ccm$ 
the spectrum is thermal only for \acp{BW} with initial momenta 
$\Gamma\beta\gtrsim1$, \ie, 
for the ejecta fast tail, whereas at
$n_{\rm ISM}=0.1\,\ccm$, emission from thermal electrons 
is seen for $\Gamma\beta\gtrsim 0.4$. 

The spectral index, $A_{\nu'}'$, and its temporal evolution 
as a function of the polar angle, $\theta$, 
for all \acp{BW} with initial momentum $\Gamma\beta=0.8$
are shown in the bottom panel of Fig.~\ref{fig:results:therm_spec}. 
As in the \ac{BNS} simulations we consider, the fastest ejecta 
is found predominantly near the equatorial plane 
(being driven by core bounces \cite{Radice:2018pdn,Nedora:2021eoj}), 
and so the emission from thermal electrons is more important 
at $\theta\gtrsim60\,$deg. 
This qualitative picture is characteristic for all ejecta in 
our \ac{BNS} merger simulation set and hence might have 
important consequences for off-axis observations of \ac{BNS} mergers.

\subsubsection{Observed emission}

For a single \ac{kN} \ac{BW}, the radio emission 
in the optically thin regime is 
characterised by the typical synchrotron 
frequency, $\nu_{\rm min}$. Using the \ac{BPL} approximation to the 
synchrotron spectrum, 
%the flux at any frequency 
the flux at $\nu_{\rm min}$ is 
$F_{\nu=\nu_{\rm min}}\propto R^3n_{\rm ISM}^{3/2}\epsilon_B^{1/2}\beta_{\rm sh} d_L^{-2}$, 
and while $\beta=$const, the flux increases. Thus, the \ac{LC} peaks on 
the deceleration timescale of the \ac{BW}  
\cite{Nakar:2011cw,Piran:2012wd}.

Combining the emission from all \ac{kN} \acp{BW}, and accounting for 
relativistic effects, we display the evolution of the observed spectrum, 
$F_{\nu}(t_{\rm obs})$, in the middle panels of 
Fig.~\ref{fig:results:therm_lc}. 
The plot shows that as \acp{BW} decelerate, a progressively 
smaller part of the spectrum remains thermal (below the dashed white line). 
This is reflected in the evolution of the spectral index $A_{\nu}$, 
shown in the bottom panels of Fig.~\ref{fig:results:therm_lc}. There, 
the \ac{BW} deceleration manifests as a decrease in the transition 
frequency in the spectrum. At a fixed frequency, however, an observer may 
trace the evolution of the spectral index and reconstruct the evolution 
of the \ac{BW} speed. 
One would see a \ac{LC} that is dominated by the emission from thermal 
electrons at first and later by the emission from non-thermal electrons, 
regardless of the \ac{ISM} density. 
Notably, the relative brightness of these two types of synchrotron emission 
depends strongly on $n_{\rm ISM}$. As shown in Fig.~\ref{fig:results:therm_lc}, 
increasing $n_{\rm ISM}$ from $0.00031\,\ccm$ to $0.1\,\ccm$ leads to 
a rise in the flux density at $1\,\rm GHz$ from thermal and non-thermal 
electrons by four and two orders of magnitude, respectively 
(see top left and top right panels in the figure). Thus, if 
thermal electrons are indeed present behind \ac{kN} shocks, their 
radio emission would be observable at early times. 
% if \ac{ISM} density is sufficiently high. 
%
For instance, for $n_{\rm ISM}=0.1\,\ccm$, the first, thermal \ac{LC} peaks at 
a few $\mu$Jy, -- slightly above the latest \ac{VLA} upper limit for \GRB{} 
\cite{Balasubramanian:2022sie}. 
For lower values of $\epsilon_T$, the contribution from 
thermal electrons is smaller. 
Thus, the presence of thermal electron population can be inferred from 
(i) a double-peak structure of the \ac{LC} and 
(ii) the characteristic evolution of the spectral index at early times. 
However, at early times the \ac{kN} afterglow emission 
will likely be overshadowed by the \ac{GRB} afterglow emission, unless the 
source is observed far off-axis.
\begin{figure}
    \centering 
    \includegraphics[width=0.49\textwidth]{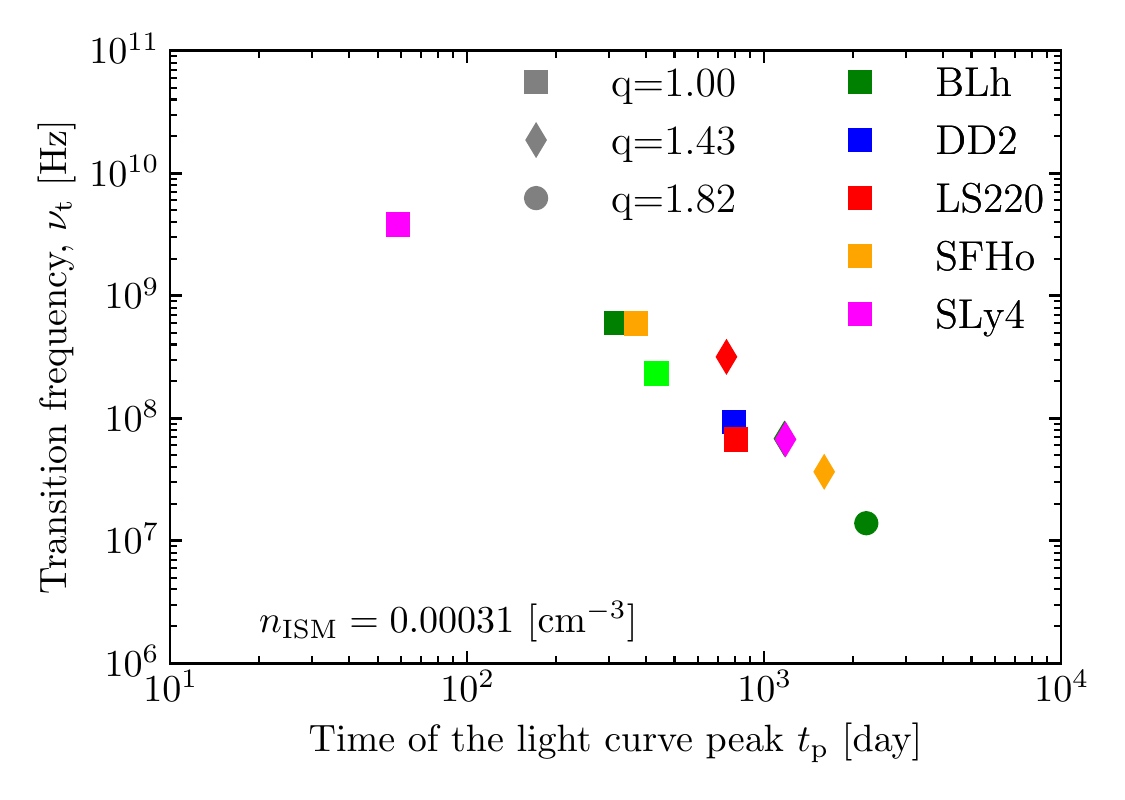}
    \caption{
        A relation between the frequency at which the spectrum 
        transitions from being dominated by thermal electrons 
        to the one dominated by non-thermal electrons, 
        $\nu_t$, and the peak time, $t_p$, at this frequency. 
        Colors indicate different 
        \ac{NS} \acp{EOS}, and markers 
        stand for various values of the \ac{BNS} \mr{}.  
        Here, $n_{\rm ISM}=0.00031\,\ccm$. 
    }
    \label{fig:results:peak_th_all}
\end{figure}

In Fig.~\ref{fig:results:therm_lc_all} the \ac{kN} afterglow \acp{LC} 
at $1\,$GHz are shown for all \ac{BNS} simulations (top panel),
as well as the evolution of the spectral index (bottom panel). 
At high density ($n_{\rm ISM}=0.1\,\ccm$), the radio \acp{LC} display 
a distinct bimodal shape with maxima corresponding to the emission from 
thermal and later from non-thermal electrons. We call them 
thermal and non-thermal peak hereafter. 
A prominent exception is the highly asymmetric model with BLh \ac{EOS}, 
in which the ejecta is of tidal origin only and lacks the fast tail 
\cite{Bernuzzi:2020txg}. 
The brightness and the peak time of the thermal peak are determined 
primarily by the ejecta velocity distribution and $n_{\rm ISM}$, 
and at sufficiently high $n_{\rm ISM}$, the \ac{LC} overall peak is 
thermal. Otherwise, the peak is non-thermal. 
The large difference in spectral index, $-0.575$ for the
non-thermal peak and ${\lesssim}-1.75$ for the thermal one, 
should permit distinguishing these scenarios. 
Similarly, if $n_{\rm ISM}$ is larger, so is the transition 
frequency, $\nu_t$. 
The relation between the transition frequency and 
the time of the \ac{LC} overall peak at this frequency 
is shown in Fig.~\ref{fig:results:peak_th_all}. 
Both, $\nu_t$ and $t_p$ depend on the model  
parameters and \ac{ISM} density. However, we find that 
the relation depends only weakly on the $n_{\rm ISM}$ and microphysical 
parameters and is primarily determined by the ejecta velocity 
distribution.
Indeed, equal mass models with soft \acp{EOS} always lie in the 
upper left corner, \ie, the spectral transition occurs at 
high frequencies, $\nu_t\gtrsim1\,$GHz, and early in time.
Meanwhile for highly asymmetric models the spectral transition 
occurs later and at lower frequency, $\mathcal{O}(50\,\rm MHz)$.

\subsection{\ac{kN} afterglow in the environment altered by a \ac{GRB} BW}\label{sec:results:interaction}

As discussed in Sec.~\ref{sec:method:grb}, we consider a Gaussian 
jet, with parameters informed by observations and 
modelling of \GRB{}. Specifically, following \citet{Hajela:2019mjy} 
and \citet{Fernandez:2021xce}, we set the jet half-opening angle 
$\theta_{\rm w} = 15\,$deg. and core half-opening angle 
$\theta_{\rm c} = 4.9\,$deg. 
The isotropic equivalent energy is $E_{\rm iso}=10^{52}\,$ergs, 
and the initial \ac{LF} of the core is $\Gamma_{\rm c} = 300$. 
The \ac{ISM} density is set to $n_{\rm ISM} = 0.00031\,\ccm$, and 
the microphiscal parameters are set as:  
$\epsilon_e=0.05$, $\epsilon_B=0.0045$, and $p=2.16$.
Luminosity distance to the source 
and the observer angle are set as $D_{\rm L} = 41.3\,$Mpc, 
$\theta_{\rm obs} = 21.5\,$deg, respectively. 
In the remainder of this section these parameters remain fixed 
unless stated otherwise.

\begin{figure}
    \centering 
    \includegraphics[width=0.49\textwidth]{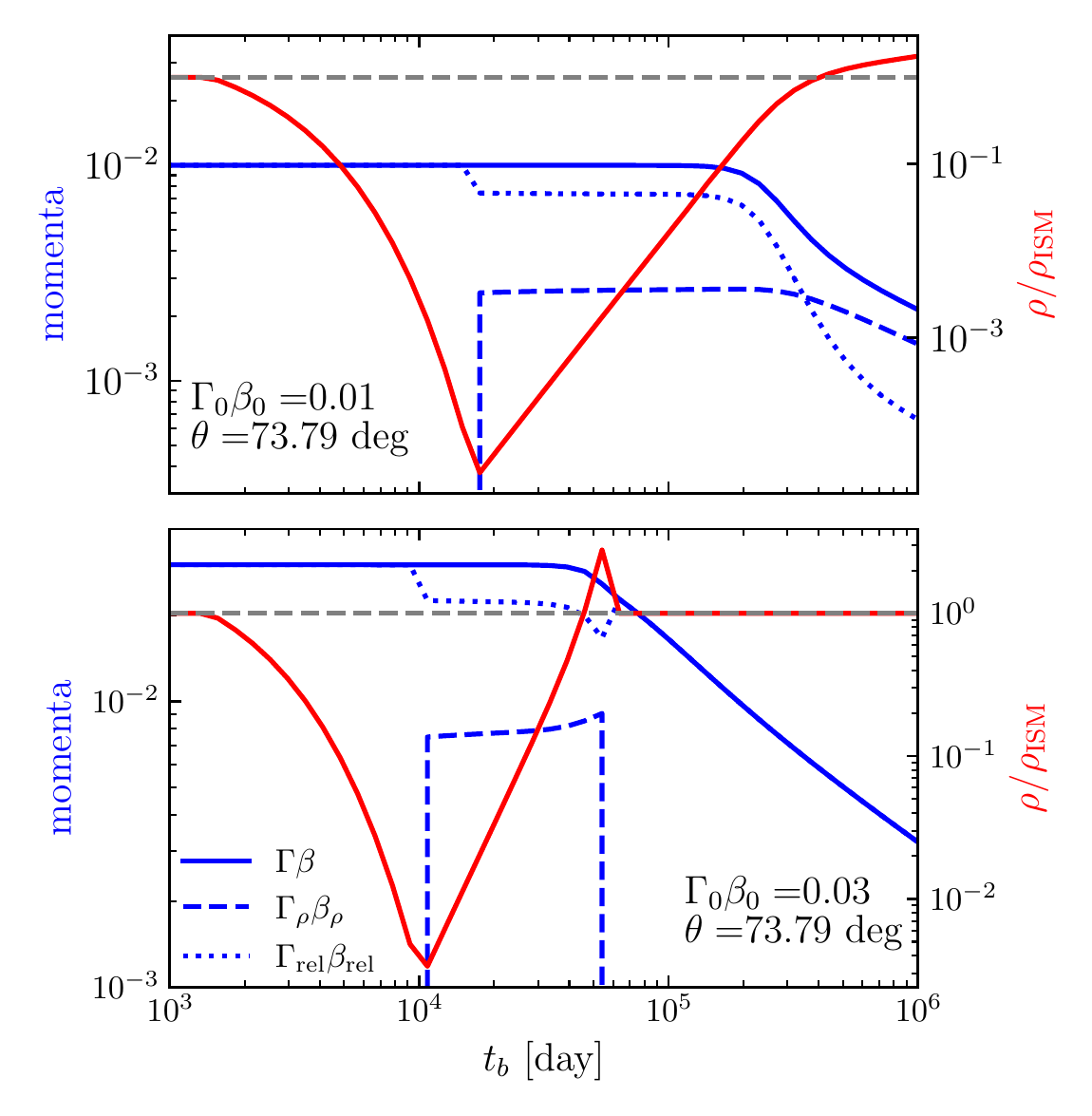}
    \caption{
        Dynamical evolution of the \ac{kN} \ac{BW}  
        with initial momentum $\Gamma_0\beta_0$, moving at an angle 
        $\theta$ through the \ac{CBM}, \ie, the medium behind the 
        laterally spreading \ac{GRB} \ac{BW}. 
        The red lines denote the density of the \ac{CBM}, immediately 
        upstream of the \ac{kN} \ac{BW}. 
        %, \ie, 
        The solid blue line indicates the \ac{kN} \ac{BW} momentum, 
        $\Gamma\beta$. The dashed blue line follows the momentum of the 
        \ac{CBM} upstream of the \ac{kN} \ac{BW}. The dotted blue line 
        corresponds to the relative momentum between the 
        \ac{CBM} and \ac{kN} \ac{BW}. 
        The gray line marks the $\rho_{\cbm} = \rho_{\rm ISM}$, 
        \ie, $1$. 
        For a low initial momentum (\textit{top panel}), the  
        \ac{kN} \ac{BW} stalls behind the overdensity at the 
        forward shock of the \ac{GRB} \ac{BW}. 
        Meanwhile, for a larger momentum \ac{kN} \ac{BW} successfully 
        breaks through the overdensity (\textit{bottom panel}). 
    }
    \label{fig:res:comb_dyn_single_cell}
\end{figure}

\begin{figure*}
    \centering 
    \includegraphics[width=0.49\textwidth]{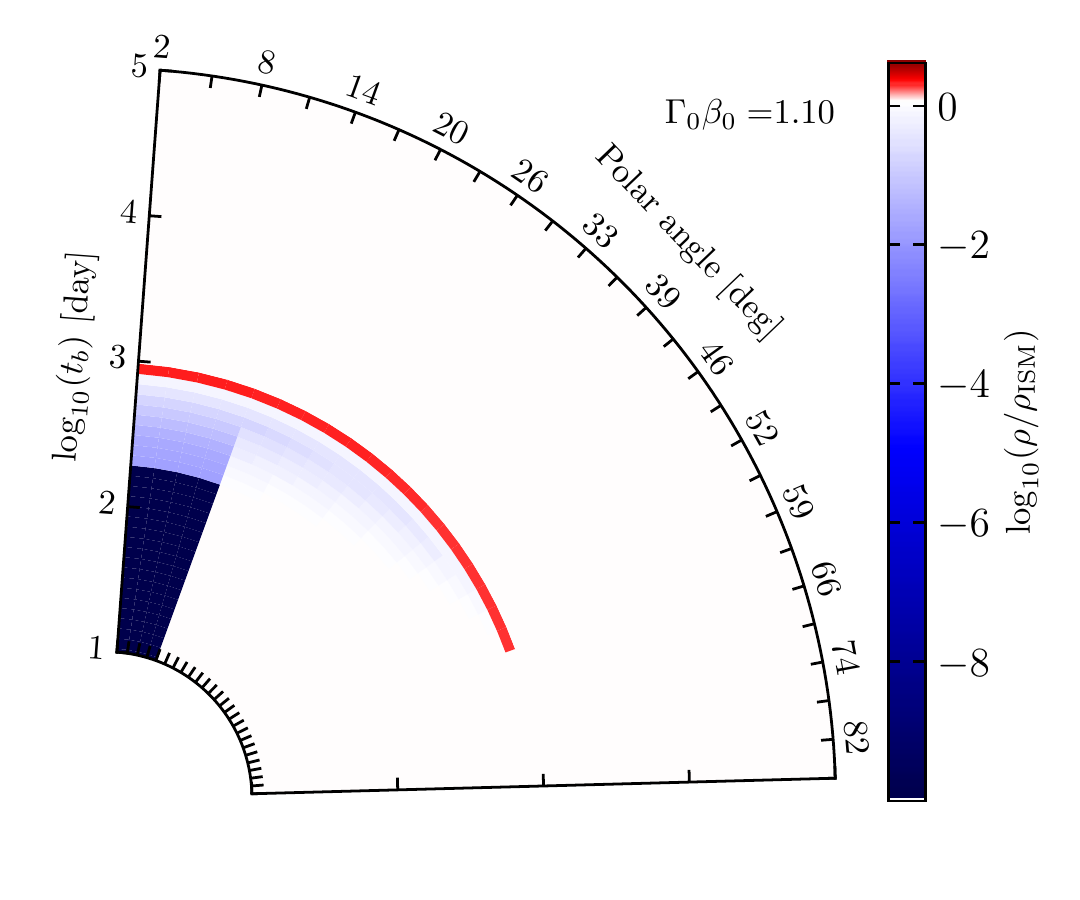}
    \includegraphics[width=0.49\textwidth]{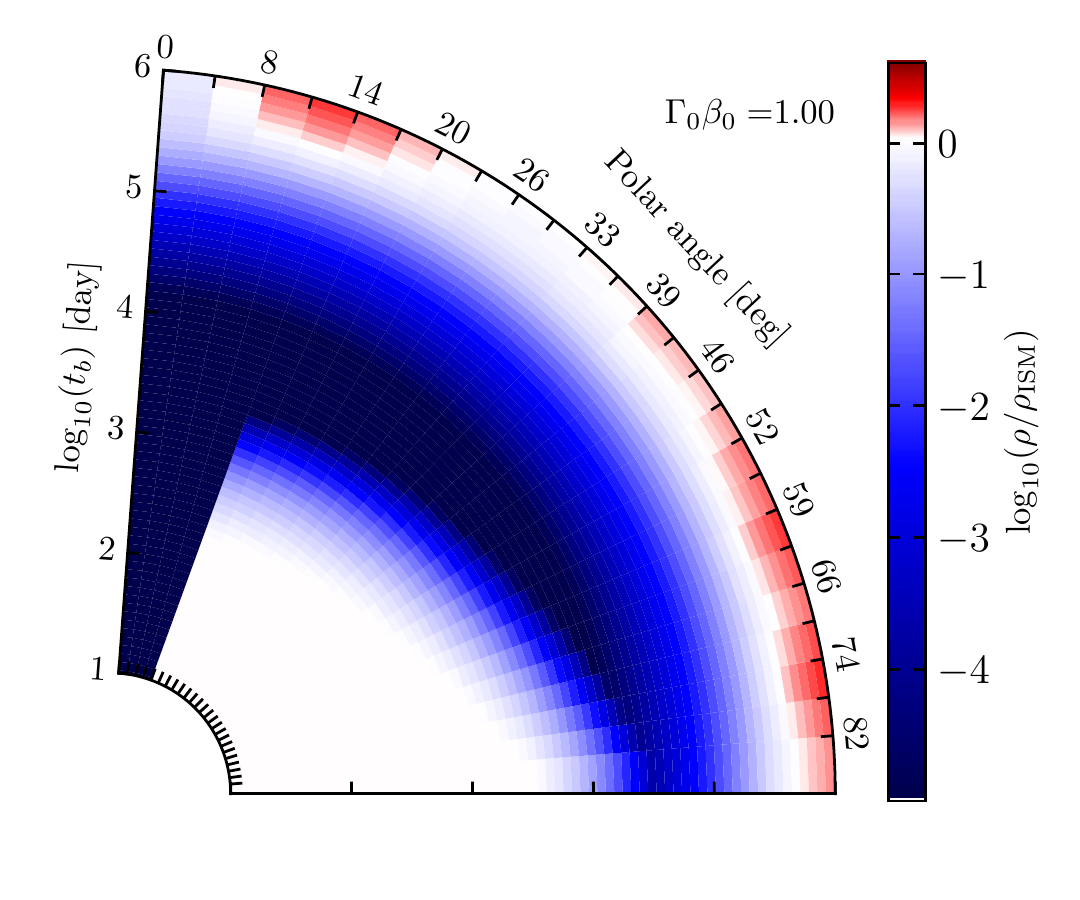}
    \caption{
        \ac{CBM} density as seen by \ac{kN} \acp{BW} with a given 
        initial momentum, $\Gamma_0\beta_0=1.10$ on the 
        \textit{left} and $\Gamma_0\beta_0=1.00$ on the \textit{right}. 
        Blue color indicates densities below that of the \ac{ISM}, 
        which is typically found far behind the \ac{GRB} \ac{BW}. 
        Red color indicates a density higher than $n_\mathrm{ISM}$, indicating
        that the \ac{kN} \ac{BW} caught up with the \ac{GRB} \ac{BW}. 
    }
    \label{fig:res:comb_dyn_two_panels}
\end{figure*}

%\red{Remove 'ISM" when it is not ISM and define clearly "ambient"}
Here, we recall the setup discussed in Sec.~\ref{sec:method} and 
shown in Fig.~\ref{fig:method:structure}. The \ac{GRB} \ac{BW} 
is moving through the \ac{ISM} with a given number density, 
$n_{\rm ISM}$. 
A \ac{kN} \ac{BW} moves through either the \ac{ISM} or the  
\ac{CBM} (see Sec.~\ref{sec:method:dens_prof}), 
depending on whether the \ac{kN} \ac{BW} polar angle is larger 
or smaller than the \ac{GRB} opening angle respectively.

In Fig.~\ref{fig:res:comb_dyn_single_cell} we show, 
for two values of initial \ac{kN} \ac{BW} momentum, the dynamics of 
this \ac{BW} moving behind the \ac{GRB} \ac{BW}, 
as well as the density profile that it encounters.  
In both cases, the \ac{kN} \ac{BW} moves outside of the \ac{GRB} initial 
opening angle, $\theta > \theta_{\rm w}$, and thus encounters the \ac{ISM} 
at the beginning. Later, when the \ac{GRB} \ac{BW} has spread, 
the \ac{kN} \ac{BW} enters the low-density region left by the passage of 
the \ac{GRB} \ac{BW}. Then the normalized upstream density, 
$\rho/\rho_{\rm ISM}$, exponentially decreases.  
Notably, if the density decreases faster than $\rho\propto R^{-3}$, 
the accumulated internal energy can be converted back into the 
bulk kinetic energy and re-accelerate the \ac{BW}  
\cite{Shapiro:1980}. 
In the case of a mildly relativistic, massive \ac{kN} \ac{BW},
however, this re-acceleration is negligible.

When the \ac{GRB} \ac{BW} slows down and the 
\ac{kN} \ac{BW} comes near, it starts to see the exponentially 
increasing density of the \acl{ST} profile, 
shown in Fig.~\ref{fig:res:comb_dyn_single_cell} at  
$t_b\gtrsim10^4\,$days. 
The upstream medium of the \ac{kN} \ac{BW}, however, moves with 
$\Gamma_{\cbm}\beta_{\cbm}$. The relative momentum, 
between the two is $\Gamma_{\rm rel} \beta_{\rm rel}$.  
When the distance between the \acp{BW} is large, both momenta 
remain relatively constant. 
The subsequent evolution depends strongly on the energy budget of 
the \ac{kN} \ac{BW}. A sufficiently fast \ac{BW} can break through the 
overdense \ac{GRB} \ac{BW}. This scenario is shown in the bottom panel of the 
Fig.~\ref{fig:res:comb_dyn_single_cell}. The increase in 
$\Gamma_{\cbm}\beta_{\cbm}$ and decrease in 
$\Gamma_{\rm rel} \beta_{\rm rel}$ before this point are due to 
the onset of \ac{kN} \ac{BW} deceleration. 
However, if the kinetic energy of the \ac{kN} \ac{BW} is insufficient, 
it stalls and $\Gamma_{\cbm}\beta_{\cbm}$ becomes larger than 
$\Gamma_{\rm rel} \beta_{\rm rel}$, meaning that the \ac{kN} \ac{BW} 
bounced off. This scenario is shown in the top panel of 
Fig.~\ref{fig:res:comb_dyn_single_cell}.

Other \acp{BW} into which the \ac{kN} ejecta is discretized follow 
similar evolutionary trajectories. Combined, they comprise the overall 
dynamics of the \ac{kN} ejecta. %\mpo{\ac{BW}}.
In Fig.~\ref{fig:res:comb_dyn_two_panels} the evolution of upstream density, 
$\rho/\rho_{\rm ISM}$ is shown as a function of the \ac{BW} polar angle 
(fixing the \ac{BW} initial momentum). 
At early times (before the lateral spreading of the \ac{GRB} \ac{BW}), 
\ac{kN} \acp{BW} that have polar angle larger than the \ac{GRB} 
opening angle ($\theta > \theta_{\rm w}$) propagate through \ac{ISM}. 
At smaller polar angles ($\theta < \theta_{\rm w}$), the
\ac{kN} \acp{BW} move almost freely through the low-density \ac{CBM}, 
indicated as a dark blue region in the figure.
As the \ac{GRB} \ac{BW} decelerates and spreads, sweeping 
progressively larger amount of \ac{ISM} at larger polar angles, 
it slows down even faster. Thus, a sufficiently fast \ac{kN} \ac{BW}  
at a large polar angle can avoid interacting with the \ac{GRB} \ac{BW} 
entirely. This is shown in the left panel of Fig.~\ref{fig:res:comb_dyn_two_panels} 
where the density remain $\rho=\rho_{\rm ISM}$ throughout the evolution.

When mildly relativistic ejecta moves through cold 
\ac{ISM}, strong shocks form naturally. When the \ac{ISM} is 
pre-accelerated and pre-heated by the \ac{GRB} \ac{BW}, 
shock formation is not guaranteed. 
Thus, not every fluid element of the \ac{kN} ejecta 
moving through \ac{CBM} can form a \ac{BW}. 
A sufficiently high sonic Mach number, 
$\mathcal{M} = \beta_{\rm REL} / c_{s} \gg 1$, the ratio of the 
relative bulk velocity to the sound speed, is required.
The upstream sound speed is $c_{s} = \sqrt{\hat{\gamma}P/\rho}$, 
where $\hat{\gamma}$, $P$ and $\rho$ are the adiabatic index, 
pressure and density of the fluid. 
\citet{Margalit:2020bdk} analytically showed that the flow of the 
\ac{kN} ejecta far behind the \ac{GRB} \ac{BW} is subsonic, 
$\mathcal{M} < 1$. 
However, right before the \ac{kN} ejecta reaches the 
\ac{GRB} \ac{BW}, $\mathcal{M}$ rises to $\mathcal{M}\simeq4$, 
and a ``shock within a shock'' can form. 
We confirm this picture on a qualitative level.
Far behind the \ac{GRB} \ac{BW}, the density 
is low with respect to the pressure, and the sound speed is high, 
exceeding the relative speed of the \ac{kN} ejecta 
($\mathcal{M} < 1$). 
Thus, \ac{kN} ejecta move through the \ac{CBM} without shocking it. 
However, close to the \ac{GRB} \ac{BW}, the density rises faster 
than the pressure, and for sufficiently 
fast part of \ac{kN} ejecta the Mach number becomes 
$\mathcal{M}\gtrsim1$ and shocks form.  
For slow elements of \ac{kN} ejecta $\mathcal{M}$ 
remains below unity, shocks do not form and the ejecta fail to break through.

It is uncertain which minimum value of $\mathcal{M}$ is needed 
for the production of non-thermal electrons 
at the shock. 
First-order Fermi acceleration relies on electrons 
having a gyro-radius much larger than the shock thickness 
(which is of order of ion gyro-radius). 
This is referred to as ``injection problem''; 
cf.~\citet{Balogh:2013} for a textbook discussion. 
Other mechanisms, such as shock drift acceleration or stochastic 
shock drift acceleration, were shown to 
energize electrons enough so 
they may participate in \ac{DSA} later \cite{Guo:2014aua,Guo:2014pka,Kang:2019rtl,Kobzar:2021zyl,Amano:2022tgr}. 
Low-$\mathcal{M}$ shocks in, \eg, galaxy clusters 
are known to produce bright synchrotron radiation from non-thermal 
electrons, likely by re-acceleration of so-called ``fossile'' electrons 
\cite{Pinzke:2013zu,Johnston:2017,Kang:2018puv}. 
In the case of a 
\ac{GRB}-\ac{kN} system such high-energy electrons may naturally 
come from the \ac{GRB} \ac{BW} \citep{Margalit:2020bdk}.  
In this paper, we assume that when a flow is supersonic, 
synchrotron radiation is produced as
described in Sec.~\ref{sec:method:synch_th}.

\begin{figure*}
    \centering 
    \includegraphics[width=0.49\textwidth]{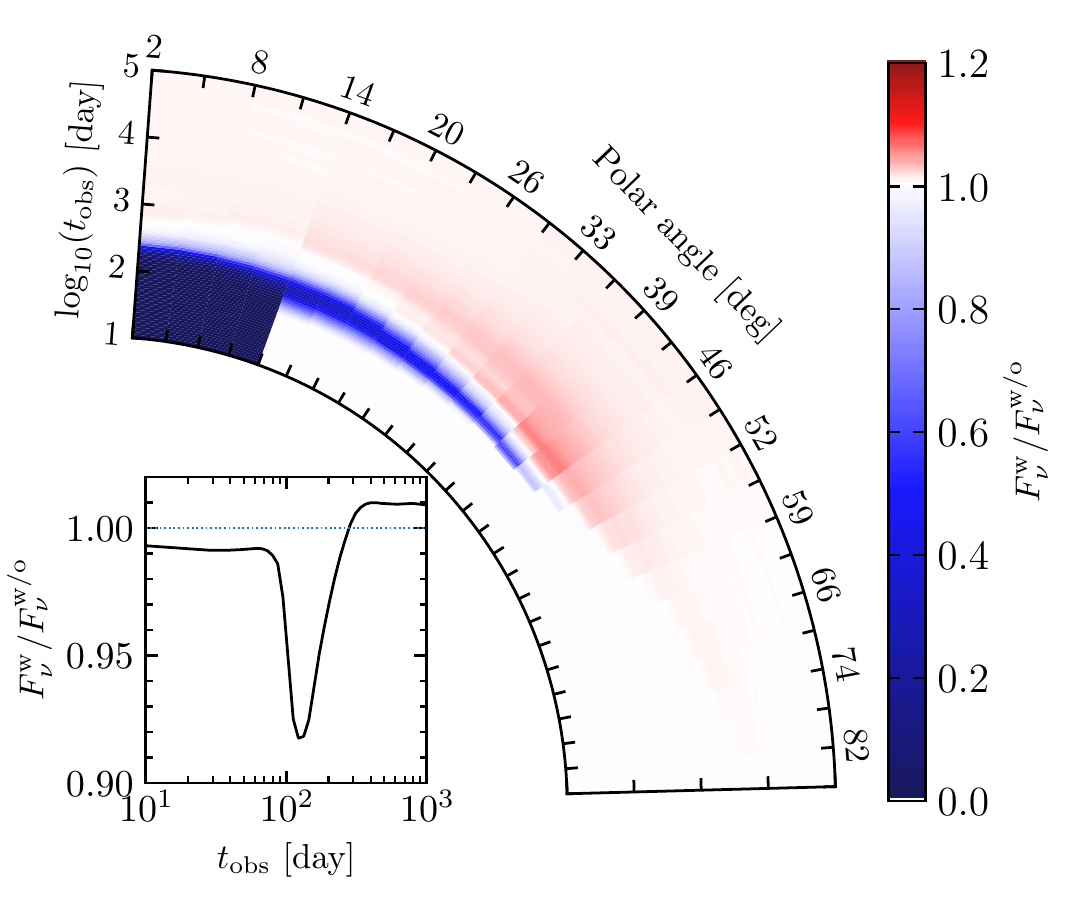}
    \includegraphics[width=0.49\textwidth]{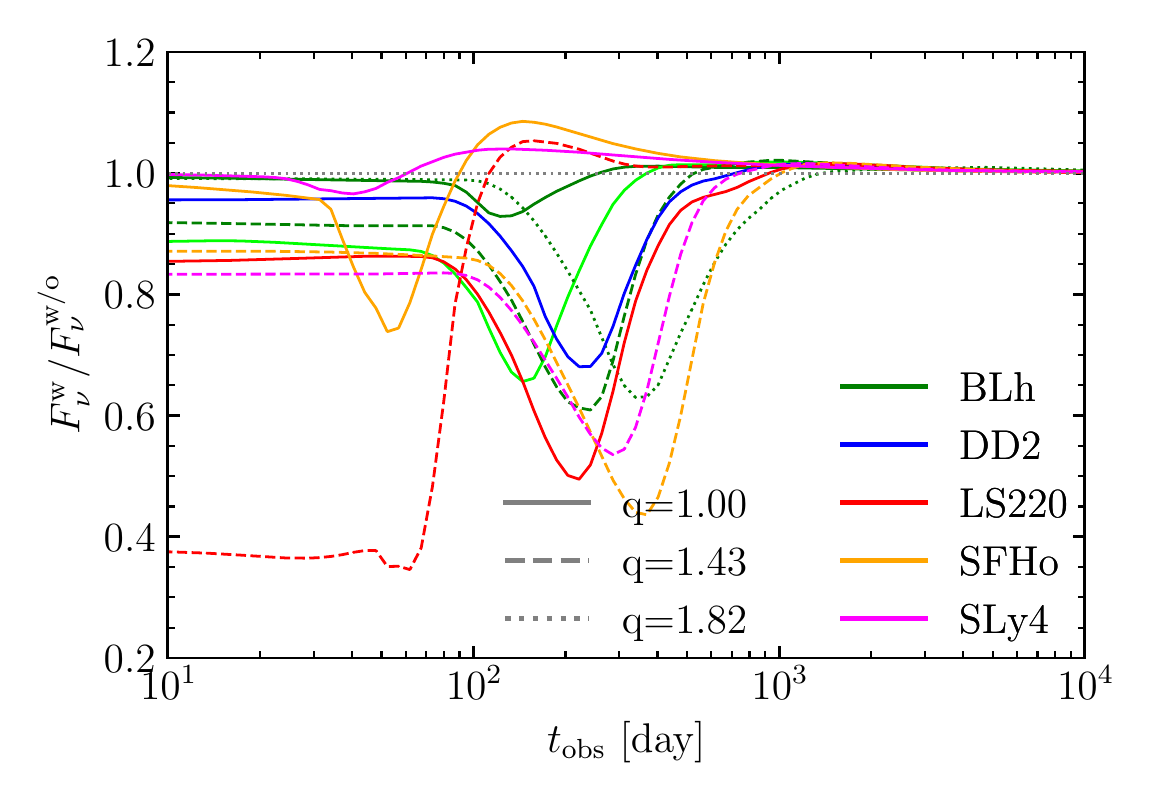}
    \caption{
        \textit{Left panel:} 
        Color-coded flux ratio at $\nu=1\,$GHz with (w) 
        and without (w/o) accounting for the presence of 
        \ac{CBM}, introduced by the passage of \ac{GRB} \ac{BW}. 
        The small bottom left panel displays the ratio of total \acp{LC}, 
        integrating the emission from all \ac{kN} \acp{BW}. 
        Here, the ejecta profile with $q=1$ and the BLh \ac{EOS} is used. 
        \textit{Right panel:} same \ac{LC} ratio but for all simulations. 
    }
    \label{fig:res:comb_lcs_layers}
\end{figure*}

\begin{figure*}
    \centering 
    \includegraphics[width=0.49\textwidth]{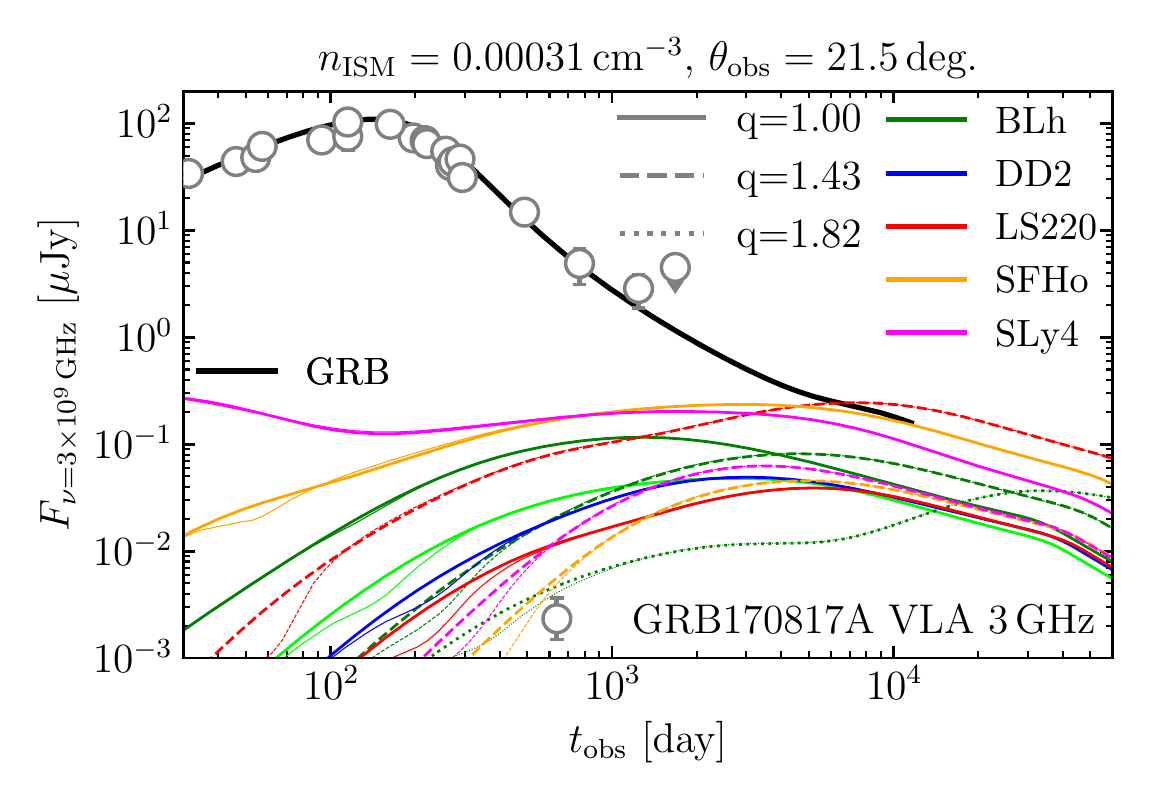}
    \includegraphics[width=0.49\textwidth]{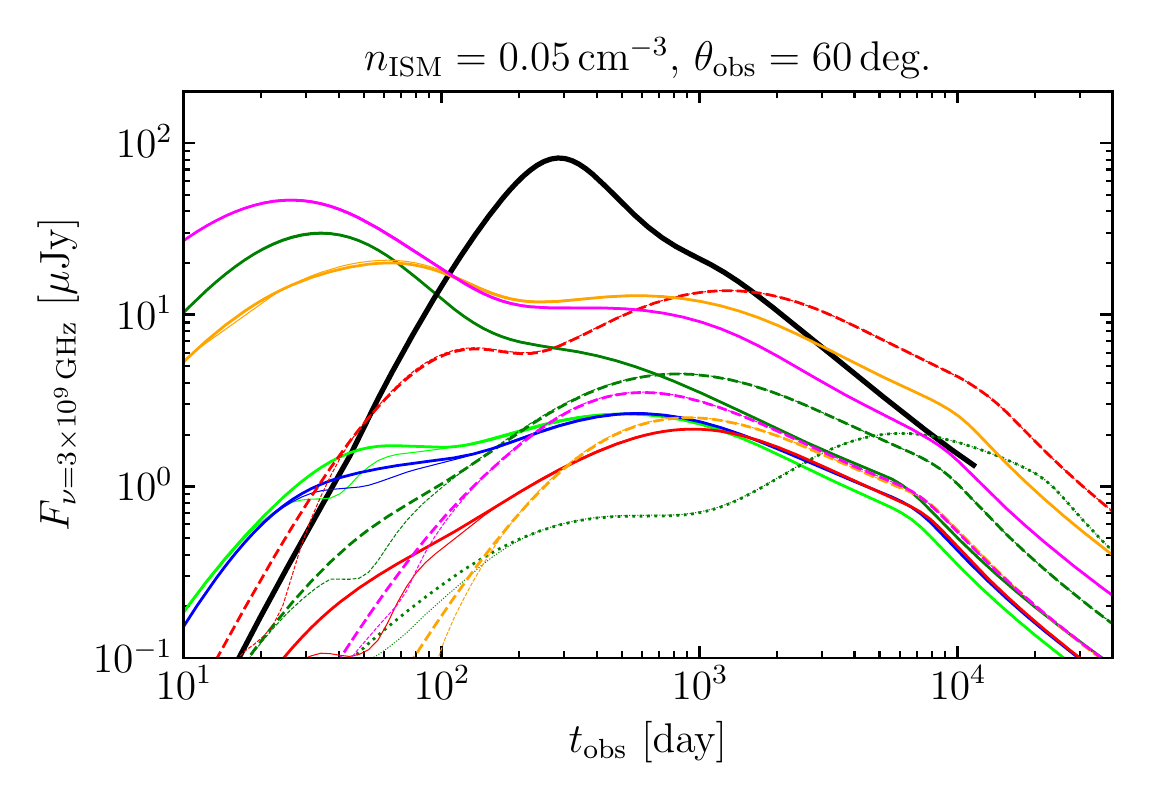}
    \caption{
        Radio ($3\,$GHz) \acp{LC} for all \ac{BNS} merger models.  
        Both, thermal and non-thermal electron populations are considered.  
        \acp{LC} with smaller line width computed accounting for the 
        presence of \ac{CBM}.  
        The \ac{GRB} afterglow \ac{LC} is shown with 
        the black line alongside the observational data 
        \citep{Hajela:2021faz,Balasubramanian:2021kny,Balasubramanian:2022sie}. 
        Left and right panels differ in the choice of 
        $n_{\rm ISM}$ and $\theta_{\rm obs}$. 
    }
    \label{fig:res:lcs_jet_ej}
\end{figure*}

The effect of the \ac{GRB}-altered \ac{CBM} on 
the \ac{kN} afterglow in 
%The effect of the interaction between \ac{kN} and \ac{GRB} ejecta 
terms of the ratio between the radio \acp{LC} 
computed with and without taking this alteration into 
account, $F_{\nu}^{\rm w} / F_{\nu}^{\rm w/o}$, 
is shown in Fig.~\ref{fig:res:comb_lcs_layers}. 
The qualitative behaviour of $F_{\nu}^{\rm w} / F_{\nu}^{\rm w/o}$ is 
similar to that suggested in  
\citet{Duran:2015cua,Margalit:2020bdk}. Early emission is 
suppressed, $F_{\nu}^{\rm w} / F_{\nu}^{\rm w/o} < 1$, 
due to the reduced \ac{CBM} density and %density of evacuated \ac{ISM} for the polar outflow and 
the low Mach number.  %in the shocked post-\ac{GRB} medium. 
The new aspect introduced here is the lateral spreading of the 
\ac{GRB} \ac{BW} and the dependency of the \ac{kN} ejecta velocity 
on the polar angle.  
Indeed, $F_{\nu}^{\rm w} / F_{\nu}^{\rm w/o}$ depends on the 
angular profile of the \ac{kN} ejecta, as the left 
panel of Fig.~\ref{fig:res:comb_lcs_layers} illustrates.  
For equatorial ejecta the flux ratio remains close to unity, as most of the 
\ac{kN} \acp{BW} either avoids interacting with post-\ac{GRB} \ac{CBM}  
entirely or passes through it too quickly to cause an appreciable change 
in the emission. 
Emission from polar ejecta is, however, largely suppressed at early times, 
and also later, if ejecta fails to form shocks and break through the 
overdensity behind the forward shock of the \ac{GRB} \ac{BW}. 
A minimum of $F_{\nu}^{\rm w} / F_{\nu}^{\rm w/o}$ is reached 
when most of the \ac{kN} ejecta resides behind the \ac{GRB} \ac{BW} 
but have not produced a shock. 
At $\theta\gtrsim45\,$ deg. the \ac{kN} outflow is fast enough to 
break through or/and to excite a shock in the \ac{CBM}, 
creating an appreciable excess in observed emission. 

This behaviour is generic and found for other \ac{BNS} models as well, 
as shown in Fig.~\ref{fig:res:comb_lcs_layers} (right panel). 
If the fast tail of the \ac{kN} ejecta is largely polar, as is the case 
for the model with LS220 \ac{EOS} and $q=1.43$ 
(see figure~2 in \citetalias{Nedora:2021eoj}), 
the flux suppression is more prominent and the minimum of 
$F_{\nu}^{\rm w} / F_{\nu}^{\rm w/o}$ is reached earlier. 
In general and across the models, however, the minimum of the 
flux ratio is seen at $t_{\rm FF; \, min}\approx 3\times10^2\,$days.  
For simulations with soft \acp{EOS} and $q=1.0$ we find, on average, 
smaller $t_{\rm FF; \, min}$, and, conversely, a larger $t_{\rm FF; \, min}$ 
we find for models with stiff \acp{EOS} and $q>1$. 
This directly reflects the strength of the core \bnc{} and the 
prominence of the fast tail in the ejecta velocity distribution. 
However, the emission suppression is generally below
$40\,\%$, as the fast tail in all our models is 
largely equatorial and evolves in the \ac{ISM}. 
The variation in flux is achromatic only if a single power-law 
electron distribution is assumed. In the presence of thermal 
electrons the spectral evolution is more complex due to 
steep dependency of $F_{\nu; \rm th}$ on the 
upstream density, as discussed in Sec.~\ref{sec:results:therm_ele}. 
The emission excess of up $10\,\%$ arises when  
\ac{kN} ejecta shocks the \ac{CBM} and is strongest in the model with 
SFHo \ac{EOS} and $q=1.00$. 
For a spherical, uniform outflow (single-shell approximation)  
\citet{Margalit:2020bdk} predicted the excess to be orders of magnitude 
larger and to be observable as ``late-time radio flare''.
We instead argue that the structure of \ac{kN} ejecta as well as the 
finite spreading time of the \ac{GRB} \ac{BW} would smear the sharp peak 
and, depending on the details of the particle 
acceleration and synchrotron emission at $\mathcal{M}\gtrsim1$ shocks, 
would produce a mild emission excess at most. 

In Fig.~\ref{fig:res:lcs_jet_ej} $3\,$GHz \acp{LC} 
are shown for both \ac{kN} and \ac{GRB} afterglows, 
for two values of $n_{\rm ISM}$ and $\theta_{\rm obs}$. 
\acp{LC} produced accounting for \ac{GRB}-\ac{kN} interaction 
are shown with thinner lines, and as expected, the difference with  
respect to those computed without including this interaction is minor 
and only present at early times. 
We reemphasize that free parameters of the \ac{kN} afterglow 
model were not tuned to fit the observations. 
At \ac{ISM} densities inferred for \GRB{} 
(left panel of Fig.~\ref{fig:res:lcs_jet_ej}), 
the \ac{kN} afterglow emission from thermal 
electrons is at most as bright as the non-thermal emission and overall 
lies below the latest upper limits on \GRB{} radio emission 
\cite{Balasubramanian:2022sie}. Thus, the \ac{kN} 
afterglow emission at early times is not bright enough to 
affect the total afterglow. 
At higher densities the emission from thermal electrons is 
significantly brighter, exceeding $10\,\mu$Jy. 
Additionally, as the fast tail of the \ac{kN} outflow is largely equatorial, 
the early emission is further enhanced for a far off-axis observer. 
Meanwhile the \ac{GRB} afterglow is dimmer, as the early emission 
from a collimated jet is beamed away from the observer \ac{LOS}. 
Such a \ac{GRB} afterglow, for which prompt emission also cannot 
be observed, is referred to as an orphan afterglow 
\citep[\eg][]{Nakar:2002ph,Ghirlanda:2015nta,Huang:2020pxr}. 
Thus, the presence of the \ac{kN} afterglow may complicate the 
orphan afterglow signature and possibly contribute to the current 
non-detection of the \ac{GRB} orphan afterglows.

\section{Discussion \& conclusion}\label{sec:conclusion}

One of the observables of \ac{BNS} mergers is the \ac{kN} 
afterglow. 
The mechanism behind this transient is similar to that of 
the \ac{GRB} afterglow, but instead of a highly relativistic 
\ac{GRB} ejecta, the mildly relativistic \ac{kN} ejecta shocks 
the ambient medium and produces the emission 
\citep[\eg][]{Nakar:2019fza}. 
The radio flux of the \ac{kN} afterglow is expected to peak 
on the deceleration timescale, which is of the order of years.
Its properties are determined primarily by the velocity and 
angular distribution of ejecta and unknown microphysical 
parameters, governing particle acceleration at mildly 
relativistic shocks. 
Thus, if detected, a \ac{kN} afterglow could provide additional 
constraints on the ejecta properties, and specifically, on the 
fast component of the dynamical ejecta. 
Such information could be used to place additional constraints on the 
properties of merging \acp{NS} and the \ac{NS} \ac{EOS}. 
In \citetalias{Nedora:2021eoj} we considered \GRB{} 
which was accompanied by the \ac{kN} \AT{}. Using the latest 
Chandra and \ac{VLA} observations 
\cite{Balasubramanian:2021kny,Hajela:2021faz} 
and dynamical ejecta profiles from ab-initio 
\ac{NR} \ac{BNS} merger simulations with advanced input physics 
\cite{Radice:2018pdn,Nedora:2020pak}, 
we illustrated how such constraints can be placed.  
In this work we considered an impact 
on the \ac{kN} afterglow of 
(i) a mixture of thermal and non-thermal electron populations 
producing synchrotron radiation,  
%in the \ac{kN} \ac{BW}  and 
(ii) an upstream medium that is altered and 
pre-accelerated by the laterally spreading \ac{GRB} 
\ac{BW}. 

Both observations and \ac{PIC} simulations support 
the presence of a significant thermal 
electron population behind mildly relativistic shocks 
\cite{Park:2014lqa,Crumley:2018kvf,Ho:2019dlw,2021MNRAS.502.5065L}. 
We find that the emission from this population can dominate 
the early \ac{kN} afterglow in radio band. At sufficiently high 
densities, $n_{\rm ISM}\simeq0.1\,\ccm$, radio \acp{LC} can have   
a double-peak structure. The strong velocity dependence of the 
emissivity from thermal electrons leads to a characteristic 
evolution of the spectra as the fastest \ac{kN} \acp{BW} decelerate 
and the contribution from thermal electron population to overall emission 
decreases. 
%from thermal to non-thermal spectra, as the fastest 
%\ac{kN} \ac{BW} elements   
%decelerate. 
Thus, a characteristic increase in the spectral 
index in the radio band may be used to constrain the ejecta 
velocity distribution. 
Additionally, we find a relation between the time of 
the \ac{LC} peak and the frequency at which one observes the 
transition of the spectrum from being dominated by the 
emission from thermal electrons to the one dominated by the 
emission from non-thermal electrons. 
This relation depends only weakly on microphysical parameters and 
$n_{\rm ISM}$, and thus can be used to constrain 
the presence of the fast tail in the ejecta velocity distribution. 

At densities similar to those inferred for \GRB{}, we find the
\ac{kN} afterglow in the radio band ($3\,$GHz) peaking at 
$10^3-10^4\,$days, reaching a flux $\lesssim0.1\,\mu$Jy, 
which is below the latest upper limits
\cite{Balasubramanian:2022sie}. 
However, as the \ac{LC} peak flux depends 
strongly on the microphysics of the shock, we 
cannot place stringent constraints in this case. 
At higher \ac{ISM} densities the early \ac{kN} afterglow 
may be observable at the distance of \GRB{} but it would be 
overshadowed by the \ac{GRB} afterglow, unless observed far 
off-axis. 
There, \ac{GRB} orphan afterglow and \ac{kN} afterglow are comparably 
bright. Thus, \ac{kN} afterglow may be an important factor 
in search strategies for \ac{GRB} orphan afterglows.

As \ac{GRB} and \ac{kN} ejecta move through the same 
environment, it is natural to expect that the former 
would affect the \ac{kN} afterglow. 
Here we considered how the dynamics of and the radiation from 
\ac{kN} \acp{BW} change when they move through the \ac{CBM} 
with density profile dependent on the position and properties 
of the laterally spreading \ac{GRB} \ac{BW} ahead. 
The early \ac{kN} afterglow is slightly (${\lesssim}20\,\%$) 
dimmer due to the lower \ac{CBM} density (with respect to the \ac{ISM}) 
behind the \ac{GRB} \ac{BW}. Later, lateral spreading of 
the \ac{GRB} \ac{BW} increases the area of low-density, 
pre-accelerated \ac{CBM} through which \ac{kN} outflow moves 
subsonically. 
This implies a more significant reduction in observed 
flux (${\lesssim}40\,\%$), followed by a slight brightening
(${\lesssim}10\,\%$), when the \ac{kN} flow excite shocks 
in the overdense part of the \ac{CBM} at the \ac{GRB} \ac{BW}. 
Thus, early-time variability in \ac{kN} afterglow \acp{LC}, 
besides the spectral evolution, may also be present due to  
the interaction with the modified upstream medium, albeit 
the former has a much stronger effect.  
If, on the other hand, the \ac{kN} ejecta velocity distribution 
is such that the fastest outflow is polar instead of equatorial, 
the suppression of emission might be much more significant, 
and, potentially, observable. 
Moreover, a system of two mildly relativistic shocks, one 
approaching another is an interesting and, to the best 
of our knowledge, unexplored setting for particle 
acceleration and synchrotron emission with seed particles. % photons. 

% ----- LIMITATIONS

The main limitations of our study relate to the 
semi-analytic models of \ac{GRB} and \ac{kN} afterglows. 
It remains to be investigated whether the qualitative results 
presented here would also be found in numerical \acl{HD} 
simulations. Such simulations, however, even with novel 
techniques like moving mesh \cite{Xie:2018vya,Akcay:2018yyh}, 
are numerically expensive. 
Additionally, the theory of particle acceleration at 
mildly relativistic shocks with very heavy ions, 
(produced in \rproc{}) is currently not well understood. 
This limits our ability to predict the properties of \ac{kN} 
afterglows. 
Nevertheless, our improved capability to localize off-axis 
\acp{GRB} using \ac{GW} detectors and the 
improved sensitivity of new radio observatories would allow 
us in the near future to follow these \acp{GRB} for longer, 
and to place constraints on the \ac{kN} afterglow properties 
and physical processes operating at shocks.

\section*{Acknowledgements}

The simulations were performed on the national supercomputer HPE Apollo Hawk at the High Performance Computing (HPC) Center Stuttgart (HLRS) under the grant number GWanalysis/44189 and on the GCS Supercomputer SuperMUC at Leibniz Supercomputing Centre (LRZ) [project pn29ba].

\textit{Software:} We are grateful to the countless developers 
contributing to open source projects that was used in the analysis 
of the simulation results of this 
work: \texttt{NumPy} \citep{numpy}, \texttt{Matplotlib} \cite{matplotlib}, and \texttt{SciPy} \cite{scipy}.

\textit{Data Availability:} 
The datasets generated during and/or analysed during the current 
study are available from the corresponding author on reasonable request.

\section*{Data avalibility}
The data underlying this article will be shared on reasonable request to the corresponding author.

% -------------------------------------------------------------------------

\bibliography{refs20230113}

\begin{thebibliography}{}
\makeatletter
\relax
\def\mn@urlcharsother{\let\do\@makeother \do\$\do\&\do\#\do\^\do\_\do\%\do\~}
\def\mn@doi{\begingroup\mn@urlcharsother \@ifnextchar [ {\mn@doi@}
  {\mn@doi@[]}}
\def\mn@doi@[#1]#2{\def\@tempa{#1}\ifx\@tempa\@empty \href
  {http://dx.doi.org/#2} {doi:#2}\else \href {http://dx.doi.org/#2} {#1}\fi
  \endgroup}
\def\mn@eprint#1#2{\mn@eprint@#1:#2::\@nil}
\def\mn@eprint@arXiv#1{\href {http://arxiv.org/abs/#1} {{\tt arXiv:#1}}}
\def\mn@eprint@dblp#1{\href {http://dblp.uni-trier.de/rec/bibtex/#1.xml}
  {dblp:#1}}
\def\mn@eprint@#1:#2:#3:#4\@nil{\def\@tempa {#1}\def\@tempb {#2}\def\@tempc
  {#3}\ifx \@tempc \@empty \let \@tempc \@tempb \let \@tempb \@tempa \fi \ifx
  \@tempb \@empty \def\@tempb {arXiv}\fi \@ifundefined
  {mn@eprint@\@tempb}{\@tempb:\@tempc}{\expandafter \expandafter \csname
  mn@eprint@\@tempb\endcsname \expandafter{\@tempc}}}

\bibitem[\protect\citeauthoryear{Abbott et~al.}{Abbott
  et~al.}{2017}]{Monitor:2017mdv}
Abbott B.~P.,  et~al., 2017, \mn@doi [Astrophys. J. Lett.]
  {10.3847/2041-8213/aa920c}, 848, L13

\bibitem[\protect\citeauthoryear{Abbott et~al.}{Abbott
  et~al.}{2018}]{Aasi:2013wya}
Abbott B.,  et~al., 2018, \mn@doi [Living Rev. Rel.]
  {10.1007/s41114-018-0012-9}, 21, 3

\bibitem[\protect\citeauthoryear{Aharonian, Kelner  \& Prosekin}{Aharonian
  et~al.}{2010}]{Aharonian:2010}
Aharonian F.~A.,  Kelner S.~R.,   Prosekin A.~Y.,  2010, \mn@doi [Physical
  Review D] {10.1103/PhysRevD.82.043002}, 82, 043002

\bibitem[\protect\citeauthoryear{Aharonian et~al.}{Aharonian
  et~al.}{2013}]{Aharonian:2013av}
Aharonian F.,  et~al., 2013

\bibitem[\protect\citeauthoryear{Ajello et~al.}{Ajello
  et~al.}{2016}]{TheFermi-LAT:2015kwa}
Ajello M.,  et~al., 2016, \mn@doi [Astrophys. J.] {10.3847/0004-637X/819/1/44},
  819, 44

\bibitem[\protect\citeauthoryear{Akcay, Bernuzzi, Messina, Nagar, Ortiz  \&
  Rettegno}{Akcay et~al.}{2019}]{Akcay:2018yyh}
Akcay S.,  Bernuzzi S.,  Messina F.,  Nagar A.,  Ortiz N.,   Rettegno P.,
  2019, \mn@doi [Phys. Rev. D] {10.1103/PhysRevD.99.044051}, 99, 044051

\bibitem[\protect\citeauthoryear{Alexander et~al.}{Alexander
  et~al.}{2017}]{Alexander:2017aly}
Alexander K.~D.,  et~al., 2017, \mn@doi [Astrophys. J.]
  {10.3847/2041-8213/aa905d}, 848, L21

\bibitem[\protect\citeauthoryear{Alexander et~al.}{Alexander
  et~al.}{2018}]{Alexander:2018dcl}
Alexander K.,  et~al., 2018, \mn@doi [Astrophys.\ J.]
  {10.3847/2041-8213/aad637}, 863, L18

\bibitem[\protect\citeauthoryear{Amano \& Hoshino}{Amano \&
  Hoshino}{2022}]{Amano:2022tgr}
Amano T.,  Hoshino M.,  2022, \mn@doi [Astrophys. J.]
  {10.3847/1538-4357/ac4f49}, 927, 132

\bibitem[\protect\citeauthoryear{Arcavi et~al.}{Arcavi
  et~al.}{2017}]{Arcavi:2017xiz}
Arcavi I.,  et~al., 2017, \mn@doi [Nature] {10.1038/nature24291}, 551, 64

\bibitem[\protect\citeauthoryear{Arnett}{Arnett}{1982}]{Arnett:1982}
Arnett W.~D.,  1982, \mn@doi [The Astrophysical Journal] {10.1086/159681},
  \href {https://ui.adsabs.harvard.edu/abs/1982ApJ...253..785A} {253, 785}

\bibitem[\protect\citeauthoryear{Ayache, van Eerten  \& Eardley}{Ayache
  et~al.}{2021}]{Ayache:2021six}
Ayache E.~H.,  van Eerten H.~J.,   Eardley R.~W.,  2021, \mn@doi [Mon. Not.
  Roy. Astron. Soc.] {10.1093/mnras/stab3509}, 510, 1315

\bibitem[\protect\citeauthoryear{Bai, Caprioli, Sironi  \& Spitkovsky}{Bai
  et~al.}{2015}]{Bai:2014kca}
Bai X.-N.,  Caprioli D.,  Sironi L.,   Spitkovsky A.,  2015, \mn@doi
  [Astrophys. J.] {10.1088/0004-637X/809/1/55}, 809, 55

\bibitem[\protect\citeauthoryear{Balasubramanian et~al.,}{Balasubramanian
  et~al.}{2021}]{Balasubramanian:2021kny}
Balasubramanian A.,  et~al., 2021, \mn@doi [Astrophys. J. Lett.]
  {10.3847/2041-8213/abfd38}, 914, L20

\bibitem[\protect\citeauthoryear{Balasubramanian et~al.,}{Balasubramanian
  et~al.}{2022}]{Balasubramanian:2022sie}
Balasubramanian A.,  et~al., 2022

\bibitem[\protect\citeauthoryear{Balogh \& Treumann}{Balogh \&
  Treumann}{2013}]{Balogh:2013}
Balogh A.,  Treumann R.~A.,  2013, Physics of Collisionless Shocks.
Space Plasma Shock Waves, Springer-Verlag New York Inc., \url
  {https://link.springer.com/book/10.1007/978-1-4614-6099-2}

\bibitem[\protect\citeauthoryear{Barnes, Kasen, Wu  \& Martínez-Pinedo}{Barnes
  et~al.}{2016}]{Barnes:2016umi}
Barnes J.,  Kasen D.,  Wu M.-R.,   Martínez-Pinedo G.,  2016, \mn@doi
  [Astrophys. J.] {10.3847/0004-637X/829/2/110}, 829, 110

\bibitem[\protect\citeauthoryear{Bauswein, Goriely  \& Janka}{Bauswein
  et~al.}{2013}]{Bauswein:2013yna}
Bauswein A.,  Goriely S.,   Janka H.-T.,  2013, \mn@doi [Astrophys.J.]
  {10.1088/0004-637X/773/1/78}, 773, 78

\bibitem[\protect\citeauthoryear{Beckers \& Beckers}{Beckers \&
  Beckers}{2012}]{Beckers:2012}
Beckers B.,  Beckers P.,  2012, \mn@doi [Computational Geometry]
  {https://doi.org/10.1016/j.comgeo.2012.01.011}, 45, 275

\bibitem[\protect\citeauthoryear{Beloborodov}{Beloborodov}{2002}]{Beloborodov:2002nkf}
Beloborodov A.~M.,  2002, eConf, C0208122, 4

\bibitem[\protect\citeauthoryear{Beloborodov}{Beloborodov}{2008}]{Beloborodov:2008nx}
Beloborodov A.~M.,  2008, \mn@doi [AIP Conf. Proc.] {10.1063/1.3002509}, 1054,
  51

\bibitem[\protect\citeauthoryear{Beniamini, Granot  \& Gill}{Beniamini
  et~al.}{2020}]{Beniamini:2020eza}
Beniamini P.,  Granot J.,   Gill R.,  2020, \mn@doi [Mon. Not. Roy. Astron.
  Soc.] {10.1093/mnras/staa538}, 493, 3521

\bibitem[\protect\citeauthoryear{Berger, Fong  \& Chornock}{Berger
  et~al.}{2013}]{Berger:2013wna}
Berger E.,  Fong W.,   Chornock R.,  2013, \mn@doi [Astrophys. J. Lett.]
  {10.1088/2041-8205/774/2/L23}, 774, L23

\bibitem[\protect\citeauthoryear{Bernuzzi}{Bernuzzi}{2020}]{Bernuzzi:2020tgt}
Bernuzzi S.,  2020, \mn@doi [Gen. Rel. Grav.] {10.1007/s10714-020-02752-5}, 52,
  108

\bibitem[\protect\citeauthoryear{Bernuzzi et~al.}{Bernuzzi
  et~al.}{2020}]{Bernuzzi:2020txg}
Bernuzzi S.,  et~al., 2020, \mn@doi [Mon. Not. Roy. Astron. Soc.]
  {10.1093/mnras/staa1860}

\bibitem[\protect\citeauthoryear{Blandford \& McKee}{Blandford \&
  McKee}{1976}]{Blandford:1976}
Blandford R.~D.,  McKee C.~F.,  1976, \mn@doi [Physics of Fluids]
  {10.1063/1.861619}, \href
  {https://ui.adsabs.harvard.edu/abs/1976PhFl...19.1130B} {19, 1130}

\bibitem[\protect\citeauthoryear{Blandford \& Ostriker}{Blandford \&
  Ostriker}{1978}]{Blandford:1978}
Blandford R.~D.,  Ostriker J.~P.,  1978, \mn@doi [The Astrophysical Journal
  Letters] {10.1086/182658}, \href
  {https://ui.adsabs.harvard.edu/abs/1978ApJ...221L..29B} {221, L29}

\bibitem[\protect\citeauthoryear{Blandford \& Znajek}{Blandford \&
  Znajek}{1977}]{Blandford:1977ds}
Blandford R.~D.,  Znajek R.~L.,  1977, \mn@doi [Mon. Not. Roy. Astron. Soc.]
  {10.1093/mnras/179.3.433}, 179, 433

\bibitem[\protect\citeauthoryear{Book}{Book}{1994}]{Book:1994}
Book D.~L.,  1994, \mn@doi [Shock Waves] {10.1007/BF01414626}, \href
  {https://ui.adsabs.harvard.edu/abs/1994ShWav...4....1B} {4, 1}

\bibitem[\protect\citeauthoryear{Bucciantini, Metzger, Thompson  \&
  Quataert}{Bucciantini et~al.}{2012}]{Bucciantini:2011kx}
Bucciantini N.,  Metzger B.,  Thompson T.,   Quataert E.,  2012, \mn@doi [Mon.
  Not. Roy. Astron. Soc.] {10.1111/j.1365-2966.2011.19810.x}, 419, 1537

\bibitem[\protect\citeauthoryear{Bulla}{Bulla}{2019}]{Bulla:2019muo}
Bulla M.,  2019, \mn@doi [Mon. Not. Roy. Astron. Soc.] {10.1093/mnras/stz2495},
  489, 5037

\bibitem[\protect\citeauthoryear{Camilletti et~al.,}{Camilletti
  et~al.}{2022}]{Camilletti:2022jms}
Camilletti A.,  et~al., 2022

\bibitem[\protect\citeauthoryear{Carilli \& Rawlings}{Carilli \&
  Rawlings}{2004}]{Carilli:2004nx}
Carilli C.~L.,  Rawlings S.,  2004, \mn@doi [New Astron. Rev.]
  {10.1016/j.newar.2004.09.001}, 48, 979

\bibitem[\protect\citeauthoryear{Cerd{\'a}-Dur{\'a}n, Obergaulinger, Aloy, Font
   \& M{\"u}ller}{Cerd{\'a}-Dur{\'a}n et~al.}{2011}]{Cerda-Duran:2011}
Cerd{\'a}-Dur{\'a}n P.,  Obergaulinger M.,  Aloy M.~A.,  Font J.~A.,
  M{\"u}ller E.,  2011, in {Journal of Physics Conference Series}. p. 012079,
  \mn@doi{10.1088/1742-6596/314/1/012079}

\bibitem[\protect\citeauthoryear{Chevalier}{Chevalier}{1982}]{Chevalier:1982}
Chevalier R.~A.,  1982, \mn@doi [Astrophysics.J] {10.1086/160126}, \href
  {https://ui.adsabs.harvard.edu/abs/1982ApJ...258..790C} {258, 790}

\bibitem[\protect\citeauthoryear{Chiaberge \& Ghisellini}{Chiaberge \&
  Ghisellini}{1999}]{Chiaberge:1998cv}
Chiaberge M.,  Ghisellini G.,  1999, \mn@doi [Mon. Not. Roy. Astron. Soc.]
  {10.1046/j.1365-8711.1999.02538.x}, 306, 551

\bibitem[\protect\citeauthoryear{Chiang \& Dermer}{Chiang \&
  Dermer}{1999}]{Chiang:1998wh}
Chiang J.,  Dermer C.~D.,  1999, \mn@doi [Astrophys. J.] {10.1086/306789}, 512,
  699

\bibitem[\protect\citeauthoryear{Corsi et~al.,}{Corsi
  et~al.}{2019}]{Corsi:2019}
Corsi A.,  et~al., 2019, Bulletin of the American Astronomical Society, \href
  {https://ui.adsabs.harvard.edu/abs/2019BAAS...51c.209C} {51, 209}

\bibitem[\protect\citeauthoryear{Coulter et~al.}{Coulter
  et~al.}{2017}]{Coulter:2017wya}
Coulter D.~A.,  et~al., 2017, \mn@doi [Science] {10.1126/science.aap9811}

\bibitem[\protect\citeauthoryear{Crumley, Caprioli, Markoff  \&
  Spitkovsky}{Crumley et~al.}{2019}]{Crumley:2018kvf}
Crumley P.,  Caprioli D.,  Markoff S.,   Spitkovsky A.,  2019, \mn@doi [Mon.
  Not. Roy. Astron. Soc.] {10.1093/mnras/stz232}, 485, 5105

\bibitem[\protect\citeauthoryear{Cusinato, Guercilena, Perego, Logoteta,
  Radice, Bernuzzi  \& Ansoldi}{Cusinato et~al.}{2021}]{Cusinato:2021zin}
Cusinato M.,  Guercilena F.~M.,  Perego A.,  Logoteta D.,  Radice D.,  Bernuzzi
  S.,   Ansoldi S.,  2021, ] {10.1140/epja/s10050-022-00743-5}

\bibitem[\protect\citeauthoryear{Damour \& Nagar}{Damour \&
  Nagar}{2009}]{Damour:2009vw}
Damour T.,  Nagar A.,  2009, \mn@doi [Phys. Rev.] {10.1103/PhysRevD.80.084035},
  D80, 084035

\bibitem[\protect\citeauthoryear{De~Colle, Ramirez-Ruiz, Granot  \&
  Lopez-Camara}{De~Colle et~al.}{2012}]{De:2012}
De~Colle F.,  Ramirez-Ruiz E.,  Granot J.,   Lopez-Camara D.,  2012, \mn@doi
  [The Astrophysical Journal] {10.1088/0004-637X/751/1/57}, \href
  {https://ui.adsabs.harvard.edu/abs/2012ApJ...751...57D} {751, 57}

\bibitem[\protect\citeauthoryear{Dermer \& Chiang}{Dermer \&
  Chiang}{1998}]{Dermer:1997pv}
Dermer C.~D.,  Chiang J.,  1998, \mn@doi [New Astron.]
  {10.1016/S1384-1076(98)00004-9}, 3, 157

\bibitem[\protect\citeauthoryear{Dermer \& Humi}{Dermer \&
  Humi}{2001}]{Dermer:2000gu}
Dermer C.~D.,  Humi M.,  2001, \mn@doi [Astrophys. J.] {10.1086/321580}, 556,
  479

\bibitem[\protect\citeauthoryear{Dermer \& Menon}{Dermer \&
  Menon}{2009}]{Dermer:2009}
Dermer C.~D.,  Menon G.,  2009, {High Energy Radiation from Black Holes: Gamma
  Rays, Cosmic Rays, and Neutrinos}

\bibitem[\protect\citeauthoryear{Desai, Metzger  \& Foucart}{Desai
  et~al.}{2019}]{Desai:2018rbc}
Desai D.,  Metzger B.~D.,   Foucart F.,  2019, \mn@doi [Mon. Not. Roy. Astron.
  Soc.] {10.1093/mnras/stz644}, 485, 4404

\bibitem[\protect\citeauthoryear{Dessart, Ott, Burrows, Rosswog  \&
  Livne}{Dessart et~al.}{2009}]{Dessart:2008zd}
Dessart L.,  Ott C.,  Burrows A.,  Rosswog S.,   Livne E.,  2009, \mn@doi
  [Astrophys.J.] {10.1088/0004-637X/690/2/1681}, 690, 1681

\bibitem[\protect\citeauthoryear{Dietrich \& Ujevic}{Dietrich \&
  Ujevic}{2017}]{Dietrich:2016fpt}
Dietrich T.,  Ujevic M.,  2017, \mn@doi [Class. Quant. Grav.]
  {10.1088/1361-6382/aa6bb0}, 34, 105014

\bibitem[\protect\citeauthoryear{Dietrich, Ujevic, Tichy, Bernuzzi  \&
  Br{\"u}gmann}{Dietrich et~al.}{2017}]{Dietrich:2016hky}
Dietrich T.,  Ujevic M.,  Tichy W.,  Bernuzzi S.,   Br{\"u}gmann B.,  2017,
  \mn@doi [Phys. Rev.] {10.1103/PhysRevD.95.024029}, D95, 024029

\bibitem[\protect\citeauthoryear{Douchin \& Haensel}{Douchin \&
  Haensel}{2001}]{Douchin:2001sv}
Douchin F.,  Haensel P.,  2001, Astron. Astrophys., 380, 151

\bibitem[\protect\citeauthoryear{Drout et~al.}{Drout
  et~al.}{2017}]{Drout:2017ijr}
Drout M.~R.,  et~al., 2017, \mn@doi [Science] {10.1126/science.aaq0049}, 358,
  1570

\bibitem[\protect\citeauthoryear{Duffell \& MacFadyen}{Duffell \&
  MacFadyen}{2013}]{Duffell:2013tha}
Duffell P.~C.,  MacFadyen A.~I.,  2013, \mn@doi [Astrophys. J.]
  {10.1088/0004-637X/775/2/87}, 775, 87

\bibitem[\protect\citeauthoryear{Duffell, Quataert, Kasen  \& Klion}{Duffell
  et~al.}{2018}]{Duffell:2018iig}
Duffell P.~C.,  Quataert E.,  Kasen D.,   Klion H.,  2018, \mn@doi [Astrophys.
  J.] {10.3847/1538-4357/aae084}, 866, 3

\bibitem[\protect\citeauthoryear{Duran \& Giannios}{Duran \&
  Giannios}{2015}]{Duran:2015cua}
Duran R.~B.,  Giannios D.,  2015, \mn@doi [Mon. Not. Roy. Astron. Soc.]
  {10.1093/mnras/stv2004}, 454, 1711

\bibitem[\protect\citeauthoryear{Eichler, Livio, Piran  \& Schramm}{Eichler
  et~al.}{1989}]{Eichler:1989ve}
Eichler D.,  Livio M.,  Piran T.,   Schramm D.~N.,  1989, \mn@doi [Nature]
  {10.1038/340126a0}, 340, 126

\bibitem[\protect\citeauthoryear{Endrizzi et~al.,}{Endrizzi
  et~al.}{2020}]{Endrizzi:2019trv}
Endrizzi A.,  et~al., 2020, \mn@doi [Eur. Phys. J. A]
  {10.1140/epja/s10050-019-00018-6}, 56, 15

\bibitem[\protect\citeauthoryear{Evans et~al.}{Evans
  et~al.}{2017}]{Evans:2017mmy}
Evans P.~A.,  et~al., 2017, \mn@doi [Science] {10.1126/science.aap9580}, 358,
  1565

\bibitem[\protect\citeauthoryear{Fahlman \& Fern{\'a}ndez}{Fahlman \&
  Fern{\'a}ndez}{2018}]{Fahlman:2018llv}
Fahlman S.,  Fern{\'a}ndez R.,  2018, \mn@doi [Astrophys. J.]
  {10.3847/2041-8213/aaf1ab}, 869, L3

\bibitem[\protect\citeauthoryear{Favata}{Favata}{2014}]{Favata:2013rwa}
Favata M.,  2014, \mn@doi [Phys.Rev.Lett.] {10.1103/PhysRevLett.112.101101},
  112, 101101

\bibitem[\protect\citeauthoryear{Fern\'{a}ndez \& Metzger}{Fern\'{a}ndez \&
  Metzger}{2013}]{Fernandez:2013tya}
Fern\'{a}ndez R.,  Metzger B.~D.,  2013, \mn@doi [Mon. Not. Roy. Astron. Soc.]
  {10.1093/mnras/stt1312}, 435, 502

\bibitem[\protect\citeauthoryear{Fern\'{a}ndez \& Metzger}{Fern\'{a}ndez \&
  Metzger}{2016}]{Fernandez:2015use}
Fern\'{a}ndez R.,  Metzger B.~D.,  2016, \mn@doi [Ann. Rev. Nucl. Part. Sci.]
  {10.1146/annurev-nucl-102115-044819}, 66, 23

\bibitem[\protect\citeauthoryear{Fern{\'a}ndez, Quataert, Schwab, Kasen  \&
  Rosswog}{Fern{\'a}ndez et~al.}{2015}]{Fernandez:2014bra}
Fern{\'a}ndez R.,  Quataert E.,  Schwab J.,  Kasen D.,   Rosswog S.,  2015,
  \mn@doi [Mon. Not. Roy. Astron. Soc.] {10.1093/mnras/stv238}, 449, 390

\bibitem[\protect\citeauthoryear{Fern{\'a}ndez, Tchekhovskoy, Quataert, Foucart
   \& Kasen}{Fern{\'a}ndez et~al.}{2019}]{Fernandez:2018kax}
Fern{\'a}ndez R.,  Tchekhovskoy A.,  Quataert E.,  Foucart F.,   Kasen D.,
  2019, \mn@doi [Mon. Not. Roy. Astron. Soc.] {10.1093/mnras/sty2932}, 482,
  3373

\bibitem[\protect\citeauthoryear{Fern\'{a}ndez, Kobayashi  \&
  Lamb}{Fern\'{a}ndez et~al.}{2021}]{Fernandez:2021xce}
Fern\'{a}ndez J.~J.,  Kobayashi S.,   Lamb G.~P.,  2021

\bibitem[\protect\citeauthoryear{Fong et~al.}{Fong et~al.}{2017}]{Fong:2017ekk}
Fong W.,  et~al., 2017, \mn@doi [Astrophys. J. Lett.]
  {10.3847/2041-8213/aa9018}, 848, L23

\bibitem[\protect\citeauthoryear{Fujibayashi, Kiuchi, Nishimura, Sekiguchi  \&
  Shibata}{Fujibayashi et~al.}{2018}]{Fujibayashi:2017puw}
Fujibayashi S.,  Kiuchi K.,  Nishimura N.,  Sekiguchi Y.,   Shibata M.,  2018,
  \mn@doi [Astrophys. J.] {10.3847/1538-4357/aabafd}, 860, 64

\bibitem[\protect\citeauthoryear{Fujibayashi, Wanajo, Kiuchi, Kyutoku,
  Sekiguchi  \& Shibata}{Fujibayashi et~al.}{2020a}]{Fujibayashi:2020dvr}
Fujibayashi S.,  Wanajo S.,  Kiuchi K.,  Kyutoku K.,  Sekiguchi Y.,   Shibata
  M.,  2020a

\bibitem[\protect\citeauthoryear{Fujibayashi, Shibata, Wanajo, Kiuchi, Kyutoku
  \& Sekiguchi}{Fujibayashi et~al.}{2020b}]{Fujibayashi:2020qda}
Fujibayashi S.,  Shibata M.,  Wanajo S.,  Kiuchi K.,  Kyutoku K.,   Sekiguchi
  Y.,  2020b, \mn@doi [Phys. Rev. D] {10.1103/PhysRevD.101.083029}, 101, 083029

\bibitem[\protect\citeauthoryear{Fujibayashi, Kiuchi, Wanajo, Kyutoku,
  Sekiguchi  \& Shibata}{Fujibayashi et~al.}{2022}]{Fujibayashi:2022ftg}
Fujibayashi S.,  Kiuchi K.,  Wanajo S.,  Kyutoku K.,  Sekiguchi Y.,   Shibata
  M.,  2022

\bibitem[\protect\citeauthoryear{Ghirlanda et~al.}{Ghirlanda
  et~al.}{2015}]{Ghirlanda:2015nta}
Ghirlanda G.,  et~al., 2015, \mn@doi [Astron. Astrophys.]
  {10.1051/0004-6361/201526112}, 578, A71

\bibitem[\protect\citeauthoryear{Ghirlanda et~al.}{Ghirlanda
  et~al.}{2019}]{Ghirlanda:2018uyx}
Ghirlanda G.,  et~al., 2019, \mn@doi [Science] {10.1126/science.aau8815}, 363,
  968

\bibitem[\protect\citeauthoryear{Giannios \& Spitkovsky}{Giannios \&
  Spitkovsky}{2009}]{Giannios:2009}
Giannios D.,  Spitkovsky A.,  2009, \mn@doi [Mon. Not. Roy. Astron. Soc.]
  {10.1111/j.1365-2966.2009.15454.x}, \href
  {https://ui.adsabs.harvard.edu/abs/2009MNRAS.400..330G} {400, 330}

\bibitem[\protect\citeauthoryear{Gill \& Granot}{Gill \&
  Granot}{2018}]{Gill:2018kcw}
Gill R.,  Granot J.,  2018, \mn@doi [Mon. Not. Roy. Astron. Soc.]
  {10.1093/mnras/sty1214}, 478, 4128

\bibitem[\protect\citeauthoryear{Gottlieb, Bromberg, Singh  \& Nakar}{Gottlieb
  et~al.}{2020}]{Gottlieb:2020mmk}
Gottlieb O.,  Bromberg O.,  Singh C.~B.,   Nakar E.,  2020, \mn@doi [Mon. Not.
  Roy. Astron. Soc.] {10.1093/mnras/staa2567}, 498, 3320

\bibitem[\protect\citeauthoryear{Gottlieb, Moseley, Ramirez-Aguilar,
  Murguia-Berthier, Liska  \& Tchekhovskoy}{Gottlieb
  et~al.}{2022}]{Gottlieb:2022sis}
Gottlieb O.,  Moseley S.,  Ramirez-Aguilar T.,  Murguia-Berthier A.,  Liska M.,
    Tchekhovskoy A.,  2022, \mn@doi [Astrophys. J. Lett.]
  {10.3847/2041-8213/ac7728}, 933, L2

\bibitem[\protect\citeauthoryear{Granot \& Kumar}{Granot \&
  Kumar}{2003}]{Granot:2002me}
Granot J.,  Kumar P.,  2003, \mn@doi [Astrophys. J.] {10.1086/375489}, 591,
  1086

\bibitem[\protect\citeauthoryear{Granot \& Piran}{Granot \&
  Piran}{2012}]{Granot:2012}
Granot J.,  Piran T.,  2012, \mn@doi [Monthly Notices of the Royal Astronomical
  Society] {10.1111/j.1365-2966.2011.20335.x}, 421, 570

\bibitem[\protect\citeauthoryear{Granot, Piran  \& Sari}{Granot
  et~al.}{1999}]{Granot:1998ek}
Granot J.,  Piran T.,   Sari R.,  1999, \mn@doi [Astrophys. J.]
  {10.1086/308052}, 527, 236

\bibitem[\protect\citeauthoryear{Granot, Cohen-Tanugi  \& do~Couto~e
  Silva}{Granot et~al.}{2008}]{Granot:2007gn}
Granot J.,  Cohen-Tanugi J.,   do~Couto~e Silva E.,  2008, \mn@doi [Astrophys.
  J.] {10.1086/526414}, 677, 92

\bibitem[\protect\citeauthoryear{Guarini, Tamborra, B\'egu\'e, Pitik  \&
  Greiner}{Guarini et~al.}{2021}]{Guarini:2021gwh}
Guarini E.,  Tamborra I.,  B\'egu\'e D.,  Pitik T.,   Greiner J.,  2021

\bibitem[\protect\citeauthoryear{Guo, Sironi  \& Narayan}{Guo
  et~al.}{2014a}]{Guo:2014aua}
Guo X.,  Sironi L.,   Narayan R.,  2014a, \mn@doi [Astrophys. J.]
  {10.1088/0004-637X/794/2/153}, 794, 153

\bibitem[\protect\citeauthoryear{Guo, Sironi  \& Narayan}{Guo
  et~al.}{2014b}]{Guo:2014pka}
Guo X.,  Sironi L.,   Narayan R.,  2014b, \mn@doi [Astrophys. J.]
  {10.1088/0004-637X/797/1/47}, 797, 47

\bibitem[\protect\citeauthoryear{Hajela et~al.}{Hajela
  et~al.}{2019}]{Hajela:2019mjy}
Hajela A.,  et~al., 2019, \mn@doi [Astrophys. J. Lett.]
  {10.3847/2041-8213/ab5226}, 886, L17

\bibitem[\protect\citeauthoryear{Hajela et~al.}{Hajela
  et~al.}{2022}]{Hajela:2021faz}
Hajela A.,  et~al., 2022, \mn@doi [Astrophys. J. Lett.]
  {10.3847/2041-8213/ac504a}, 927, L17

\bibitem[\protect\citeauthoryear{Hallinan et~al.}{Hallinan
  et~al.}{2017}]{Hallinan:2017woc}
Hallinan G.,  et~al., 2017, \mn@doi [Science] {10.1126/science.aap9855}, 358,
  1579

\bibitem[\protect\citeauthoryear{Harris et~al.,}{Harris et~al.}{2020}]{numpy}
Harris C.~R.,  et~al., 2020, \mn@doi [Nature] {10.1038/s41586-020-2649-2}, 585,
  357

\bibitem[\protect\citeauthoryear{Hempel \& Schaffner-Bielich}{Hempel \&
  Schaffner-Bielich}{2010}]{Hempel:2009mc}
Hempel M.,  Schaffner-Bielich J.,  2010, \mn@doi [Nucl. Phys.]
  {10.1016/j.nuclphysa.2010.02.010}, A837, 210

\bibitem[\protect\citeauthoryear{Ho et~al.}{Ho et~al.}{2019a}]{Ho:2019dlw}
Ho A. Y.~Q.,  et~al., 2019a, ] {10.3847/1538-4357/ab55ec}

\bibitem[\protect\citeauthoryear{Ho et~al.}{Ho et~al.}{2019b}]{Ho:2018emo}
Ho A. Y.~Q.,  et~al., 2019b, \mn@doi [Astrophys. J.]
  {10.3847/1538-4357/aaf473}, 871, 73

\bibitem[\protect\citeauthoryear{Ho et~al.,}{Ho et~al.}{2022}]{Ho:2021bjl}
Ho A. Y.~Q.,  et~al., 2022, \mn@doi [Astrophys. J.] {10.3847/1538-4357/ac4e97},
  932, 116

\bibitem[\protect\citeauthoryear{Hotokezaka \& Piran}{Hotokezaka \&
  Piran}{2015}]{Hotokezaka:2015eja}
Hotokezaka K.,  Piran T.,  2015, \mn@doi [Mon. Not. Roy. Astron. Soc.]
  {10.1093/mnras/stv620}, 450, 1430

\bibitem[\protect\citeauthoryear{Hotokezaka, Kiuchi, Kyutoku, Okawa, Sekiguchi,
  Shibata  \& Taniguchi}{Hotokezaka et~al.}{2013}]{Hotokezaka:2013b}
Hotokezaka K.,  Kiuchi K.,  Kyutoku K.,  Okawa H.,  Sekiguchi Y.-i.,  Shibata
  M.,   Taniguchi K.,  2013, \mn@doi [Physical Review D]
  {10.1103/PhysRevD.87.024001}, 87.2, 024001

\bibitem[\protect\citeauthoryear{Hotokezaka, Kiuchi, Shibata, Nakar  \&
  Piran}{Hotokezaka et~al.}{2018}]{Hotokezaka:2018gmo}
Hotokezaka K.,  Kiuchi K.,  Shibata M.,  Nakar E.,   Piran T.,  2018, \mn@doi
  [Astrophys. J.] {10.3847/1538-4357/aadf92}, 867, 95

\bibitem[\protect\citeauthoryear{Huang, Dai  \& Lu}{Huang
  et~al.}{1999}]{Huang:1999di}
Huang Y.,  Dai Z.,   Lu T.,  1999, \mn@doi [Mon. Not. Roy. Astron. Soc.]
  {10.1046/j.1365-8711.1999.02887.x}, 309, 513

\bibitem[\protect\citeauthoryear{Huang, Gou, Dai  \& Lu}{Huang
  et~al.}{2000}]{Huang:1999qa}
Huang Y.,  Gou L.,  Dai Z.,   Lu T.,  2000, \mn@doi [Astrophys. J.]
  {10.1086/317076}, 543, 90

\bibitem[\protect\citeauthoryear{Huang et~al.}{Huang
  et~al.}{2020}]{Huang:2020pxr}
Huang Y.-J.,  et~al., 2020, \mn@doi [Astrophys. J.] {10.3847/1538-4357/ab8f9a},
  897, 69

\bibitem[\protect\citeauthoryear{Hunter}{Hunter}{2007}]{matplotlib}
Hunter J.~D.,  2007, \mn@doi [Computing in Science \& Engineering]
  {10.1109/MCSE.2007.55}, 9, 90

\bibitem[\protect\citeauthoryear{Jin et~al.,}{Jin et~al.}{2016}]{Jin:2016pnm}
Jin Z.-P.,  et~al., 2016, \mn@doi [Nature Commun.] {10.1038/ncomms12898}, 7,
  12898

\bibitem[\protect\citeauthoryear{Jin et~al.,}{Jin et~al.}{2018}]{Jin:2017hle}
Jin Z.-P.,  et~al., 2018, \mn@doi [Astrophys. J.] {10.3847/1538-4357/aab76d},
  857, 128

\bibitem[\protect\citeauthoryear{Jin, Covino, Liao, Li, D'Avanzo, Fan  \&
  Wei}{Jin et~al.}{2020}]{Jin:2019uqr}
Jin Z.-P.,  Covino S.,  Liao N.-H.,  Li X.,  D'Avanzo P.,  Fan Y.-Z.,   Wei
  D.-M.,  2020, Nature Astron., 4, 77

\bibitem[\protect\citeauthoryear{Johannesson, Bjornsson  \&
  Gudmundsson}{Johannesson et~al.}{2006}]{Johannesson:2006zs}
Johannesson G.,  Bjornsson G.,   Gudmundsson E.~H.,  2006, \mn@doi [Astrophys.
  J.] {10.1086/505520}, 647, 1238

\bibitem[\protect\citeauthoryear{Johnston-Hollitt}{Johnston-Hollitt}{2017}]{Johnston:2017}
Johnston-Hollitt M.,  2017, \mn@doi [Nature Astronomy]
  {10.1038/s41550-016-0014}, \href
  {https://ui.adsabs.harvard.edu/abs/2017NatAs...1E..14J} {1, 0014}

\bibitem[\protect\citeauthoryear{Just, Bauswein, Pulpillo, Goriely  \&
  Janka}{Just et~al.}{2015}]{Just:2014fka}
Just O.,  Bauswein A.,  Pulpillo R.~A.,  Goriely S.,   Janka H.~T.,  2015,
  \mn@doi [Mon. Not. Roy. Astron. Soc.] {10.1093/mnras/stv009}, 448, 541

\bibitem[\protect\citeauthoryear{Kang}{Kang}{2018}]{Kang:2018puv}
Kang H.,  2018, \mn@doi [J. Korean Astron. Soc.] {10.5303/JKAS.2018.51.6.185},
  51, 185

\bibitem[\protect\citeauthoryear{Kang, Ryu  \& Ha}{Kang
  et~al.}{2019}]{Kang:2019rtl}
Kang H.,  Ryu D.,   Ha J.-H.,  2019, \mn@doi [Astrophys. J.]
  {10.3847/1538-4357/ab16d1}, 876, 79

\bibitem[\protect\citeauthoryear{Kasen, Fern{\'a}ndez  \& Metzger}{Kasen
  et~al.}{2015}]{Kasen:2014toa}
Kasen D.,  Fern{\'a}ndez R.,   Metzger B.,  2015, \mn@doi [Mon. Not. Roy.
  Astron. Soc.] {10.1093/mnras/stv721}, 450, 1777

\bibitem[\protect\citeauthoryear{Kasen, Metzger, Barnes, Quataert  \&
  Ramirez-Ruiz}{Kasen et~al.}{2017}]{Kasen:2017sxr}
Kasen D.,  Metzger B.,  Barnes J.,  Quataert E.,   Ramirez-Ruiz E.,  2017,
  \mn@doi [Nature] {10.1038/nature24453}

\bibitem[\protect\citeauthoryear{Kasliwal et~al.}{Kasliwal
  et~al.}{2017}]{Kasliwal:2017ngb}
Kasliwal M.~M.,  et~al., 2017, \mn@doi [Science] {10.1126/science.aap9455},
  358, 1559

\bibitem[\protect\citeauthoryear{Kathirgamaraju, Tchekhovskoy, Giannios  \&
  Barniol~Duran}{Kathirgamaraju et~al.}{2019}]{Kathirgamaraju:2018mac}
Kathirgamaraju A.,  Tchekhovskoy A.,  Giannios D.,   Barniol~Duran R.,  2019,
  \mn@doi [Mon. Not. Roy. Astron. Soc.] {10.1093/mnrasl/slz012}, 484, L98

\bibitem[\protect\citeauthoryear{Kawaguchi, Shibata  \& Tanaka}{Kawaguchi
  et~al.}{2018}]{Kawaguchi:2018ptg}
Kawaguchi K.,  Shibata M.,   Tanaka M.,  2018, \mn@doi [Astrophys. J.]
  {10.3847/2041-8213/aade02}, 865, L21

\bibitem[\protect\citeauthoryear{Keshet \& Waxman}{Keshet \&
  Waxman}{2005}]{Keshet:2004ch}
Keshet U.,  Waxman E.,  2005, \mn@doi [Phys. Rev. Lett.]
  {10.1103/PhysRevLett.94.111102}, 94, 111102

\bibitem[\protect\citeauthoryear{Kirk \& Duffy}{Kirk \&
  Duffy}{1999}]{Kirk:1999km}
Kirk J.~G.,  Duffy P.,  1999, \mn@doi [J. Phys. G]
  {10.1088/0954-3899/25/8/201}, 25, R163

\bibitem[\protect\citeauthoryear{Klose et~al.,}{Klose
  et~al.}{2019}]{Klose:2019amd}
Klose S.,  et~al., 2019, ] {10.3847/1538-4357/ab528a}

\bibitem[\protect\citeauthoryear{Komissarov \& Barkov}{Komissarov \&
  Barkov}{2009}]{Komissarov:2009}
Komissarov S.~S.,  Barkov M.~V.,  2009, \mn@doi [Mon. Not. Roy. Astron. Soc.]
  {10.1111/j.1365-2966.2009.14831.x}, \href
  {https://ui.adsabs.harvard.edu/abs/2009MNRAS.397.1153K} {397, 1153}

\bibitem[\protect\citeauthoryear{Kr{\"u}ger \& Foucart}{Kr{\"u}ger \&
  Foucart}{2020}]{Kruger:2020gig}
Kr{\"u}ger C.~J.,  Foucart F.,  2020, \mn@doi [Phys. Rev. D]
  {10.1103/PhysRevD.101.103002}, 101, 103002

\bibitem[\protect\citeauthoryear{Kumar \& Granot}{Kumar \&
  Granot}{2003}]{Kumar:2003yt}
Kumar P.,  Granot J.,  2003, \mn@doi [Astrophys. J.] {10.1086/375186}, 591,
  1075

\bibitem[\protect\citeauthoryear{Kumar \& Zhang}{Kumar \&
  Zhang}{2014}]{Kumar:2014upa}
Kumar P.,  Zhang B.,  2014, \mn@doi [Phys. Rept.]
  {10.1016/j.physrep.2014.09.008}, 561, 1

\bibitem[\protect\citeauthoryear{Lamb \& Kobayashi}{Lamb \&
  Kobayashi}{2017}]{Lamb:2017ych}
Lamb G.~P.,  Kobayashi S.,  2017, \mn@doi [Mon. Not. Roy. Astron. Soc.]
  {10.1093/mnras/stx2345}, 472, 4953

\bibitem[\protect\citeauthoryear{Lamb, Mandel  \& Resmi}{Lamb
  et~al.}{2018}]{Lamb:2018ohw}
Lamb G.~P.,  Mandel I.,   Resmi L.,  2018, \mn@doi [Mon. Not. Roy. Astron.
  Soc.] {10.1093/mnras/sty2196}, 481, 2581

\bibitem[\protect\citeauthoryear{Lamb et~al.}{Lamb
  et~al.}{2019a}]{Lamb:2019lao}
Lamb G.~P.,  et~al., 2019a, ] {10.3847/1538-4357/ab38bb}

\bibitem[\protect\citeauthoryear{Lamb et~al.}{Lamb
  et~al.}{2019b}]{Lamb:2018qfn}
Lamb G.~P.,  et~al., 2019b, \mn@doi [Astrophys. J. Lett.]
  {10.3847/2041-8213/aaf96b}, 870, L15

\bibitem[\protect\citeauthoryear{Lamb, Levan  \& Tanvir}{Lamb
  et~al.}{2020}]{Lamb:2020ccz}
Lamb G.~P.,  Levan A.~J.,   Tanvir N.~R.,  2020, \mn@doi [Astrophys. J.]
  {10.3847/1538-4357/aba75a}, 899, 105

\bibitem[\protect\citeauthoryear{Lamb, Nativi, Rosswog, Kann, Levan, Lundman
  \& Tanvir}{Lamb et~al.}{2022}]{Lamb:2022pvr}
Lamb G.~P.,  Nativi L.,  Rosswog S.,  Kann D.~A.,  Levan A.,  Lundman C.,
  Tanvir N.,  2022

\bibitem[\protect\citeauthoryear{Lattimer \& Swesty}{Lattimer \&
  Swesty}{1991}]{Lattimer:1991nc}
Lattimer J.~M.,  Swesty F.~D.,  1991, \mn@doi [Nucl. Phys.]
  {10.1016/0375-9474(91)90452-C}, A535, 331

\bibitem[\protect\citeauthoryear{Lee, Ramirez-Ruiz  \& Diego-Lopez-Camara}{Lee
  et~al.}{2009}]{Lee:2009uc}
Lee W.~H.,  Ramirez-Ruiz E.,   Diego-Lopez-Camara 2009, \mn@doi [Astrophys.J.]
  {10.1088/0004-637X/699/2/L93}, 699, L93

\bibitem[\protect\citeauthoryear{Lemoine \& Pelletier}{Lemoine \&
  Pelletier}{2010}]{Lemoine:2010}
Lemoine M.,  Pelletier G.,  2010, \mn@doi [Mon. Not. Roy. Astron. Soc.]
  {10.1111/j.1365-2966.2009.15869.x}, \href
  {https://ui.adsabs.harvard.edu/abs/2010MNRAS.402..321L} {402, 321}

\bibitem[\protect\citeauthoryear{Leung et~al.}{Leung
  et~al.}{2021}]{Leung:2021ebv}
Leung J.~K.,  et~al., 2021, \mn@doi [Mon. Not. Roy. Astron. Soc.]
  {10.1093/mnras/stab326}, 503, 1847

\bibitem[\protect\citeauthoryear{Ligorini et~al.,}{Ligorini
  et~al.}{2021}]{Ligorini:2021lbj}
Ligorini A.,  et~al., 2021, \mn@doi [Mon. Not. Roy. Astron. Soc.]
  {10.1093/mnras/stab220}, 502, 5065

\bibitem[\protect\citeauthoryear{Lippuner, Fern\'{a}ndez, Roberts, Foucart,
  Kasen, Metzger  \& Ott}{Lippuner et~al.}{2017}]{Lippuner:2017bfm}
Lippuner J.,  Fern\'{a}ndez R.,  Roberts L.~F.,  Foucart F.,  Kasen D.,
  Metzger B.~D.,   Ott C.~D.,  2017, \mn@doi [Mon. Not. Roy. Astron. Soc.]
  {10.1093/mnras/stx1987}, 472, 904

\bibitem[\protect\citeauthoryear{Lloyd-Ronning, Fryer, Hartmann  \&
  Wiggins}{Lloyd-Ronning et~al.}{2017}]{Lloyd-Ronning:2017dci}
Lloyd-Ronning N.~M.,  Fryer C.~L.,  Hartmann D.~H.,   Wiggins B.,  2017

\bibitem[\protect\citeauthoryear{Logoteta, Perego  \& Bombaci}{Logoteta
  et~al.}{2021}]{Logoteta:2020yxf}
Logoteta D.,  Perego A.,   Bombaci I.,  2021, \mn@doi [Astron. Astrophys.]
  {10.1051/0004-6361/202039457}, 646, A55

\bibitem[\protect\citeauthoryear{Lu, Beniamini  \& McDowell}{Lu
  et~al.}{2020}]{Lu:2020hka}
Lu W.,  Beniamini P.,   McDowell A.,  2020

\bibitem[\protect\citeauthoryear{Lyman et~al.}{Lyman
  et~al.}{2018}]{Lyman:2018qjg}
Lyman J.~D.,  et~al., 2018, \mn@doi [Nat. Astron.] {10.1038/s41550-018-0511-3},
  2, 751

\bibitem[\protect\citeauthoryear{Mahadevan, Narayan  \& Yi}{Mahadevan
  et~al.}{1996}]{Mahadevan:1996cc}
Mahadevan R.,  Narayan R.,   Yi I.,  1996, \mn@doi [Astrophys. J.]
  {10.1086/177422}, 465, 327

\bibitem[\protect\citeauthoryear{Marcowith, Ferrand, Grech, Meliani, Plotnikov
  \& Walder}{Marcowith et~al.}{2020}]{Marcowith:2020vho}
Marcowith A.,  Ferrand G.,  Grech M.,  Meliani Z.,  Plotnikov I.,   Walder R.,
  2020, \mn@doi [Liv. Rev. Comput. Astrophys.] {10.1007/s41115-020-0007-6}, 6,
  1

\bibitem[\protect\citeauthoryear{Margalit \& Piran}{Margalit \&
  Piran}{2020}]{Margalit:2020bdk}
Margalit B.,  Piran T.,  2020, ] {10.1093/mnras/staa1486}

\bibitem[\protect\citeauthoryear{Margalit \& Quataert}{Margalit \&
  Quataert}{2021}]{Margalit:2021kuf}
Margalit B.,  Quataert E.,  2021, \mn@doi [Astrophys. J. Lett.]
  {10.3847/2041-8213/ac3d97}, 923, L14

\bibitem[\protect\citeauthoryear{Margalit, Quataert  \& Ho}{Margalit
  et~al.}{2022}]{Margalit:2021bqe}
Margalit B.,  Quataert E.,   Ho A. Y.~Q.,  2022, \mn@doi [Astrophys. J.]
  {10.3847/1538-4357/ac53b0}, 928, 122

\bibitem[\protect\citeauthoryear{Margutti et~al.}{Margutti
  et~al.}{2018}]{Margutti:2018xqd}
Margutti R.,  et~al., 2018, \mn@doi [Astrophys. J.] {10.3847/2041-8213/aab2ad},
  856, L18

\bibitem[\protect\citeauthoryear{Martin, Perego, Arcones, Thielemann, Korobkin
  \& Rosswog}{Martin et~al.}{2015}]{Martin:2015hxa}
Martin D.,  Perego A.,  Arcones A.,  Thielemann F.-K.,  Korobkin O.,   Rosswog
  S.,  2015, \mn@doi [Astrophys. J.] {10.1088/0004-637X/813/1/2}, 813, 2

\bibitem[\protect\citeauthoryear{Medvedev \& Loeb}{Medvedev \&
  Loeb}{1999}]{Medvedev:1999tu}
Medvedev M.~V.,  Loeb A.,  1999, \mn@doi [Astrophys. J.] {10.1086/308038}, 526,
  697

\bibitem[\protect\citeauthoryear{Metzger}{Metzger}{2017}]{Metzger:2016pju}
Metzger B.~D.,  2017, \mn@doi [Living Rev. Rel.] {10.1007/s41114-017-0006-z},
  20, 3

\bibitem[\protect\citeauthoryear{Metzger}{Metzger}{2020}]{Metzger:2019zeh}
Metzger B.~D.,  2020, \mn@doi [Living Rev. Rel.] {10.1007/s41114-019-0024-0},
  23, 1

\bibitem[\protect\citeauthoryear{Metzger \& Fern\'{a}ndez}{Metzger \&
  Fern\'{a}ndez}{2014}]{Metzger:2014ila}
Metzger B.~D.,  Fern\'{a}ndez R.,  2014, \mn@doi [Mon.Not.Roy.Astron.Soc.]
  {10.1093/mnras/stu802}, 441, 3444

\bibitem[\protect\citeauthoryear{Metzger et~al.,}{Metzger
  et~al.}{2010}]{Metzger:2010}
Metzger B.~D.,  et~al., 2010, \mn@doi [Monthly Notices of the Royal
  Astronomical Society] {10.1111/j.1365-2966.2010.16864.x}, \href
  {https://ui.adsabs.harvard.edu/abs/2010MNRAS.406.2650M} {406, 2650}

\bibitem[\protect\citeauthoryear{Metzger, Bauswein, Goriely  \& Kasen}{Metzger
  et~al.}{2015}]{Metzger:2014yda}
Metzger B.~D.,  Bauswein A.,  Goriely S.,   Kasen D.,  2015, \mn@doi [Mon. Not.
  Roy. Astron. Soc.] {10.1093/mnras/stu2225}, 446, 1115

\bibitem[\protect\citeauthoryear{Miceli \& Nava}{Miceli \&
  Nava}{2022}]{Miceli:2022efx}
Miceli D.,  Nava L.,  2022, \mn@doi [Galaxies] {10.3390/galaxies10030066}, 10,
  66

\bibitem[\protect\citeauthoryear{Mignone, Plewa  \& Bodo}{Mignone
  et~al.}{2005}]{Mignone:2005ns}
Mignone A.,  Plewa T.,   Bodo G.,  2005, \mn@doi [Astrophys. J. Suppl.]
  {10.1086/430905}, 160, 199

\bibitem[\protect\citeauthoryear{Mignone, Bodo, Vaidya  \& Mattia}{Mignone
  et~al.}{2018}]{Mignone:2018per}
Mignone A.,  Bodo G.,  Vaidya B.,   Mattia G.,  2018, \mn@doi [Astrophys. J.]
  {10.3847/1538-4357/aabccd}, 859, 13

\bibitem[\protect\citeauthoryear{Mihalas}{Mihalas}{1978}]{Mihalas:1978}
Mihalas D.,  1978, {Stellar atmospheres}

\bibitem[\protect\citeauthoryear{Miller et~al.,}{Miller
  et~al.}{2019}]{Miller:2019dpt}
Miller J.~M.,  et~al., 2019, \mn@doi [Phys. Rev.]
  {10.1103/PhysRevD.100.023008}, D100, 023008

\bibitem[\protect\citeauthoryear{Mooley et~al.,}{Mooley
  et~al.}{2018}]{Mooley:2018dlz}
Mooley K.~P.,  et~al., 2018, \mn@doi [Nature] {10.1038/s41586-018-0486-3}, 561,
  355

\bibitem[\protect\citeauthoryear{Nakar}{Nakar}{2020}]{Nakar:2019fza}
Nakar E.,  2020, \mn@doi [Phys. Rept.] {10.1016/j.physrep.2020.08.008}, 886, 1

\bibitem[\protect\citeauthoryear{Nakar \& Piran}{Nakar \&
  Piran}{2011}]{Nakar:2011cw}
Nakar E.,  Piran T.,  2011, \mn@doi [Nature] {10.1038/nature10365}, 478, 82

\bibitem[\protect\citeauthoryear{Nakar, Piran  \& Granot}{Nakar
  et~al.}{2002}]{Nakar:2002ph}
Nakar E.,  Piran T.,   Granot J.,  2002, \mn@doi [Astrophys. J.]
  {10.1086/342791}, 579, 699

\bibitem[\protect\citeauthoryear{Nathanail, Gill, Porth, Fromm  \&
  Rezzolla}{Nathanail et~al.}{2021}]{Nathanail:2020hkx}
Nathanail A.,  Gill R.,  Porth O.,  Fromm C.~M.,   Rezzolla L.,  2021, \mn@doi
  [Mon. Not. Roy. Astron. Soc.] {10.1093/mnras/stab115}, 502, 1843

\bibitem[\protect\citeauthoryear{Nava, Sironi, Ghisellini, Celotti  \&
  Ghirlanda}{Nava et~al.}{2013}]{Nava:2013}
Nava L.,  Sironi L.,  Ghisellini G.,  Celotti A.,   Ghirlanda G.,  2013,
  \mn@doi [Monthly Notices of the Royal Astronomical Society]
  {10.1093/mnras/stt872}, \href
  {https://ui.adsabs.harvard.edu/abs/2013MNRAS.433.2107N} {433, 2107}

\bibitem[\protect\citeauthoryear{Nedora, Bernuzzi, Radice, Perego, Endrizzi  \&
  Ortiz}{Nedora et~al.}{2019}]{Nedora:2019jhl}
Nedora V.,  Bernuzzi S.,  Radice D.,  Perego A.,  Endrizzi A.,   Ortiz N.,
  2019, \mn@doi [Astrophys. J.] {10.3847/2041-8213/ab5794}, 886, L30

\bibitem[\protect\citeauthoryear{Nedora et~al.,}{Nedora
  et~al.}{2020}]{Nedora:2020qtd}
Nedora V.,  et~al., 2020

\bibitem[\protect\citeauthoryear{Nedora, Radice, Bernuzzi, Perego, Daszuta,
  Endrizzi, Prakash  \& Schianchi}{Nedora et~al.}{2021a}]{Nedora:2021eoj}
Nedora V.,  Radice D.,  Bernuzzi S.,  Perego A.,  Daszuta B.,  Endrizzi A.,
  Prakash A.,   Schianchi F.,  2021a

\bibitem[\protect\citeauthoryear{Nedora et~al.,}{Nedora
  et~al.}{2021b}]{Nedora:2020pak}
Nedora V.,  et~al., 2021b, \mn@doi [Astrophys. J.] {10.3847/1538-4357/abc9be},
  906, 98

\bibitem[\protect\citeauthoryear{Nicholl et~al.}{Nicholl
  et~al.}{2017}]{Nicholl:2017ahq}
Nicholl M.,  et~al., 2017, \mn@doi [Astrophys. J.] {10.3847/2041-8213/aa9029},
  848, L18

\bibitem[\protect\citeauthoryear{Nynka, Ruan, Haggard  \& Evans}{Nynka
  et~al.}{2018}]{Nynka:2018vup}
Nynka M.,  Ruan J.~J.,  Haggard D.,   Evans P.~A.,  2018, \mn@doi [Astrophys.
  J. Lett.] {10.3847/2041-8213/aad32d}, 862, L19

\bibitem[\protect\citeauthoryear{Ozel, Psaltis  \& Narayan}{Ozel
  et~al.}{2000}]{Ozel:2000}
Ozel F.,  Psaltis D.,   Narayan R.,  2000, \mn@doi [The Astrophysical Journal]
  {10.1086/309396}, 541, 234

\bibitem[\protect\citeauthoryear{Pacholczyk}{Pacholczyk}{1970}]{Pacholczyk:1970}
Pacholczyk A.~G.,  1970, {Radio astrophysics. Nonthermal processes in galactic
  and extragalactic sources}

\bibitem[\protect\citeauthoryear{Park, Caprioli  \& Spitkovsky}{Park
  et~al.}{2015}]{Park:2014lqa}
Park J.,  Caprioli D.,   Spitkovsky A.,  2015, \mn@doi [Phys. Rev. Lett.]
  {10.1103/PhysRevLett.114.085003}, 114, 085003

\bibitem[\protect\citeauthoryear{Pe'er}{Pe'er}{2012}]{Peer:2012}
Pe'er A.,  2012, \mn@doi [The Astrophysical Journal Letters]
  {10.1088/2041-8205/752/1/L8}, \href
  {https://ui.adsabs.harvard.edu/abs/2012ApJ...752L...8P} {752, L8}

\bibitem[\protect\citeauthoryear{Perego, Rosswog, Cabezon, Korobkin, Kaeppeli
  et~al.}{Perego et~al.}{2014}]{Perego:2014fma}
Perego A.,  Rosswog S.,  Cabezon R.,  Korobkin O.,  Kaeppeli R.,   et~al.,
  2014, \mn@doi [Mon.Not.Roy.Astron.Soc.] {10.1093/mnras/stu1352}, 443, 3134

\bibitem[\protect\citeauthoryear{Perego, Radice  \& Bernuzzi}{Perego
  et~al.}{2017}]{Perego:2017wtu}
Perego A.,  Radice D.,   Bernuzzi S.,  2017, \mn@doi [Astrophys. J.]
  {10.3847/2041-8213/aa9ab9}, 850, L37

\bibitem[\protect\citeauthoryear{Perego, Bernuzzi  \& Radice}{Perego
  et~al.}{2019}]{Perego:2019adq}
Perego A.,  Bernuzzi S.,   Radice D.,  2019, \mn@doi [Eur. Phys. J.]
  {10.1140/epja/i2019-12810-7}, A55, 124

\bibitem[\protect\citeauthoryear{Petrosian}{Petrosian}{1981}]{Petrosian:1981}
Petrosian V.,  1981, \mn@doi [The Astrophysical Journal] {10.1086/159517},
  \href {https://ui.adsabs.harvard.edu/abs/1981ApJ...251..727P} {251, 727}

\bibitem[\protect\citeauthoryear{Pinzke, Oh  \& Pfrommer}{Pinzke
  et~al.}{2013}]{Pinzke:2013zu}
Pinzke A.,  Oh S.~P.,   Pfrommer C.,  2013, \mn@doi [Mon. Not. Roy. Astron.
  Soc.] {10.1093/mnras/stt1308}, 435, 1061

\bibitem[\protect\citeauthoryear{Piran, Nakar  \& Rosswog}{Piran
  et~al.}{2013}]{Piran:2012wd}
Piran T.,  Nakar E.,   Rosswog S.,  2013, \mn@doi [Mon. Not. Roy. Astron. Soc.]
  {10.1093/mnras/stt037}, 430, 2121

\bibitem[\protect\citeauthoryear{Pohl, Hoshino  \& Niemiec}{Pohl
  et~al.}{2020}]{Pohl:2019nvw}
Pohl M.,  Hoshino M.,   Niemiec J.,  2020, \mn@doi [Prog. Part. Nucl. Phys.]
  {10.1016/j.ppnp.2019.103751}, 111, 103751

\bibitem[\protect\citeauthoryear{Prince \& Dormand}{Prince \&
  Dormand}{1981}]{Prince:1981}
Prince P.~J.,  Dormand J.~R.,  1981, \mn@doi [Journal of Computational and
  Applied Mathematics] {https://doi.org/10.1016/0771-050X(81)90010-3}, 7, 67

\bibitem[\protect\citeauthoryear{Radice}{Radice}{2017}]{Radice:2017zta}
Radice D.,  2017, \mn@doi [Astrophys. J.] {10.3847/2041-8213/aa6483}, 838, L2

\bibitem[\protect\citeauthoryear{Radice}{Radice}{2020}]{Radice:2020ids}
Radice D.,  2020, \mn@doi [Symmetry] {10.3390/sym12081249}, 12, 1249

\bibitem[\protect\citeauthoryear{Radice \& Rezzolla}{Radice \&
  Rezzolla}{2012}]{Radice:2012cu}
Radice D.,  Rezzolla L.,  2012, \mn@doi [Astron. Astrophys.]
  {10.1051/0004-6361/201219735}, 547, A26

\bibitem[\protect\citeauthoryear{Radice, Rezzolla  \& Galeazzi}{Radice
  et~al.}{2014a}]{Radice:2013xpa}
Radice D.,  Rezzolla L.,   Galeazzi F.,  2014a, \mn@doi [Class.Quant.Grav.]
  {10.1088/0264-9381/31/7/075012}, 31, 075012

\bibitem[\protect\citeauthoryear{Radice, Rezzolla  \& Galeazzi}{Radice
  et~al.}{2014b}]{Radice:2013hxh}
Radice D.,  Rezzolla L.,   Galeazzi F.,  2014b, \mn@doi
  [Mon.Not.Roy.Astron.Soc.] {10.1093/mnrasl/slt137}, 437, L46

\bibitem[\protect\citeauthoryear{Radice, Rezzolla  \& Galeazzi}{Radice
  et~al.}{2015}]{Radice:2015nva}
Radice D.,  Rezzolla L.,   Galeazzi F.,  2015, ASP Conf. Ser., 498, 121

\bibitem[\protect\citeauthoryear{Radice, Galeazzi, Lippuner, Roberts, Ott  \&
  Rezzolla}{Radice et~al.}{2016}]{Radice:2016dwd}
Radice D.,  Galeazzi F.,  Lippuner J.,  Roberts L.~F.,  Ott C.~D.,   Rezzolla
  L.,  2016, \mn@doi [Mon. Not. Roy. Astron. Soc.] {10.1093/mnras/stw1227},
  460, 3255

\bibitem[\protect\citeauthoryear{Radice, Perego, Bernuzzi  \& Zhang}{Radice
  et~al.}{2018a}]{Radice:2018xqa}
Radice D.,  Perego A.,  Bernuzzi S.,   Zhang B.,  2018a, \mn@doi [Mon. Not.
  Roy. Astron. Soc.] {10.1093/mnras/sty2531}, 481, 3670

\bibitem[\protect\citeauthoryear{Radice, Perego, Hotokezaka, Bernuzzi, Fromm
  \& Roberts}{Radice et~al.}{2018b}]{Radice:2018ghv}
Radice D.,  Perego A.,  Hotokezaka K.,  Bernuzzi S.,  Fromm S.~A.,   Roberts
  L.~F.,  2018b, \mn@doi [Astrophys. J. Lett.] {10.3847/2041-8213/aaf053}, 869,
  L35

\bibitem[\protect\citeauthoryear{Radice, Perego, Hotokezaka, Fromm, Bernuzzi
  \& Roberts}{Radice et~al.}{2018c}]{Radice:2018pdn}
Radice D.,  Perego A.,  Hotokezaka K.,  Fromm S.~A.,  Bernuzzi S.,   Roberts
  L.~F.,  2018c, \mn@doi [Astrophys. J.] {10.3847/1538-4357/aaf054}, 869, 130

\bibitem[\protect\citeauthoryear{Radice, Bernuzzi  \& Perego}{Radice
  et~al.}{2020}]{Radice:2020ddv}
Radice D.,  Bernuzzi S.,   Perego A.,  2020, \mn@doi [Ann. Rev. Nucl. Part.
  Sci.] {10.1146/annurev-nucl-013120-114541}, 70, 95

\bibitem[\protect\citeauthoryear{Rastinejad et~al.}{Rastinejad
  et~al.}{2022}]{Rastinejad:2022zbg}
Rastinejad J.~C.,  et~al., 2022

\bibitem[\protect\citeauthoryear{Resmi et~al.}{Resmi
  et~al.}{2018}]{Resmi:2018wuc}
Resmi L.,  et~al., 2018, \mn@doi [Astrophys. J.] {10.3847/1538-4357/aae1a6},
  867, 57

\bibitem[\protect\citeauthoryear{Ressler \& Laskar}{Ressler \&
  Laskar}{2017}]{Ressler:2017qjo}
Ressler S.~M.,  Laskar T.,  2017, \mn@doi [Astrophys. J.]
  {10.3847/1538-4357/aa8268}, 845, 150

\bibitem[\protect\citeauthoryear{Reville, Kirk  \& Duffy}{Reville
  et~al.}{2006}]{Reville:2006px}
Reville B.,  Kirk J.~G.,   Duffy P.,  2006, \mn@doi [Plasma Phys. Control.
  Fusion] {10.1088/0741-3335/48/12/004}, 48, 1741

\bibitem[\protect\citeauthoryear{Rezzolla \& Zanotti}{Rezzolla \&
  Zanotti}{2013}]{Rezzolla:2013}
Rezzolla L.,  Zanotti O.,  2013, {Relativistic Hydrodynamics}, 1 edn.
Mathematics, Oxford University Press, Oxford

\bibitem[\protect\citeauthoryear{Rocha~da Silva, Falceta-Gon\c{c}alves, Kowal
  \& de Gouveia Dal~Pino}{Rocha~da Silva et~al.}{2015}]{RochadaSilva:2014ehi}
Rocha~da Silva G.,  Falceta-Gon\c{c}alves D.,  Kowal G.,   de Gouveia Dal~Pino
  E.~M.,  2015, \mn@doi [Mon. Not. Roy. Astron. Soc.] {10.1093/mnras/stu2104},
  446, 104

\bibitem[\protect\citeauthoryear{Rolfs, Rodney  \& Fowler}{Rolfs
  et~al.}{1988}]{Rolfs:1988}
Rolfs C.~E.,  Rodney W.~S.,   Fowler W.~A.,  1988, {Cauldrons in the cosmos :
  nuclear astrophysics}.
Theoretical astrophysics, University of Chicago Press, Chicago u.a., \url
  {https://repository.gsi.de/record/66499}

\bibitem[\protect\citeauthoryear{Rossi, Lazzati, Salmonson  \&
  Ghisellini}{Rossi et~al.}{2004}]{Rossi:2004dz}
Rossi E.~M.,  Lazzati D.,  Salmonson J.~D.,   Ghisellini G.,  2004, \mn@doi
  [Mon. Not. Roy. Astron. Soc.] {10.1111/j.1365-2966.2004.08165.x}, 354, 86

\bibitem[\protect\citeauthoryear{Ruan, Nynka, Haggard, Kalogera  \& Evans}{Ruan
  et~al.}{2018}]{Ruan:2017bha}
Ruan J.~J.,  Nynka M.,  Haggard D.,  Kalogera V.,   Evans P.,  2018, \mn@doi
  [Astrophys.\ J.] {10.3847/2041-8213/aaa4f3}, 853, L4

\bibitem[\protect\citeauthoryear{Ruiz, Lang, Paschalidis  \& Shapiro}{Ruiz
  et~al.}{2016}]{Ruiz:2016rai}
Ruiz M.,  Lang R.~N.,  Paschalidis V.,   Shapiro S.~L.,  2016, \mn@doi
  [Astrophys. J.] {10.3847/2041-8205/824/1/L6}, 824, L6

\bibitem[\protect\citeauthoryear{Ryan, van Eerten, Piro  \& Troja}{Ryan
  et~al.}{2020}]{Ryan:2019fhz}
Ryan G.,  van Eerten H.,  Piro L.,   Troja E.,  2020, \mn@doi [Astrophys. J.]
  {10.3847/1538-4357/ab93cf}, 896, 166

\bibitem[\protect\citeauthoryear{Rybicki \& Lightman}{Rybicki \&
  Lightman}{1986}]{RybickiLightman:1985}
Rybicki G.~B.,  Lightman A.~P.,  1986, {Radiative Processes in Astrophysics}

\bibitem[\protect\citeauthoryear{Sadeh, Guttman  \& Waxman}{Sadeh
  et~al.}{2022}]{Sadeh:2022enp}
Sadeh G.,  Guttman O.,   Waxman E.,  2022

\bibitem[\protect\citeauthoryear{Salafia, Ghisellini, Pescalli, Ghirlanda  \&
  Nappo}{Salafia et~al.}{2015}]{Salafia:2015vla}
Salafia O.~S.,  Ghisellini G.,  Pescalli A.,  Ghirlanda G.,   Nappo F.,  2015,
  \mn@doi [Mon. Not. Roy. Astron. Soc.] {10.1093/mnras/stv766}, 450, 3549

\bibitem[\protect\citeauthoryear{Samuelsson, B\'egu\'e, Ryde, Pe'er  \&
  Murase}{Samuelsson et~al.}{2020}]{Samuelsson:2020upt}
Samuelsson F.,  B\'egu\'e D.,  Ryde F.,  Pe'er A.,   Murase K.,  2020, \mn@doi
  [Astrophys. J.] {10.3847/1538-4357/abb60c}, 902, 148

\bibitem[\protect\citeauthoryear{Sari, Piran  \& Narayan}{Sari
  et~al.}{1998}]{Sari:1997qe}
Sari R.,  Piran T.,   Narayan R.,  1998, \mn@doi [Astrophys. J. Lett.]
  {10.1086/311269}, 497, L17

\bibitem[\protect\citeauthoryear{Savchenko et~al.}{Savchenko
  et~al.}{2017}]{Savchenko:2017ffs}
Savchenko V.,  et~al., 2017, \mn@doi [Astrophys. J. Lett.]
  {10.3847/2041-8213/aa8f94}, 848, L15

\bibitem[\protect\citeauthoryear{Schneider, Roberts  \& Ott}{Schneider
  et~al.}{2017}]{daSilvaSchneider:2017jpg}
Schneider A.~S.,  Roberts L.~F.,   Ott C.~D.,  2017, \mn@doi [Phys. Rev.]
  {10.1103/PhysRevC.96.065802}, C96, 065802

\bibitem[\protect\citeauthoryear{Sedov}{Sedov}{1959}]{Sedov:1959}
Sedov L.~I.,  1959, {Similarity and Dimensional Methods in Mechanics}

\bibitem[\protect\citeauthoryear{Sekiguchi, Kiuchi, Kyutoku  \&
  Shibata}{Sekiguchi et~al.}{2015}]{Sekiguchi:2015dma}
Sekiguchi Y.,  Kiuchi K.,  Kyutoku K.,   Shibata M.,  2015, \mn@doi [Phys.Rev.]
  {10.1103/PhysRevD.91.064059}, D91, 064059

\bibitem[\protect\citeauthoryear{Sekiguchi, Kiuchi, Kyutoku, Shibata  \&
  Taniguchi}{Sekiguchi et~al.}{2016}]{Sekiguchi:2016bjd}
Sekiguchi Y.,  Kiuchi K.,  Kyutoku K.,  Shibata M.,   Taniguchi K.,  2016,
  \mn@doi [Phys. Rev.] {10.1103/PhysRevD.93.124046}, D93, 124046

\bibitem[\protect\citeauthoryear{Selina et~al.,}{Selina
  et~al.}{2018}]{Selina:2018vla}
Selina R.~J.,  et~al., 2018, ] {10.1117/12.2312089}, \href
  {https://ui.adsabs.harvard.edu/abs/2018SPIE10700E..1OS} {10700, 107001O}

\bibitem[\protect\citeauthoryear{Shapiro}{Shapiro}{1980}]{Shapiro:1980}
Shapiro P.~R.,  1980, \mn@doi [The Astrophysical Journal] {10.1086/157823},
  \href {https://ui.adsabs.harvard.edu/abs/1980ApJ...236..958S} {236, 958}

\bibitem[\protect\citeauthoryear{Shibata \& Hotokezaka}{Shibata \&
  Hotokezaka}{2019}]{Shibata:2019wef}
Shibata M.,  Hotokezaka K.,  2019, \mn@doi [Ann. Rev. Nucl. Part. Sci.]
  {10.1146/annurev-nucl-101918-023625}, 69, 41

\bibitem[\protect\citeauthoryear{Shibata, Fujibayashi, Hotokezaka, Kiuchi,
  Kyutoku, Sekiguchi  \& Tanaka}{Shibata et~al.}{2017}]{Shibata:2017xdx}
Shibata M.,  Fujibayashi S.,  Hotokezaka K.,  Kiuchi K.,  Kyutoku K.,
  Sekiguchi Y.,   Tanaka M.,  2017, \mn@doi [Phys. Rev.]
  {10.1103/PhysRevD.96.123012}, D96, 123012

\bibitem[\protect\citeauthoryear{Siegel}{Siegel}{2019}]{Siegel:2019mlp}
Siegel D.~M.,  2019, \mn@doi [Eur. Phys. J. A] {10.1140/epja/i2019-12888-9},
  55, 203

\bibitem[\protect\citeauthoryear{Siegel \& Metzger}{Siegel \&
  Metzger}{2017}]{Siegel:2017nub}
Siegel D.~M.,  Metzger B.~D.,  2017, \mn@doi [Phys. Rev. Lett.]
  {10.1103/PhysRevLett.119.231102}, 119, 231102

\bibitem[\protect\citeauthoryear{Sironi \& Giannios}{Sironi \&
  Giannios}{2013}]{Sironi:2013tva}
Sironi L.,  Giannios D.,  2013, \mn@doi [Astrophys. J.]
  {10.1088/0004-637X/778/2/107}, 778, 107

\bibitem[\protect\citeauthoryear{Sironi \& Spitkovsky}{Sironi \&
  Spitkovsky}{2009}]{Sironi:2009}
Sironi L.,  Spitkovsky A.,  2009, \mn@doi [The Astrophysical Journal]
  {10.1088/0004-637X/698/2/1523}, \href
  {https://ui.adsabs.harvard.edu/abs/2009ApJ...698.1523S} {698, 1523}

\bibitem[\protect\citeauthoryear{Sironi \& Spitkovsky}{Sironi \&
  Spitkovsky}{2011}]{Sironi:2011}
Sironi L.,  Spitkovsky A.,  2011, \mn@doi [The Astrophysical Journal]
  {10.1088/0004-637X/726/2/75}, \href
  {https://ui.adsabs.harvard.edu/abs/2011ApJ...726...75S} {726, 75}

\bibitem[\protect\citeauthoryear{Sironi, Keshet  \& Lemoine}{Sironi
  et~al.}{2015}]{Sironi:2015oza}
Sironi L.,  Keshet U.,   Lemoine M.,  2015, \mn@doi [Space Sci. Rev.]
  {10.1007/s11214-015-0181-8}, 191, 519

\bibitem[\protect\citeauthoryear{Smartt et~al.}{Smartt
  et~al.}{2017}]{Smartt:2017fuw}
Smartt S.~J.,  et~al., 2017, \mn@doi [Nature] {10.1038/nature24303}

\bibitem[\protect\citeauthoryear{Soares-Santos et~al.}{Soares-Santos
  et~al.}{2017}]{Soares-santos:2017lru}
Soares-Santos M.,  et~al., 2017, \mn@doi [Astrophys. J.]
  {10.3847/2041-8213/aa9059}, 848, L16

\bibitem[\protect\citeauthoryear{Spitkovsky}{Spitkovsky}{2008}]{Spitkovsky:2008fi}
Spitkovsky A.,  2008, \mn@doi [Astrophys. J. Lett.] {10.1086/590248}, 682, L5

\bibitem[\protect\citeauthoryear{Steiner, Hempel  \& Fischer}{Steiner
  et~al.}{2013}]{Steiner:2012rk}
Steiner A.~W.,  Hempel M.,   Fischer T.,  2013, \mn@doi [Astrophys. J.]
  {10.1088/0004-637X/774/1/17}, 774, 17

\bibitem[\protect\citeauthoryear{Takahashi \& Ioka}{Takahashi \&
  Ioka}{2021}]{Takahashi:2020jcc}
Takahashi K.,  Ioka K.,  2021, \mn@doi [Mon. Not. Roy. Astron. Soc.]
  {10.1093/mnras/stab032}, 501, 5746

\bibitem[\protect\citeauthoryear{Tanaka et~al.}{Tanaka
  et~al.}{2017}]{Tanaka:2017qxj}
Tanaka M.,  et~al., 2017, \mn@doi [Publ. Astron. Soc. Jap.]
  {10.1093/pasj/psx121}

\bibitem[\protect\citeauthoryear{Tanvir, Levan, Fruchter, Hjorth, Wiersema,
  Tunnicliffe  \& de Ugarte~Postigo}{Tanvir et~al.}{2013}]{Tanvir:2013pia}
Tanvir N.,  Levan A.,  Fruchter A.,  Hjorth J.,  Wiersema K.,  Tunnicliffe R.,
   de Ugarte~Postigo A.,  2013, \mn@doi [Nature] {10.1038/nature12505}, 500,
  547

\bibitem[\protect\citeauthoryear{Tanvir et~al.}{Tanvir
  et~al.}{2017}]{Tanvir:2017pws}
Tanvir N.~R.,  et~al., 2017, \mn@doi [Astrophys. J.]
  {10.3847/2041-8213/aa90b6}, 848, L27

\bibitem[\protect\citeauthoryear{Tomita \& Ohira}{Tomita \&
  Ohira}{2016}]{Tomita:2016yib}
Tomita S.,  Ohira Y.,  2016, \mn@doi [Astrophys. J.]
  {10.3847/0004-637X/825/2/103}, 825, 103

\bibitem[\protect\citeauthoryear{Troja et~al.}{Troja
  et~al.}{2017}]{Troja:2017nqp}
Troja E.,  et~al., 2017, \mn@doi [Nature] {10.1038/nature24290}

\bibitem[\protect\citeauthoryear{Troja et~al.}{Troja
  et~al.}{2018}]{Troja:2018ybt}
Troja E.,  et~al., 2018, \mn@doi [Nature Commun.] {10.1038/s41467-018-06558-7},
  9, 4089

\bibitem[\protect\citeauthoryear{Typel, Ropke, Klahn, Blaschke  \&
  Wolter}{Typel et~al.}{2010}]{Typel:2009sy}
Typel S.,  Ropke G.,  Klahn T.,  Blaschke D.,   Wolter H.~H.,  2010, \mn@doi
  [Phys. Rev.] {10.1103/PhysRevC.81.015803}, C81, 015803

\bibitem[\protect\citeauthoryear{Uhm \& Beloborodov}{Uhm \&
  Beloborodov}{2006}]{Uhm:2006qk}
Uhm Z.,  Beloborodov A.~M.,  2006, \mn@doi [AIP Conf. Proc.]
  {10.1063/1.2207887}, 836, 189

\bibitem[\protect\citeauthoryear{Virtanen et~al.,}{Virtanen
  et~al.}{2020}]{scipy}
Virtanen P.,  et~al., 2020, \mn@doi [Nature Methods]
  {10.1038/s41592-019-0686-2}, \href {https://rdcu.be/b08Wh} {17, 261}

\bibitem[\protect\citeauthoryear{Wanajo, Sekiguchi, Nishimura, Kiuchi, Kyutoku
  \& Shibata}{Wanajo et~al.}{2014}]{Wanajo:2014wha}
Wanajo S.,  Sekiguchi Y.,  Nishimura N.,  Kiuchi K.,  Kyutoku K.,   Shibata M.,
   2014, \mn@doi [Astrophys. J.] {10.1088/2041-8205/789/2/L39}, 789, L39

\bibitem[\protect\citeauthoryear{Warren, Barkov, Ito, Nagataki  \&
  Laskar}{Warren et~al.}{2018}]{Warren:2018lyx}
Warren D.~C.,  Barkov M.~V.,  Ito H.,  Nagataki S.,   Laskar T.,  2018, \mn@doi
  [Mon. Not. Roy. Astron. Soc.] {10.1093/mnras/sty2138}, 480, 4060

\bibitem[\protect\citeauthoryear{Wei \& Jin}{Wei \& Jin}{2003}]{Wei:2002gc}
Wei D.-m.,  Jin Z.~P.,  2003, \mn@doi [Astron. Astrophys.]
  {10.1051/0004-6361:20030007}, 400, 415

\bibitem[\protect\citeauthoryear{Wijers \& Galama}{Wijers \&
  Galama}{1999}]{Wijers:1998st}
Wijers R.,  Galama T.,  1999, \mn@doi [Astrophys. J.] {10.1086/307705}, 523,
  177

\bibitem[\protect\citeauthoryear{Winkler, Diehl, Ubertini  \& Wilms}{Winkler
  et~al.}{2011}]{Winkler:2011}
Winkler C.,  Diehl R.,  Ubertini P.,   Wilms J.,  2011, \mn@doi [Space Science
  Reviews] {10.1007/s11214-011-9846-0}, \href
  {https://ui.adsabs.harvard.edu/abs/2011SSRv..161..149W} {161, 149}

\bibitem[\protect\citeauthoryear{Wu, Fern\'{a}ndez, Martínez-Pinedo  \&
  Metzger}{Wu et~al.}{2016}]{Wu:2016pnw}
Wu M.-R.,  Fern\'{a}ndez R.,  Martínez-Pinedo G.,   Metzger B.~D.,  2016,
  \mn@doi [Mon. Not. Roy. Astron. Soc.] {10.1093/mnras/stw2156}, 463, 2323

\bibitem[\protect\citeauthoryear{Xie, Zrake  \& MacFadyen}{Xie
  et~al.}{2018}]{Xie:2018vya}
Xie X.,  Zrake J.,   MacFadyen A.,  2018, \mn@doi [Astrophys. J.]
  {10.3847/1538-4357/aacf9c}, 863, 58

\bibitem[\protect\citeauthoryear{Yang et~al.,}{Yang
  et~al.}{2015}]{Yang:2015pha}
Yang B.,  et~al., 2015, \mn@doi [Nature Commun.] {10.1038/ncomms8323}, 6, 7323

\bibitem[\protect\citeauthoryear{Zhang}{Zhang}{2018}]{Zhang:2018book}
Zhang B.,  2018, {The Physics of Gamma-Ray Bursts},
  \mn@doi{10.1017/9781139226530.
}

\bibitem[\protect\citeauthoryear{Zhang \& Meszaros}{Zhang \&
  Meszaros}{2001}]{Zhang:2000wx}
Zhang B.,  Meszaros P.,  2001, \mn@doi [Astrophys. J. Lett.] {10.1086/320255},
  552, L35

\bibitem[\protect\citeauthoryear{Zhang \& Meszaros}{Zhang \&
  Meszaros}{2002}]{Zhang:2002jt}
Zhang B.,  Meszaros P.,  2002, \mn@doi [Astrophys. J.] {10.1086/344338}, 581,
  1236

\bibitem[\protect\citeauthoryear{Zhang \& Shu}{Zhang \& Shu}{2011}]{Zhang:2011}
Zhang X.,  Shu C.-W.,  2011, \mn@doi [Proceedings of The Royal Society A:
  Mathematical, Physical and Engineering Sciences] {10.1098/rspa.2011.0153},
  467

\bibitem[\protect\citeauthoryear{van Eerten, Leventis, Meliani, Wijers  \&
  Keppens}{van Eerten et~al.}{2010}]{vanEerten:2009pa}
van Eerten H.,  Leventis K.,  Meliani Z.,  Wijers R.,   Keppens R.,  2010,
  \mn@doi [Mon. Not. Roy. Astron. Soc.] {10.1111/j.1365-2966.2009.16109.x},
  403, 300

\makeatother
\end{thebibliography}

\appendix

\section{Synchrotron spectrum approximants} \label{sec:app:synch}

\begin{figure}
    \centering 
    \includegraphics[width=0.49\textwidth]{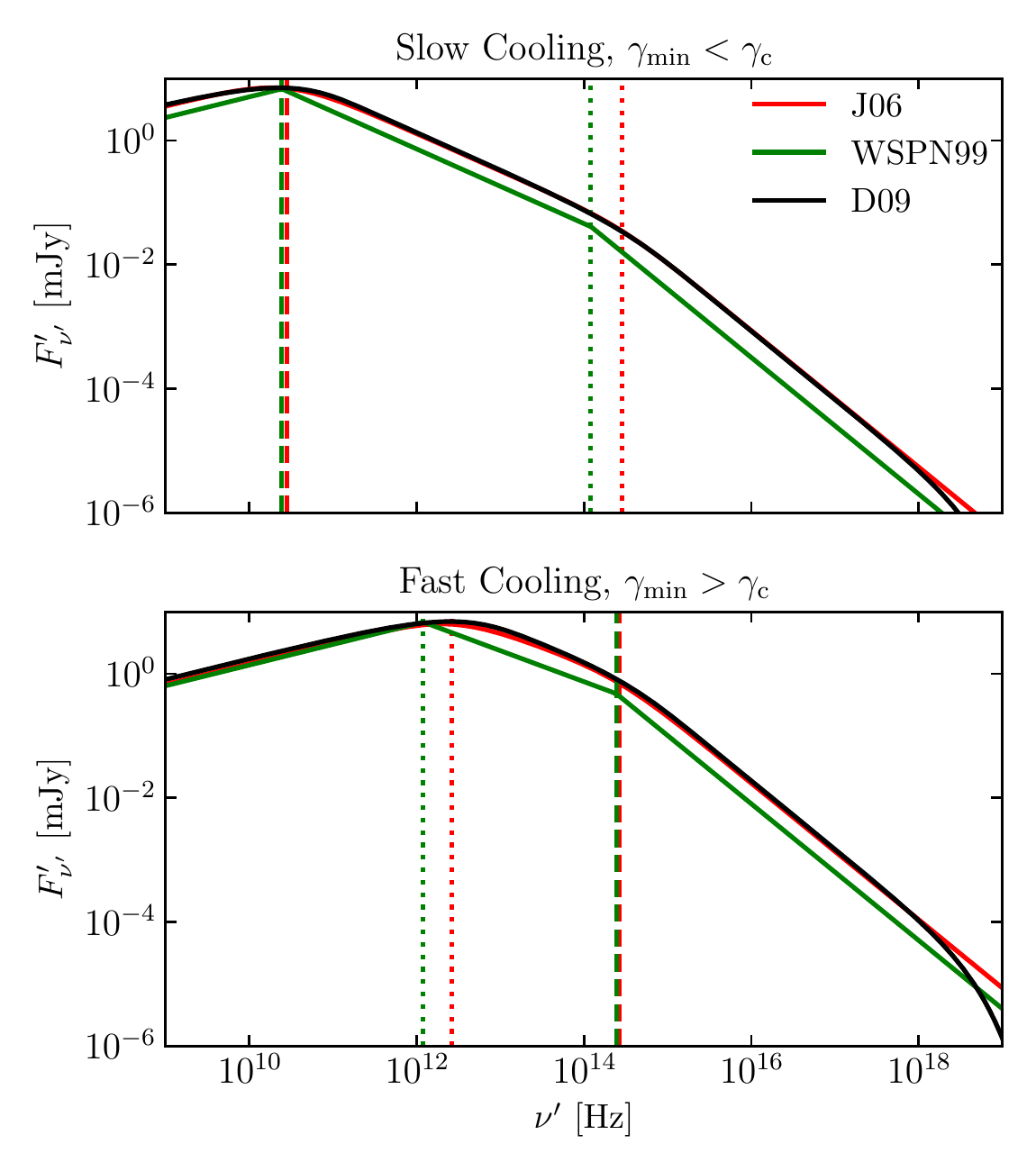}
    \caption{
        Comparison between several approximants to the 
        synchrotron radiation from power-law distribution of 
        electrons with $p=2.2$, $\gamma_{e;\, \rm min}'=10^2$, $\gamma_{e;\,\rm c}'=10^4$ 
        in the magnetic fields $B=1\,$G. The emitting region has radius 
        $R=10^{12}\,$cm, mass $m_2=10^{20}\,$g, moving through the \ac{ISM} 
        with number density $n_{\rm ISM}=10^{-1}\,\ccm$.
        We compare the \ac{BPL} approximants from \citet{Johannesson:2006zs}
        (red line); from \citet{Sari:1997qe} and \citet{Wijers:1998st} 
        (green line); and numeric integration of the approximated 
        synchrotron function from \citet{Dermer:2009} (black line).
    }
    \label{fig:app:synchrotron}
\end{figure}

Several approximants to the synchrotron emission from a power 
law distribution of electrons exist in the literature. 
In the main text we focused on the formulation proposed by 
\citet{Johannesson:2006zs} for \ac{GRB} afterglows, that we 
label as J06 in this section. 
To motivate this choice we compare this formulation with 
widely used model by \citet{Sari:1997qe} and with more 
direct integration of a synchrotron function 
\cite{RybickiLightman:1985} given in \citet{Dermer:2009}.
In the former, the comoving emissivity is given as 
\begin{equation}
    \label{eq:add:syn:sari_sed_sc}
    j_{\rm pl}'(\nu') = j_{\rm pl;\, max}'
    \begin{cases}
        \Big(\frac{\nu'}{\nu_{\rm min}'})^{1/3} \text{ if } \nu' < \nu_{\rm min}' \, , \\
        \Big(\frac{\nu'}{\nu_{\rm min}'}\Big)^{(1-p)/2} \text{ if } 
        \nu_{\rm min}' < \nu' < \nu_{\rm c}'\, , \\
        %        \nu' > \nu_{\rm min}' \:\&\: \nu' < \nu_{\rm c}'\, , \\
        \Big(\frac{\nu'}{\nu_{\rm min}'}\Big)^{(1-p)/2}\Big(\frac{\nu'}{\nu_{\rm c}'}\Big)^{-p/2} \text{ if } \nu' > \nu_{\rm c}'\, ,
    \end{cases}
\end{equation}
in the slow cooling regime and 
%and in case of fast cooling, \textit{e.g.,} where $\nu_{min}' > \nu_c'$,
%
\begin{equation}
    \label{eq:add:syn:sari_sed_fc}
    j_{\rm pl}'(\nu') = j_{\rm pl;\, max}'
    \begin{cases}
        \Big(\frac{\nu'}{\nu_{\rm c}'})^{1/3} \text{ if } \nu' < \nu_{\rm c}'\, , \\
        \Big(\frac{\nu'}{\nu_{\rm c}'}\Big)^{-1/2} \text{ if } 
        \nu_{\rm c}' < \nu' < \nu_{\rm min}' \, , \\
        %        \nu' > \nu_{\rm c}' \:\:\&\:\: \nu' < \nu_{\rm min}' \, , \\
        \Big(\frac{\nu_{\rm c}'}{\nu_{\rm min}'}\Big)^{-1/2}\Big(\frac{\nu'}{\nu_{\rm c}'}\Big)^{-p/2} \text{ if } \nu' > \nu_{\rm min}'\, , 
    \end{cases}
\end{equation}
in the fast cooling regime. 
In the calculation of the spectral breaks and 
$j'_{\rm pl;\,max}$ the integration over the emission 
angle has to be included which gives a correction 
factor of $3/4\pi$ \cite{Wijers:1998st}. 
Then, the spectral breaks read
\begin{equation}
    \nu_{\rm min}' = \chi_p \frac{3}{4\pi} \gamma_{e;\,\rm min}^2 \frac{q_e B'}{m_e c}\, ,
\end{equation}
and
\begin{equation}
    \nu_{\rm c} ' = 0.286 \frac{3}{4\pi} \gamma_{e;\,\rm c}^2 \frac{q_e B'}{m_e c}\, .
\end{equation}
The maximum of the spectrum is 
\begin{equation}
    j_{\rm pl;\, max}' = \phi_p \sqrt{3} \frac{q_{e}^3 B'}{m_e c^2} \, ,
\end{equation}
where $\chi_p$ and $\phi_p$ are electron spectrum 
dimensionless maximum and corresponding dimensionless flux. 
They account for the isotropic distribution 
of angles between the electron velocity and the magnetic field. %\red{Check.}
They are tabulated in \citet{Wijers:1998st}. 
%\todo{Complete this by using Lamb et al 18 paper}
%
We label this formulation as WSPN99 in Fig.~\ref{fig:app:synchrotron}.

In \citet{Dermer:2009} approximations to modified Bessel 
functions are provided for a more numerically efficient 
calculation of a synchrotron emission from an arbitrary 
electron distribution. 
We consider the \ac{BPL} electron distribution, 
\begin{equation}
    \begin{aligned}
        n_e(\gamma') =& k_e \Bigg[
        \left(\frac{\gamma_e'}{\gamma'_{e;\,\rm c}}\right)^{-p_1} \, H(\gamma_e'; \gamma'_{e;\,\rm min}, \gamma'_{e;\,\rm c}) \\
        &+ \left(\frac{\gamma_e'}{\gamma'_{e;\,\rm c}}\right)^{-p_2} \, H(\gamma_e'; \gamma'_{e;\,\rm c}, \gamma'_{e;\,\rm max}) 
        \Bigg]\, ,
    \end{aligned}
\end{equation}
where $k_e$ is the spectral normalization, $H(...)$ 
is the Heaviside step function, $p_1 = p$ if 
$\gamma_{e;\,\rm min}' < \gamma_{e;\,\rm c}'$ 
and $p_1 = 2$ otherwise, $p_2 = p+1$, accounting for 
the slow and fast cooling regimes respectively. 
The angle-averaged integrand of the radiated power, 
$R(x)$, is approximated with 
equation D7 of \citet{Aharonian:2010} where the 
ratio of the frequency to the critical synchrotron frequency, 
$x$, is computed with equation 7.34 in \citet{Dermer:2009}. 
We label this formulation as D09 in 
Fig.~\ref{fig:app:synchrotron} and 
consider it as a reference point.

Comparing the spectra, we observe that while the spectral 
peaks and slopes in different regimes are captured by the 
analytic approximants, WSPN99 and D09, the value of the 
flux density $F_{\nu'}'$ between the spectral breaks is 
generally underestimated by the WSPN99 formulation. 
In \texttt{PyBlastAfterglow}, where radiation from a large 
number of \acp{BW} in combined to obtain the observed flux, 
this might lead to lower fluxes.
Meanwhile, the spectra produced by J06 formulation is 
in a good agreement with the reference, especially in 
the slow cooling regime which is of prime 
importance for this work. 
Thus, due to the high computational efficiency of 
analytic methods, we consider J06 formulation in the main text.

\section{Blastwave dynamics approximants} \label{sec:app:dyn}

\begin{figure}
    \centering 
    \includegraphics[width=0.49\textwidth]{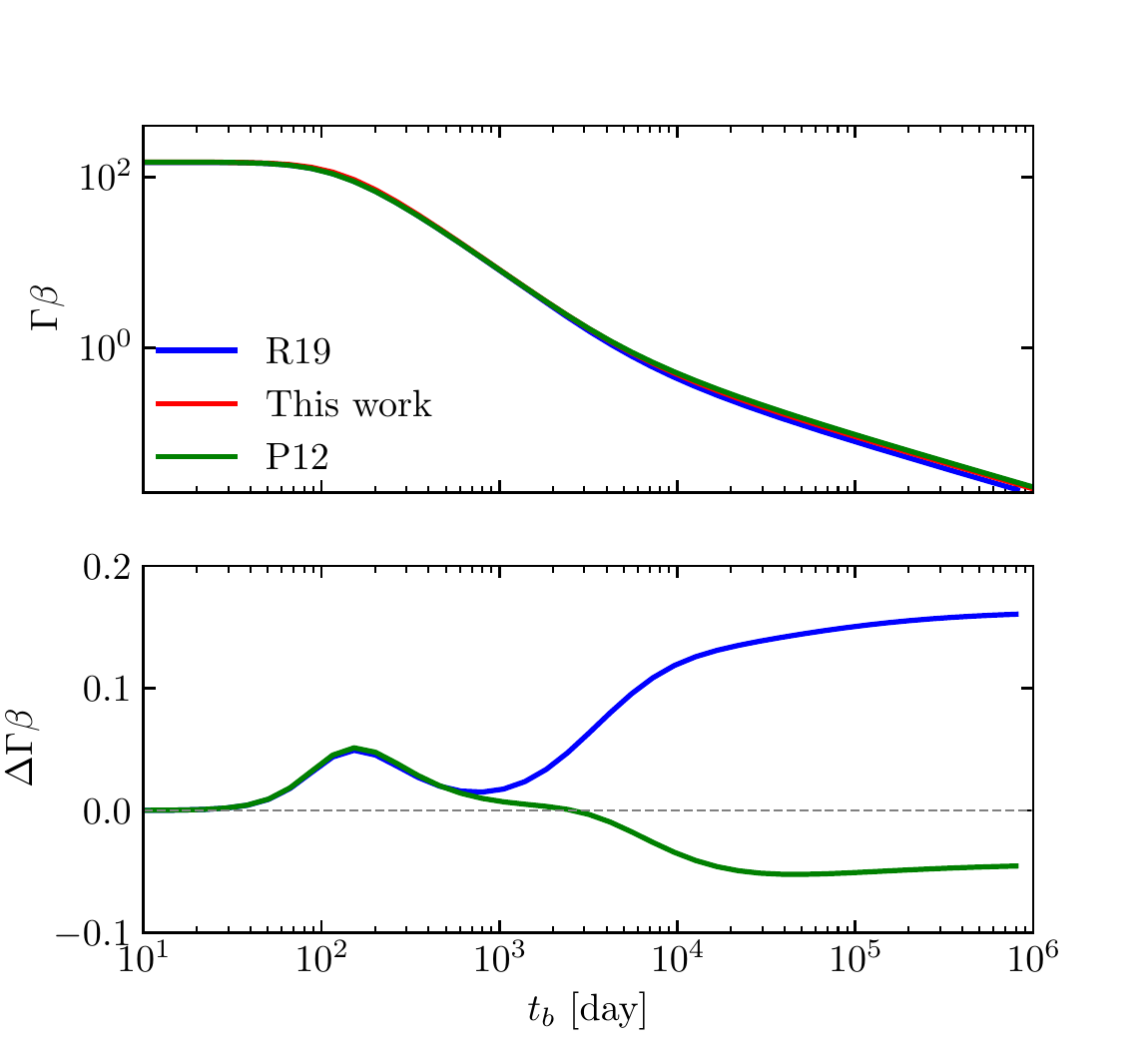}
    \caption{
        Dynamical evolution of a \ac{BW} with 
        $\Gamma_0=150$, $E_{0} = 10^{52}\,$erg, and 
        half-opening angle, $\theta_0=0.1\,$rad 
        propagating through $n_{\rm ISM} = 10^{-3}\,\ccm$. 
        The red line indicates to the evolution computed with \texttt{PyBlastAfterglow}, 
        Eq.~\eqref{eq:method:dGdRold}.
        The while blue line corresponds to the model of \citet{Ryan:2019fhz}. 
        The green line denotes the formulation of \citet{Peer:2012}. 
        The relative difference is shown in the bottom panel. 
    }
    \label{fig:app:dynamics}
\end{figure}

There are several formulations for the dynamics of a 
transrelativistc \ac{BW} propagating through a cold 
\ac{ISM} under the ``thin-shell'' approximation in the literature. 
It is instructive to compare the evolution of a \ac{BW} 
computed with \texttt{PyBlastAfterglow} with other 
formulations in the literature. 
First, we consider the formulation proposed by \citet{Peer:2012}, 
where the adiabatic losses are neglected % \cite{Huang:1999di,Peer:2012} 
which we label here P12. 
The evolution equation for the bulk \ac{LF} for P12 reads
\begin{equation}\label{eq:app:dgdm_peer}
    \frac{d\Gamma}{dm_2} = \frac{-(\hat{\gamma} (\Gamma^2 - 1) - (\hat{\gamma} - 1) \Gamma \beta^2)}{M_0 + m_2 (2 \hat{\gamma} \Gamma - (\hat{\gamma} - 1) (1 + \Gamma^{-2}))}\, ,
\end{equation}
where $M_0$ is the initial mass of the fireball 
and $m_2$ is the swept-up mass.
The adiabatic index, $\hat{\gamma}$, is computed with the same, 
Eq.~\eqref{eq:method:eos}, as in \texttt{PyBlastAfterglow}. 
%We label this formulation \Peer{} in Fig.~\ref{fig:app:dynamics}.

Additionally, we consider the formulation proposed by 
\citet{Ryan:2019fhz} that is implemented in the publicly 
available code \texttt{afterglowpy}. 
There, the \ac{EOS} is the ``TM'' variant presented 
in \citet{Mignone:2005ns}. We label this formulation as 
R19 in the Fig.~\eqref{fig:app:dynamics}.

Overall, the evolution of a \ac{BW} consists of three stages:  
free-coasting, deceleration in the \acl{BM} regime and deceleration 
in the \acl{ST} regime. 
Both the R19 and the P12 formulations display these stages 
and show an overall good agreement with \texttt{PyBlastAfterglow} 
at early times. 
At late times, however, there is a small discrepancy, 
given primarily by the different \ac{EOS}, when comparing with R19 
and different treatment of the internal energy transformation 
when comparing with the P12 formulation.

\section{Blastwave lateral expansion approximants} \label{sec:app:spread}

\begin{figure}
    \centering 
    \includegraphics[width=0.49\textwidth]{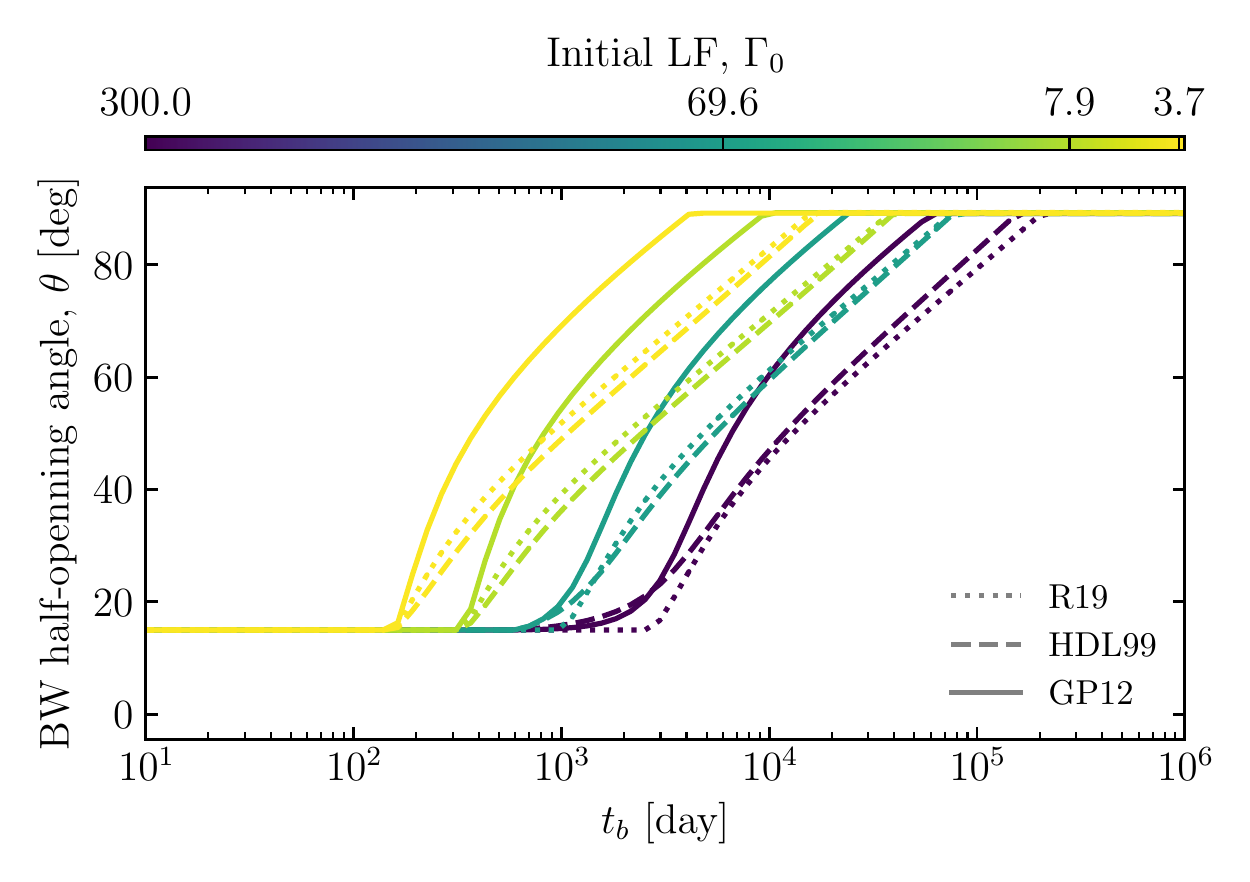}
    \includegraphics[width=0.49\textwidth]{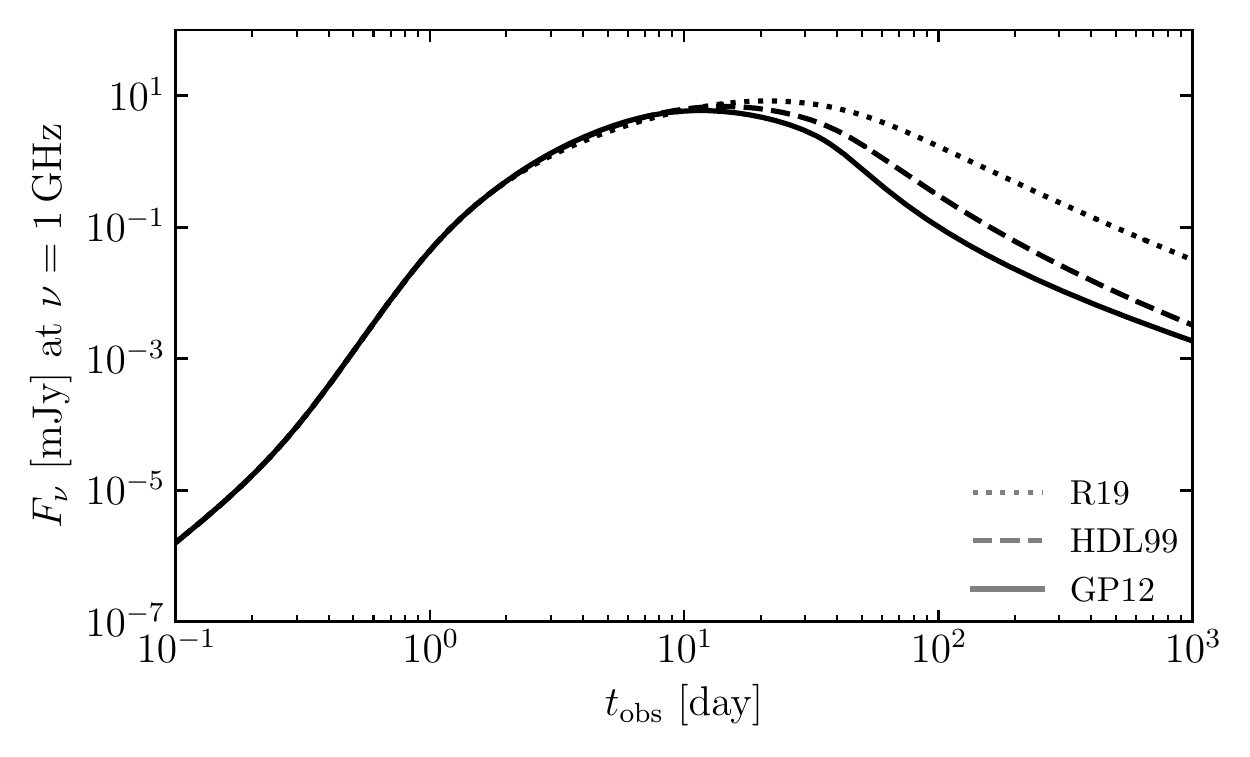}
    \caption{
        \textit{Top panel}: evolution of the \ac{BW} half-opening angle 
        for several initial \acp{LF} (color-coded).
        Several lateral expansion prescriptions are considered. 
        The solid line denotes $d\omega/dR$ from \citet{Granot:2012}, 
        The dashed line denotes the model of \cite{Huang:1999qa} and 
        the dotted line corresponds to the prescription from \cite{Ryan:2019fhz}. 
        The color of the line indicate the initial \ac{LF} of the \ac{BW}. 
        \textit{Bottom panel}: radio \acp{LC} for a top-hat jet 
        observed off-axis, $\theta_{\rm obs}=0.16$, 
        for the three aforementioned lateral 
        spreading prescriptions. Geometry and microphsyics 
        of the \ac{GRB} model are discussed in Sec.~\ref{sec:app:comp_grb}. 
    }
    \label{fig:app:spread}
\end{figure}

In most semi-analytic models of the \ac{BW} evolution 
that employ the thin-shell approximation, lateral spreading 
cannot be incorporated in a self-consistent way 
(see, however \citet{Lu:2020hka}). 
Here we compare the lateral spreading prescription from \citet{Granot:2012}, 
the default option in \texttt{PyBlastAfterglow}, with other 
prescriptions available in the literature (and implemented in 
\texttt{PyBlastAfterglow}).

Lateral expansion is determined by the co-moving sound speed, 
$c_s^2 = dp'/de'|_s$, at a shock \cite{Kirk:1999km}
\begin{equation}\label{eq:app:cs_shock}
    c_{s}^2 = \frac{\hat{\gamma}p'}{\rho'}\Big[ \frac{(\hat{\gamma}-1)\rho'}{(\hat{\gamma}-1)\rho' + \hat{\gamma}\rho'} \Big] c^2 = 
    \frac{\hat{\gamma} (\hat{\gamma} - 1) (\Gamma - 1)}{1 + \hat{\gamma} (\Gamma - 1)} c^2\, ,
\end{equation}
where in the last equation we expressed $\hat{\gamma}$ through the 
\ac{EOS}, Eq.~\eqref{eq:method:eos}. 

Assuming that the expanding fluid element interacts only with its  
immediate vicinity, the lateral and radial components of the velocity 
are related as $\beta_{r}/\beta_{\omega}=\partial \omega / \partial\ln R $.  
Furthermore, assuming that the spreading proceeds at the sound speed, 
$\upsilon_{\omega} = c_s$, the lateral expansion can be written as 
\cite{Huang:1999qa} 
\begin{equation} \label{eq:method:dthetadr_hang}
    \frac{d\omega}{dR} = \frac{\upsilon_{\omega}}{R \Gamma \beta c}\, .
\end{equation}
This formulation, labeled as HDL99, has been broadly used in the early 
semi-analytic \ac{GRB} afterglow models \citep[\eg][]{Rossi:2004dz}.

More recently, \citet{Ryan:2019fhz} proposed a ``conical'' 
spreading model, where at a given time, all material that has 
been swept up affects the spreading. 
The tangential component of the velocity then reads  
\begin{equation}
    \upsilon_{\perp} = c \Gamma \sqrt{(1 - \beta\beta_{\rm sh})c_2^2 - (\beta_{\rm sh}-\beta)^2}
\end{equation}
where both $c_s^2$ and $\beta_{\rm sh} = \dot{R_{\rm sh}}$ 
are evaluated using the ``TM'' \ac{EOS} \cite{Mignone:2005ns}. 
The spreading is allowed once 
$\Gamma\beta > 1/(3\sqrt{2}\omega_{\rm c})$, 
where $\omega_{\rm c}$ is the half-opening angle of the jet core. 
The spreading is given as $\dot{\omega}=\upsilon_{\perp}/R$. 
We label this prescription as R19.

The result of the comparison is shown in Fig.~\ref{fig:app:spread} for 
several \ac{GRB} layers with different initial \acp{LF}, $\Gamma_0$ 
(\textit{top panel}). The difference in \acp{LC} for an off-axis 
top-hat jet discussin in the next section, Sec.~\ref{sec:app:comp_grb}, 
are shown in the bottom panel of the figure.
For the largest $\Gamma_0$, for all prescriptions, the lateral spreading 
starts smoothly when the \ac{BW} enters mildly relativistic regime.  
For low values of $\Gamma_0$, however, the onset of spreading is sharp, 
as sound speed is relatively low. %most of the requirements for it are fulfilled from the start. 

The subsequent evolution of the \ac{BW} 
half-opening angle proceeds similar for the HDL99 
and the R19 formulations. Notably, we did not use the 
final equation for $d\omega/dR$ from \citet{Ryan:2019fhz}, 
as it implicitly assumes the ``TM'' \ac{EOS}, that is different 
from the one adopted here. 
Moreover, the formulation designed in that work is tailored to the 
specific structured model and jet discretization, 
that differs considerably from the one used in 
\texttt{PyBlastAfterglow}. This contributes to the large 
difference in radio \acp{LC}. 
The lateral spreading computed with GP12 formulation proceeds  
faster. Fast spreading has been observed in the 
number of numerical studies of jet spreading 
\cite{Xie:2018vya,vanEerten:2009pa,Granot:2012,Duffell:2018iig}. 
It results in a reduced late time emission, as the faster 
spreading leads to larger accreted mass and earlier \ac{BW} 
deceleration. 
As this formulation has been used in semi-analytic models with 
similar jet structure and discretization as ours 
\cite{Fernandez:2021xce}, we employ it as a default option. 
Additionally, we find that qualitative results discussed 
in the main text do not depend on the exact formulation of the 
lateral spreading.

\section{\ac{GRB} afterglow comparison with \texttt{afterglowpy}} \label{sec:app:comp_grb}

\begin{figure}
    \centering 
    \includegraphics[width=0.49\textwidth]{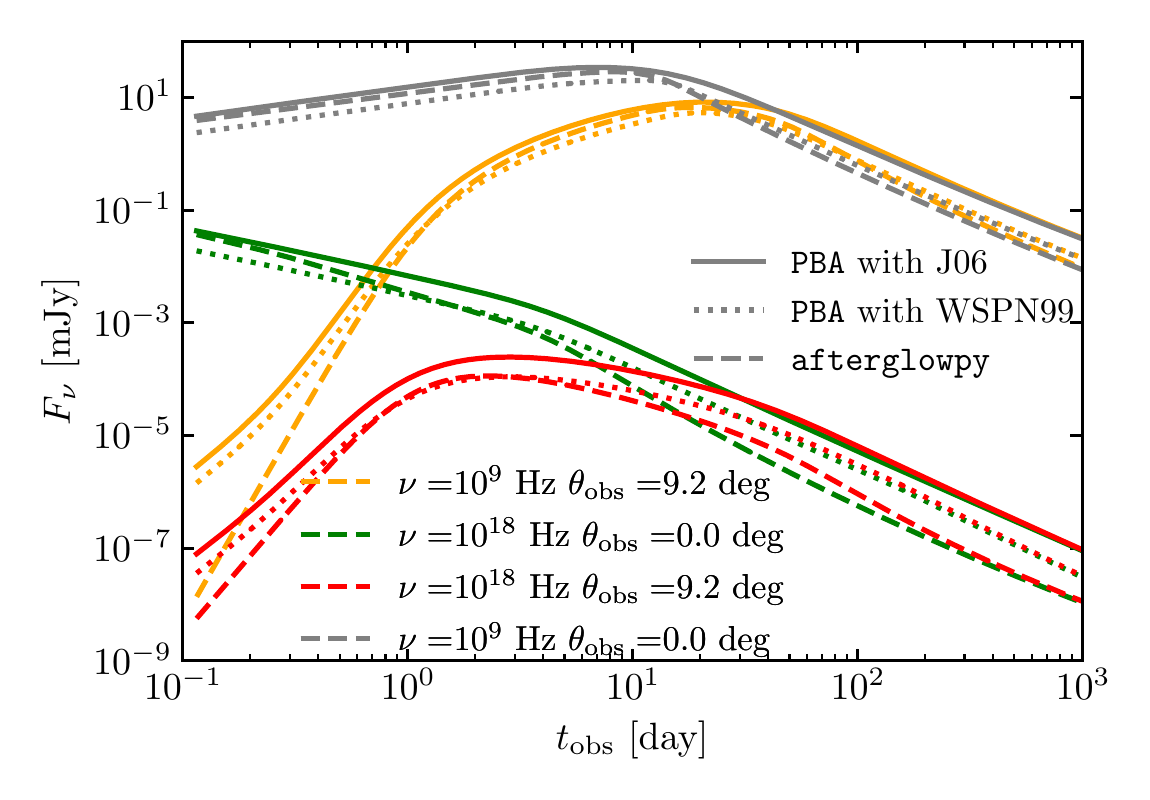}
    \caption{
        Comparison between \acp{LC} from a top-hat jet 
        between \texttt{PyBlastAfterglow} with two different synchrotron 
        radiation approximations (sold and dotted lines) and 
        \texttt{afterglowpy}. This is analogous to the figure~2 of 
        \citet{Ryan:2019fhz}. 
    }
    \label{fig:app:lcs}
\end{figure}

\begin{figure}
    \centering 
    \includegraphics[width=0.49\textwidth]{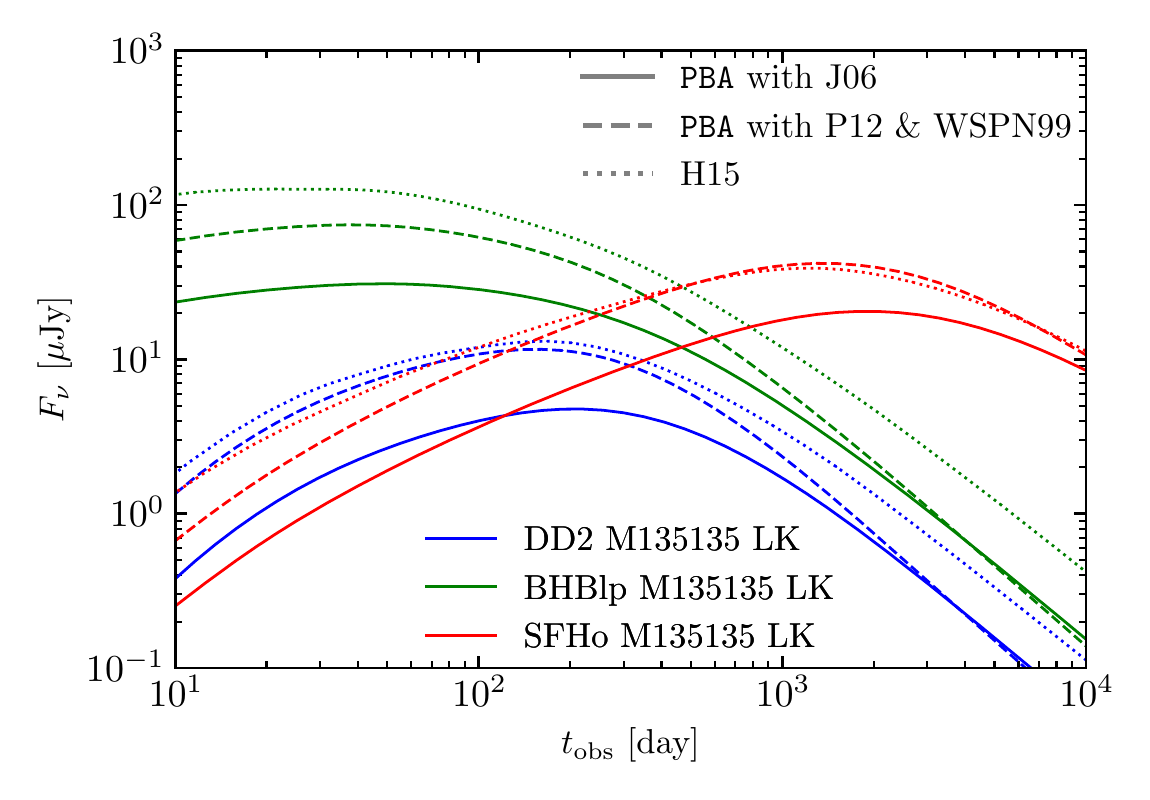}
    \caption{
        Comparison between \ac{kN} afterglow \acp{LC}, computed with 
        \texttt{PyBlastAfterglow} 
        (using two different input physics settings, denoted with the 
        solid and dashed lines) and the afterglow code of \citet{Hotokezaka:2015eja}. 
        The ejecta profiles from three \ac{NR} simulations, presented in 
        \citet{Radice:2018pdn} (see their figures 30 and 31) were used. 
    }
    \label{fig:app:lcs_ejecta}
\end{figure}

Here we compare the \ac{GRB} afterglow \acp{LC} generated with \texttt{PyBlastAfterglow} and those computed with \texttt{afterglowpy}. 
As in the latter the analytic synchrotron radiation formulation 
of \citet{Sari:1997qe} was is, we compare \acp{LC} computed using 
the WSPN99 and the J06 formulations (see Sec.~\ref{sec:app:synch}) 
separately.
%(dotted and solid lines in Fig.~\ref{fig:app:lcs}).
%
The \ac{GRB} parameters are: 
$\Gamma_0 =150$, 
$E_{\rm iso}=10^{52}\,$ergs, 
$\theta_{\rm w} = 0.1\,$rad, 
$n_{\rm ISM}=10^{-3}\,\ccm$, 
$\epsilon_e=0.1$, $\epsilon_b=0.001$, $p=2.2$, 
$d_{\rm L}=3.09\times10^{26}\,$cm,
and $z = 0.028$.  

The result is shown in Fig.~\ref{fig:app:lcs}. 
Overall we find a reasonably good agreement between the 
\acp{LC} produced with \texttt{PyBlastAfterglow} and 
\texttt{afterglowpy}. The differences stem largely from 
different \ac{EATS} integration methods. Especially, 
at early times, as the \ac{GRB} is observed off-axis. 
At late times the differences in dynamics 
formulations (see Sec.~\ref{sec:app:dyn}) also contribute.

\section{Method comparison for kN afterglow} \label{sec:app:comp_kn}

In this section we compare \ac{kN} afterglow \acp{LC} 
computed with \texttt{PyBlastAfterglow} and 
with the code of \citet{Hotokezaka:2015eja}.
We label the latter as H15.
Specifically, we consider ejecta profiles from three \ac{NR} \ac{BNS} 
merger simulations, described in \citet{Radice:2018pdn}, 
the radio \ac{LC} for which are shown in 
figures 30 and 31 in that work. 
The data for these simulations is publicly available\footnote{
    Data is available on Zenodo: \url{https://doi.org/10.5281/zenodo.3588344}. 
}.
To the best of our knowledge, this is the first direct comparison 
between two different models for \ac{kN} afterglows. Although, these 
models are semi-analytic and approximate, such comparisons are 
necessary in order to assess systematic uncertainties. 
However, no detailed information regarding the \ac{BW} dynamics 
formulation and \ac{EATS} integration procedure are available in 
\citet{Hotokezaka:2015eja}. 
%Thus, we can only assume where the differences between our results may stem from. 
%
Comparing the radio \acp{LC}, shown in Fig.~\ref{fig:app:lcs_ejecta}, 
we observe that the overall \ac{LC} shape and the time of the peak 
are well reproduced by \texttt{PyBlastAfterglow}. 
This implies that the dynamics of different ejecta elements 
is similarly modelled.  
However, \acp{LC} computed with \texttt{PyBlastAfterglow} 
are systematically dimmer, especially if the J06 formulation 
for synchrotron radiation is used. 
The best agreement is found when the P12 formulation for dynamics 
(see Sec.~\ref{sec:app:dyn}), and the WSPN99 formulations for radiation 
(see Sec.~\ref{sec:app:synch}) are used.  
The remaining discrepancy may stem from 
different \ac{EATS} integration methods.

\end{document}